\documentclass[prl,aps,floats,nofootinbib,showpacs,superscriptaddress,twocolumn]{revtex4-1}

\usepackage{amsmath,amssymb}
\usepackage{dsfont,verbatim,varwidth}
\usepackage{graphicx,epsfig,epstopdf}
\usepackage{amsopn,amsthm,sidecap,enumerate}
\usepackage[usenames,dvipsnames]{color}
\usepackage{subfiles}

\def\ua{{\uparrow}}
\def\da{{\downarrow}}
\def\beq{\begin{equation}}
\def\eeq{\end{equation}}
\def\beqr{\begin{eqnarray}}
\def\eeqr{\end{eqnarray}}
\def\bq{{\vec q}}

\def\br{{\vec r}}

\def\mrho{{\mathbf\rho}}

\def\tw{{\tilde w}}
\def\tEc{{\tilde E_c}}

\begin{document}

\title{Spin Mode-Switching at the Edge of a Quantum Hall System}
\author{Udit Khanna} 
\affiliation{\protect\begin{varwidth}[t]{\linewidth} {\fontsize{9}{10.2}\selectfont {Harish-Chandra Research Institute, Chhatnag Road, Jhunsi, Allahabad 211019, India}}\protect\end{varwidth}}
\affiliation{\protect\begin{varwidth}[t]{\linewidth} {\fontsize{9}{10.2}\selectfont {Homi Bhabha National Institute, Training School Complex, Anushaktinagar, Mumbai, Maharastra 400085, India}}\protect\end{varwidth}}
\author{Ganpathy Murthy}
\affiliation{\protect\begin{varwidth}[t]{\linewidth} {\fontsize{9}{10.2}\selectfont {Department of Physics and Astronomy, University of Kentucky, Lexington KY 40506-0055, USA}}\protect\end{varwidth}}
\author{Sumathi Rao}
\affiliation{\protect\begin{varwidth}[t]{\linewidth} {\fontsize{9}{10.2}\selectfont {Harish-Chandra Research Institute, Chhatnag Road, Jhunsi, Allahabad 211019, India}}\protect\end{varwidth}}
\affiliation{\protect\begin{varwidth}[t]{\linewidth} {\fontsize{9}{10.2}\selectfont {Homi Bhabha National Institute, Training School Complex, Anushaktinagar, Mumbai, Maharastra 400085, India}}\protect\end{varwidth}}
\author{Yuval Gefen} 
\affiliation{\protect\begin{varwidth}[t]{\linewidth} {\fontsize{9}{10.2}\selectfont {Department of Condensed Matter Physics, Weizmann Institute, 76100 Rehovot, Israel}}\protect\end{varwidth}}

\date{\today}

\begin{abstract}
Quantum Hall states can be characterized by their chiral edge
modes. Upon softening the edge potential, the edge has long been known
to undergo spontaneous reconstruction driven by charging effects. In
this paper we demonstrate a qualitatively distinct phenomenon driven
by exchange effects, in which the {\it ordering} of the edge modes at
$\nu=3$ switches abruptly as the edge potential is made softer, while
the ordering in the bulk remains intact. We demonstrate that this
phenomenon is robust, and has many verifiable experimental signatures
in transport.

\end{abstract}
%73.21.-b: Electron states and collective excitations in multilayers, quantum wells, mesoscopic, and nanoscale systems; 73.22.Gk: Broken symmetry phases;
%73.43.Lp: Collective excitations (in context of QHE); 72.80.Vp: Electronic transport in graphene; 75.10.Pq: Spin chain models;
%75.10.Jm: Quantized spin models, including quantum spin frustration
\maketitle

Shortly after the discovery of the integer quantum Hall effect (QHE)
it was realized that the edges of an incompressible electron gas play
a crucial role in transport \cite{halperin82}. In a quantum Hall
state, the bulk has a charge gap. Near a sharp edge, gapless chiral
modes (described as chiral Luttinger liquids \cite{wen-chiral-LL})
carry the current between the contacts, consistent with the
topologically protected transport observables of the QHE.

In the early 90s it was realized that both integer
 \cite{chklovskii,dempsey,chamon-wen} and fractional
 \cite{macdonald90,meir} edges reconstruct as the slope of the edge
confining potential $V_{\text{edge}}(y)$ is made smoother. Reconstruction is
the modification of the position and/or the number and nature of the
edge modes \cite{chklovskii,dempsey,chamon-wen,macdonald90,meir,kane-fisher-polchinski,kane-fisher}. Subsequently,
various manifestations of edge reconstruction have been observed in
the QHE regime
 \cite{klein-chamon-wind,zhitenev-melloch,venkatachalam,grivnin-heiblum,sabo-heiblum},
and theoretically studied in many QHE states 
 \cite{Wan2002,Yang2003,Zhang2013,kun-confine,kun-quantum-wires,wang-meir-gefen-13}
and in time reversal invariant topological insulators
 \cite{meir-gefen}.

%The physics driving 
Edge reconstruction is driven by charge effects
\cite{footnote-charge}, as seen by the work of Dempsey {\it et al.}
\cite{dempsey} who studied the unpolarized filling factor $\nu = 2$.
For a sharp edge, the $n=(0\ua)$ and the $n=(0\da)$
single-particle levels cross the chemical potential $\mu$ at the same
location, with a sharp change in electron density there. As
$V_{\text{edge}}(y)$ is made smoother, the $\ua$ and $\da$ crossing points
spontaneously move away from each other. The occupations now go from
$\nu=2\to\nu=1\to\nu=0$ as one moves towards the edge, resulting in a
smoother change in electron density, which is better able to
neutralize the positive background.

In this work we focus on the edge of a $\nu=3$ quantum Hall state and
uncover edge phenomena driven by spin exchange rather than charge
effects \cite{footnote-charge}. The bulk remains inert at the 
parameters we consider and only the edge shows a phase transition. 
We find that, depending on parameters,
the order of the two inner or the two outer edge channels switches as
$V_{\text{edge}}(y)$ becomes smoother. The charge density does not change
significantly through the transition; no charge reconstruction is
observed in the regime where {\it spin-mode-switching} occurs. Our
(approximate) theoretical analyses indicate that the phase transitions
are first-order. In designed geometries with controlled edge steepness
and quantum point contacts (QPCs), a host of phenomena can serve as
``smoking gun'' tests of spin-mode-switching. These include a change
in the nature of the spin transport through a single QPC system with and
without spin-mode-switching, and a qualitative change in the way
disorder affects transport following a spin-mode-switching transition.

To set
the stage for our model, we define the cyclotron
energy $\hbar\omega_c=\frac{\hbar eB}{m_b}$ ($m_b$ is the band mass), 
the interaction scale $E_c=\hbar\omega_c \tEc=\frac{e^2}{4\pi\epsilon \ell}$ 
where $\epsilon$ includes the dielectric constant of the 
  medium, and $\ell=\sqrt{\hbar/eB}$, the magnetic length.  We will work at tiny Zeeman
coupling $E_z \ll E_c$.

%--------------- Fig 1  ----------
\begin{figure*}[t]
 \includegraphics[width=0.329\textwidth]{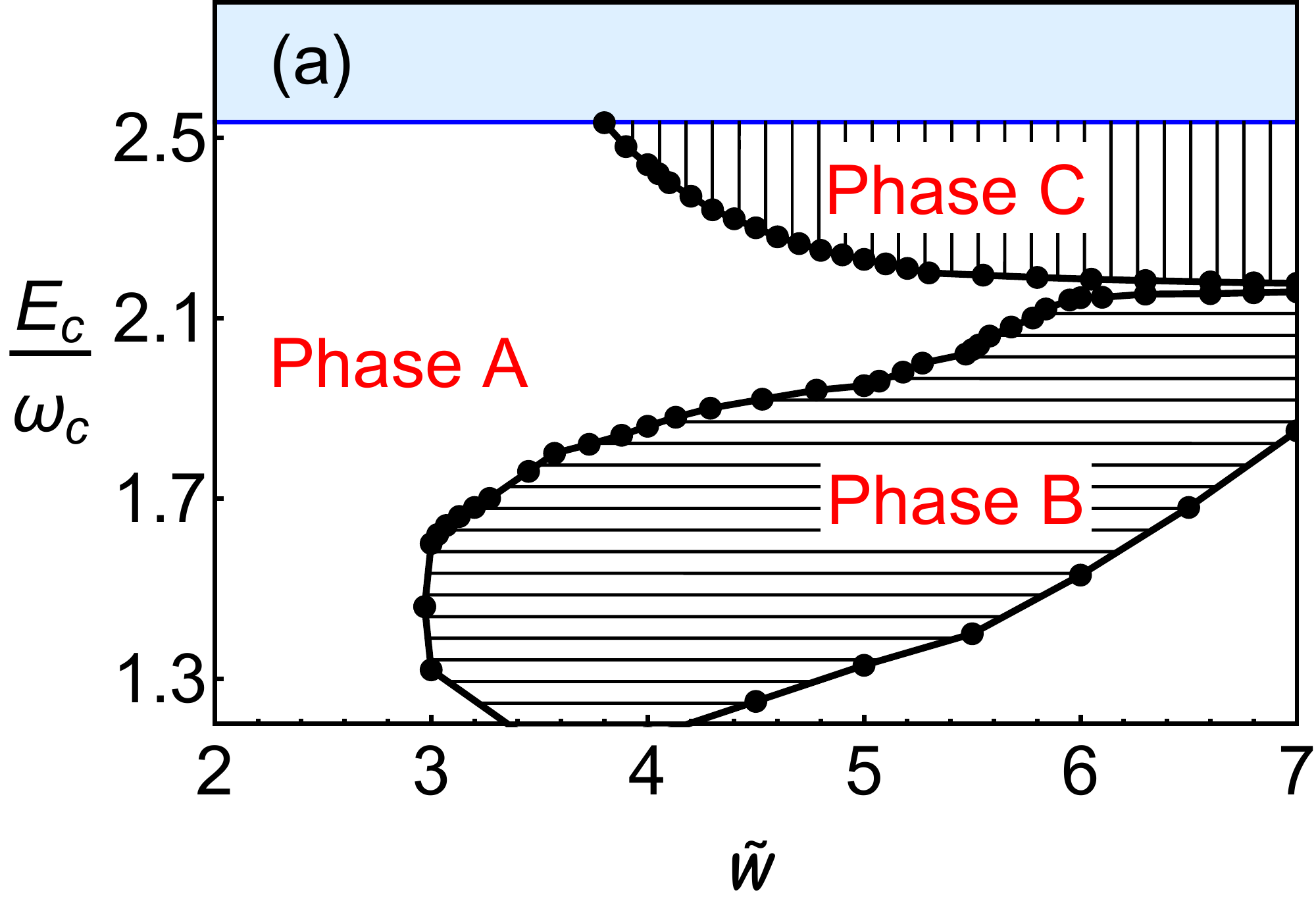}
  \includegraphics[width=0.329\textwidth]{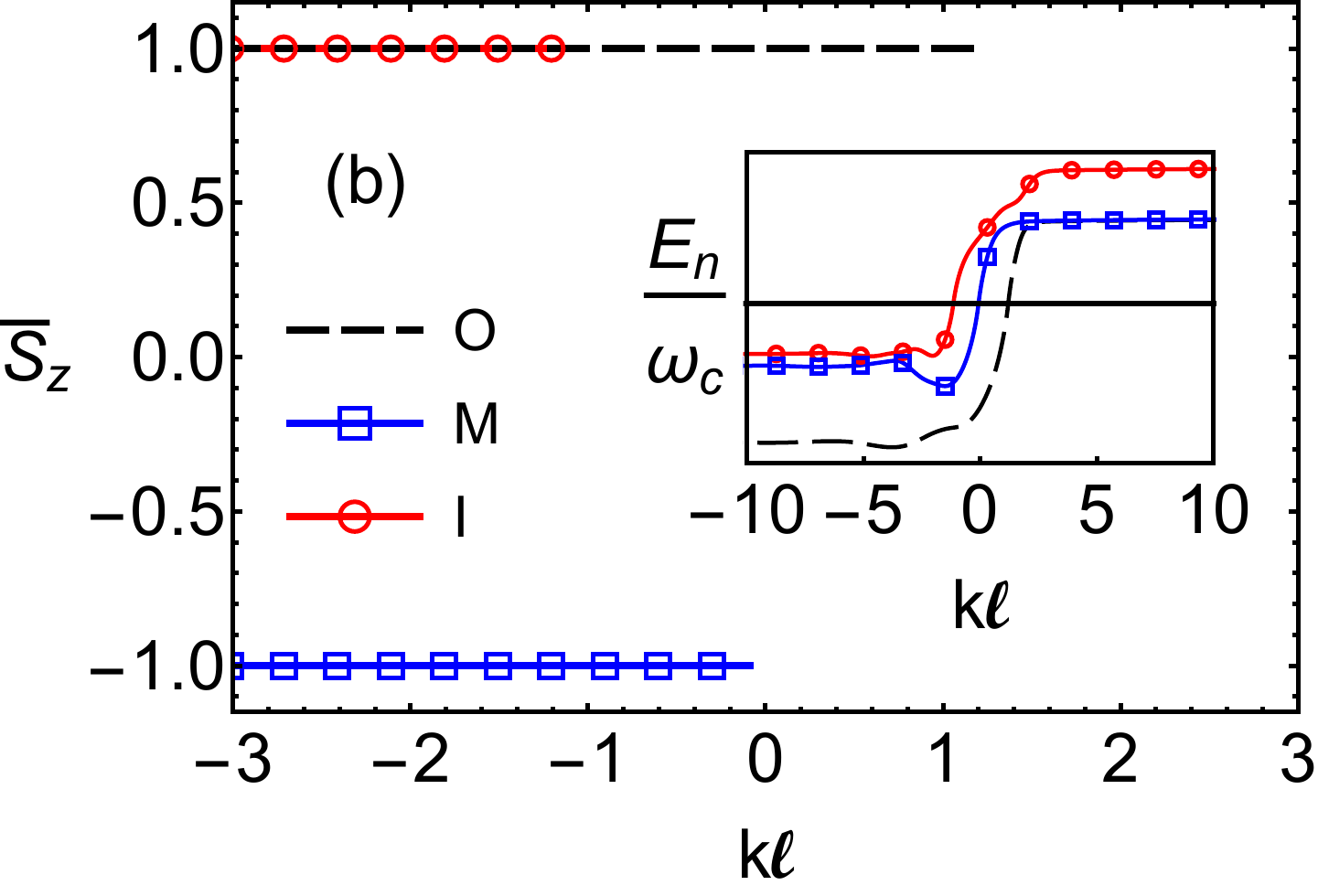}
   \includegraphics[width=0.329\textwidth]{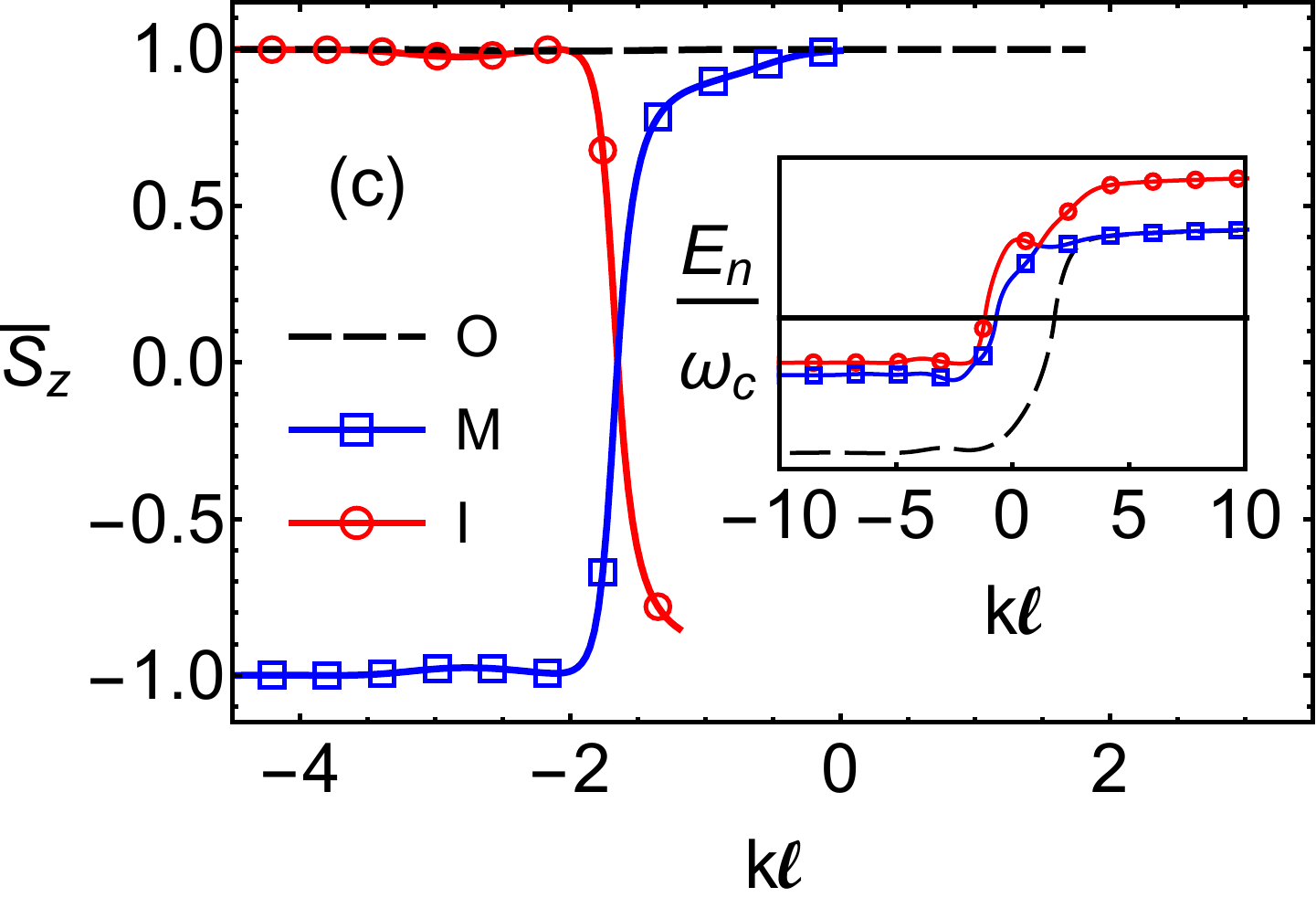}
    \includegraphics[width=0.329\textwidth]{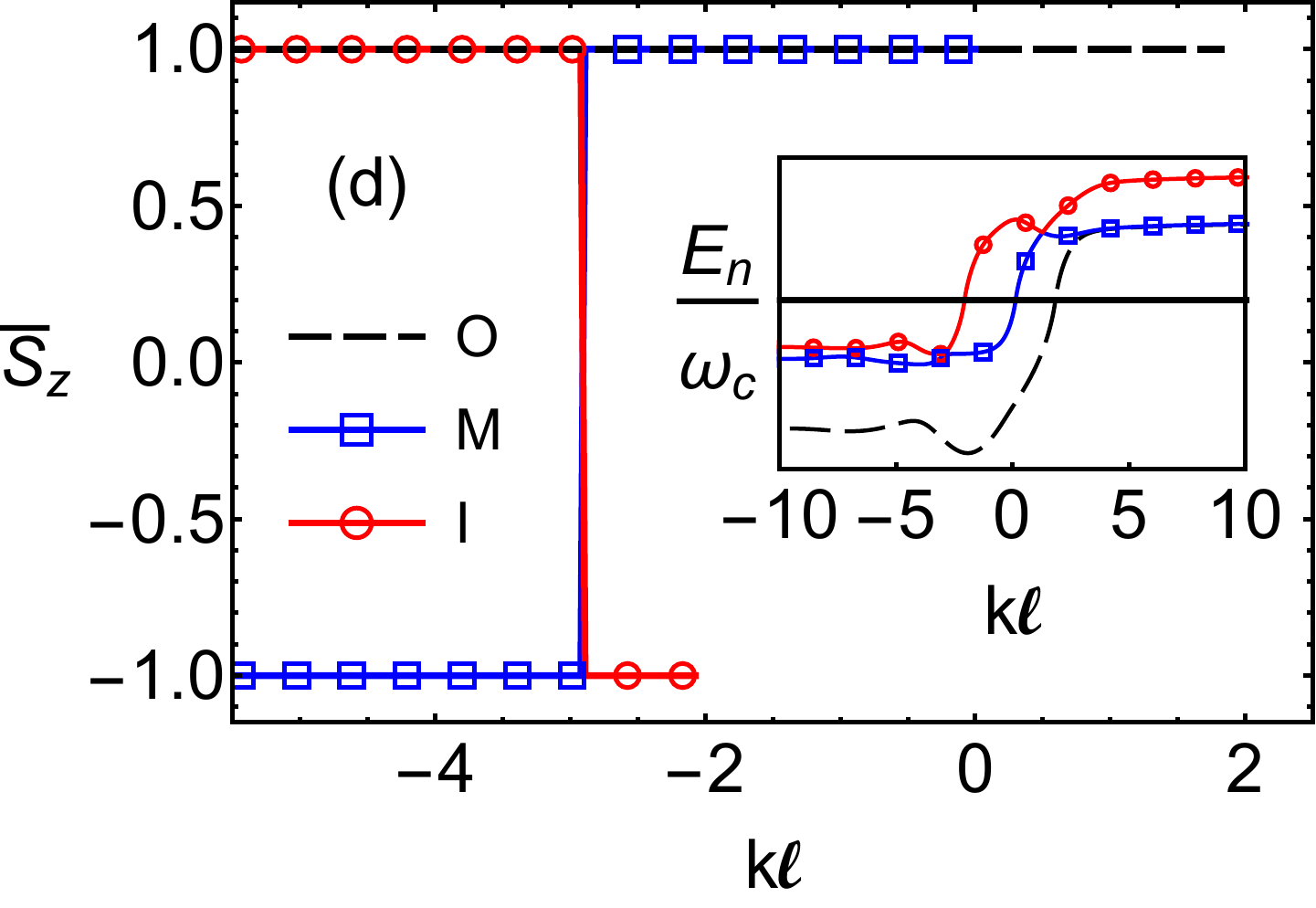}
    \includegraphics[width=0.329\textwidth]{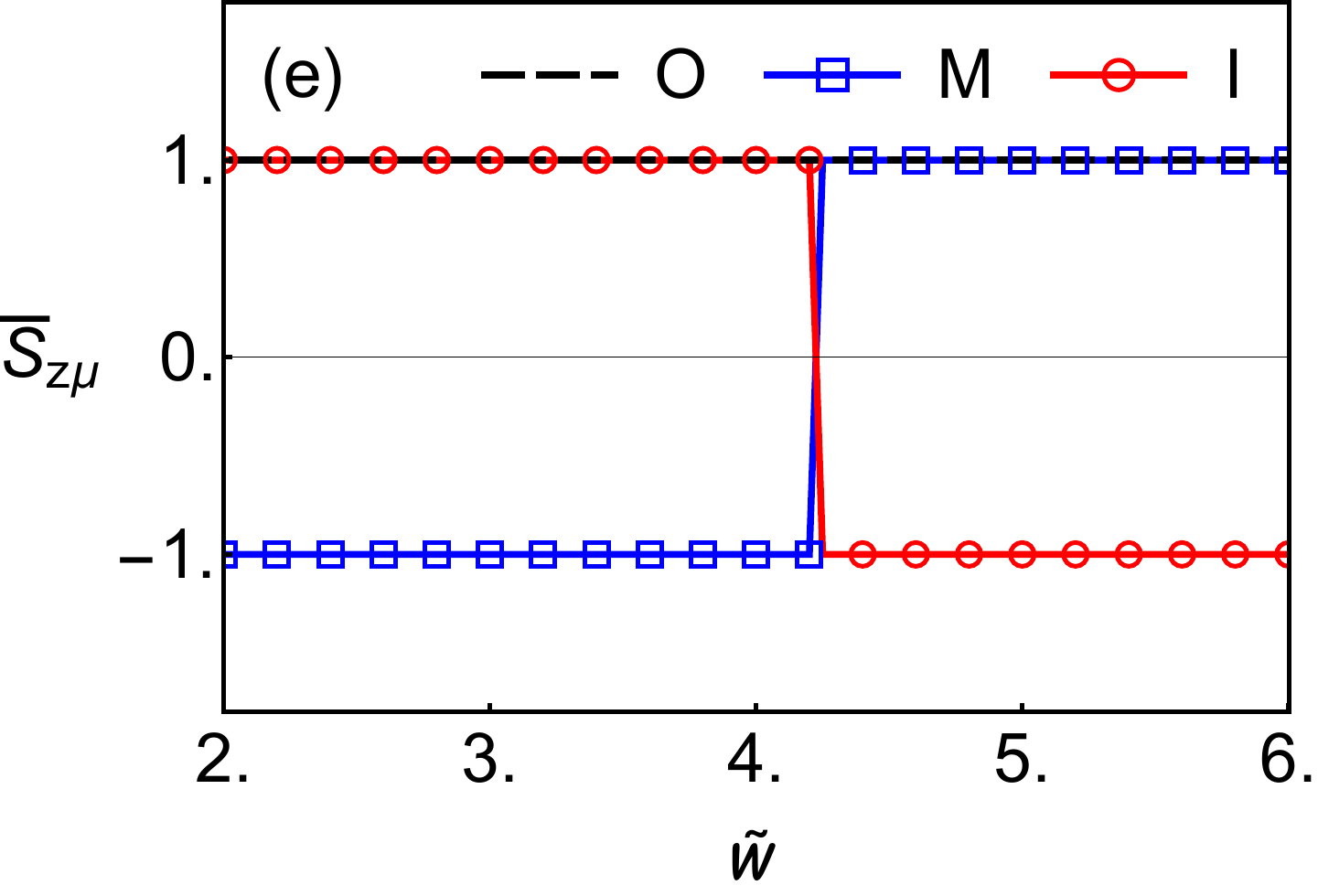}
    \includegraphics[width=0.329\textwidth]{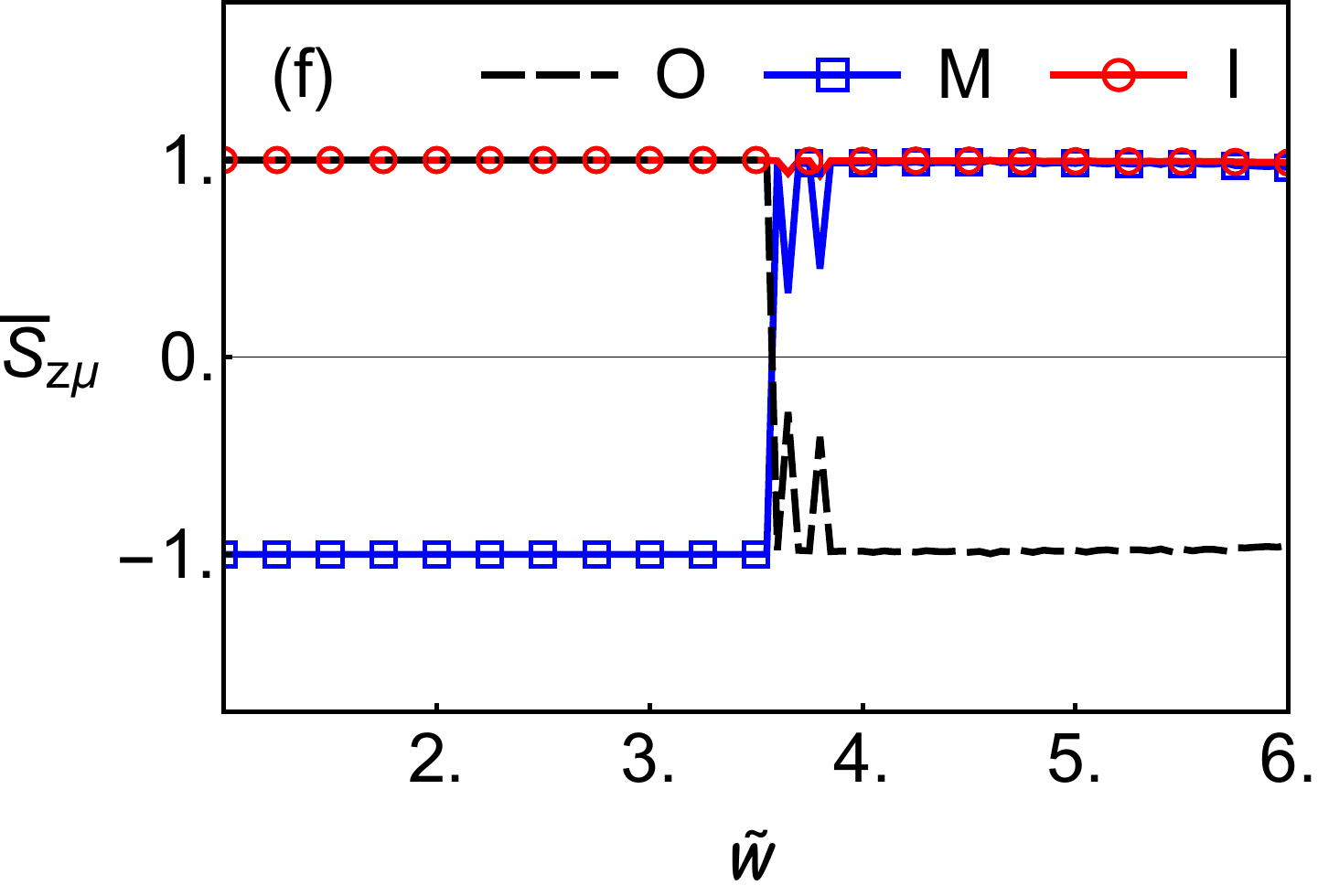}
  \caption{(Color online) (a) The phase diagram. The background color
    represents the bulk phase, white being partially polarized
    ($|0\ua,0\da,1\ua\rangle$) and blue being fully polarized
    ($|0\ua,1\ua,2\ua\rangle$). 
    States are labelled $i=$ O (outermost), M (middle) and I (innermost).   
    The plain white region also denotes
    edge Phase A (O=$0\ua$, M=$0\da$ and I=$1\ua$). 
    Edge Phase B (O=$0\da$, M=$0\ua$ and I=$1\ua$) is horizontally
    hatched, while Phase C (O=$0\ua$, M=$1\ua$ and I=$0\da$) 
    is vertically hatched. Due to poor convergence of the HF, 
    for $6\leq \tw\leq 7$, $\tEc\simeq2.13$ it is not
    clear whether there is a direct transition between Phases B and C,
    or whether Phase A intervenes. In Figs.~\ref{fig1}(b),~\ref{fig1}(c) and~\ref{fig1}(d) we
    depict $\bar{S}_z(i,k)$ of the occupied single-particle states
    versus $k\ell$ at $\tEc=2.3$. Only occupied
    levels are depicted. The line for level $i$ terminates where the
    level $i$ crosses $\mu$, with $\bar{S}_z(i,k)$ at the $\mu$-crossing
    defined as $\bar{S}_{z\mu}(i)$. The insets depict the energy
    dispersions of the HF single-particle states vs. $k\ell$, with the
    horizontal black line being $\mu$. In Fig.~\ref{fig1}(b) we are in Phase A
    ($\tw=2.0$) where no spin rotations occur. Fig.~\ref{fig1}(c) shows
    $\bar{S}_z(i,k)$ vs. $k\ell$ at the transition ($\tw=4.28$), with
    spin rotations occurring over a scale $\ell$, where the
    corresponding energy level dispersions come close together in an
    avoided crossing (inset). Fig.~\ref{fig1}(d) shows $\bar{S}_z(i,k)$
    vs. $k\ell$ in Phase C ($\tw=5.0$). The spin rotations are quite
    abrupt, and occur where the corresponding dispersions undergo a
    sharp avoided crossing. In Fig.~\ref{fig1}(e) we plot $\bar{S}_{z\mu}(i)$
    vs. $\tw$ at $\tEc=2.3$. A discontinuous change in
    $\bar{S}_{z\mu}$ for the M and I levels is seen at the 
    transition between Phases A and C.  Similar results hold at $\tEc=1.8$ for the Phase A to
    Phase B transition, with $\bar{S}_{z\mu}$ showing a discontinuous
    change for the O and M levels, as shown in Fig.~\ref{fig1}(f). }
 \label{fig1}
	\end{figure*}
%--------------------------------------- 

In the Landau gauge $eA_x=-\frac{y}{\ell^2},\ eA_y=0$, the
single-particle wavefunctions of the $n^{th}$ Landau level (LL) in a system with
periodic boundary conditions in $x$ can be written as \cite{prange-girvin}
\beq \Phi_{nk}(x,y)=\frac{e^{ikx}e^{-\frac{(y-k\ell^2)^2}{2\ell^2}}}{\sqrt{2^nn!L_x\ell\sqrt{\pi}}} H_n\bigg(\frac{y-k\ell^2}{\ell}\bigg) ,
\label{wavefunction}\eeq
where $k=2\pi m/L_x$ determines the position of the guiding
center along the y-axis. 
The Hamiltonian of the system $H = H_{\text{b}} + H_{\text{e\,-bg}}$ 
can be split into a electronic bulk part $H_{\text{b}}$ and 
the electron-background interaction $H_{\text{e\,-bg}}$ responsible for the 
confining potential $V_{\text{edge}}$ at the edge.
The bulk Hamiltonian is
\beq
H_{\text{b}}=\hbar\omega_c\sum\limits_{nks}nc_{nks}^{\dagger}c_{nks}+ \frac{1}{2L_xL_y}\sum\limits_{\bq}v(q):\mrho_e(\bq)\mrho_e(-\bq):
\eeq
where the electron density operator $\mrho_e(x,y)=\sum\limits_{s}\Psi_s^\dagger(x,y)\Psi_s(x,y)$,
$\Psi_{s}(x,y)=\sum\limits_{n,k}\Phi_{nk}(x,y) c_{nks}$, with $c_{nks}$ being
canonical fermion operators, and $v(q)$ and $\mrho_e(\bq)$ are the
Fourier transforms of the interaction $v(\br-\br')$ and
$\mrho_e(x,y)$.  The possible translation-invariant ground states of
the $\nu=3$ bulk are $|\psi_1\rangle=|0\ua,0\da,1\ua\rangle$
(partially polarized) and $|\psi_2\rangle=|0\ua,1\ua,2\ua\rangle$
(fully polarized), where we write only the spin-labelled LLs that are
occupied.  As $\tEc$ increases there is a bulk first-order transition
 \cite{Jungwirth1,Jungwirth2,polyakov2} driven by exchange from $|\psi_1\rangle$
to $|\psi_2\rangle$. In the Hartree-Fock (HF) approximation this
occurs at $\tEc\approx2.5$ for the Coulomb interaction.

The electron-background interaction is 
\beq
H_{\text{e\,-bg}} = -\int d^2r d^2r' \rho_{b}(y') v(\br-\br')\mrho_e(x,y) 
\eeq
where $\rho_b(y)$ is the positive background density 
which gives rise to the edge confining potential 
$V_{\text{edge}}(y) = -\int d^2r' \rho_{b}(y') v(\br-\br') $.  
In our model the background density decreases linearly to zero
over a distance $W$ at the edge \cite{chamon-wen}. The dimensionless
parameter $\tw=W/\ell$ characterizes the slope of $V_{\text{edge}}$, \beq
%\rho_b(x)=\left\{\begin{array}{cc}
\rho_b(y)=\bigg\{\begin{array}{cc}
                \rho_0 & y<-\frac{W}{2}\\
                \rho_0\frac{\frac{W}{2} - y}{W} & -\frac{W}{2}<y<\frac{W}{2}\\
                0 & y>\frac{W}{2} 
\end{array}.
\eeq

The rest of the paper is devoted to our theoretical evidence for spin
mode-switching and its robustness to mixing with higher LLs, 
Zeeman coupling, and varying the interaction 
parameters.  We also propose a set of charge and spin transport 
experiments to detect spin-mode-switched phases.

{\it Theoretical Analysis}: Our primary tool is the spin-unrestricted
Hartree-Fock (HF) approximation keeping up to 6 spin resolved LLs to
include the effect of LL-mixing and spin-mixing. 
In the HF approximation, the many-body state is replaced by a 
variational Slater determinant, characterized by all 
possible averages $\langle c^{\dagger}_ic_j\rangle$. 
We confine ourselves to translation invariant states:
\beq
\langle c^{\dagger}_{nks}c_{n'k's'}\rangle=\delta_{kk'} \Delta_{ns,n's'}(k).
\eeq
In the bulk the matrix $\Delta_{ns,n's'}$ is independent of $k$ and
diagonal in $n$ as well as in $s$ (no LL-mixing or spin-mixing). Near
the edge $\Delta_{ns,n's'}$ acquires a $k$-dependence, and
LL-mixing/spin-mixing will occur. The optimal Slater determinant
that minimizes the variational energy is found by an iterative procedure
carried out to self-consistency. At each step, a one-body Hamiltonian
(where the interaction term has been replaced by effective one-body
terms dependent on $\Delta_{ns,n's'}(k)$)
is solved and the energy levels filled up to a chemical potential chosen to
satisfy overall charge neutrality. The new state enables the 
computation of a new set of $\Delta$, giving the seed for the next
iterative step \cite{Supplemental}.  
The results of the HF calculation are shown in
Fig.~\ref{fig1}. We use a screened Coulomb interaction of the
form $v(\bq)=\frac{2\pi E_c}{q+q_{sc}}$, where $q_{sc}$ is the inverse
screening length. The results shown are for $q_{sc}\ell=10^{-2}$,
though spin-mode-switching persists at least up to $q_{sc}\ell=0.5$. In
unrestricted HF single-particle levels generically cannot be labelled
by spin and cannot cross due to level repulsion. We therefore label
the edge modes by their location as $i$ = O (outermost), M(middle) and
I(innermost). To proceed further, we compute 
the quantum expectation value $\bar{S}_{z}(i,k)$ for each
occupied single-particle state $i$ at position $k\ell$. The spin
character of the chiral edge modes transporting current are determined
by the $\bar{S}_{z}(i,k)$ of the corresponding single-particle levels at the
crossing with the chemical potential, $\bar{S}_{z\mu}(i)$. This allows us
to label an edge mode with a spin.

Fig.~\ref{fig1}(a) shows two edge-mode-switched
phases. For $\tw\lesssim3$, there is no spin-mixing, and the edges
follow the bulk order: O=$0\ua$, M=$0\da$ and I=$1\ua$. This is Phase
$A$. For $1.5\lesssim\tEc\lesssim2.13$ and $\tw>3$, the system enters
Phase $B$ where the order of the edge modes is O=$0\da$, M=$0\ua$ and
I=$1\ua$. Edge Phase $C$ occurs for $2.13<\tEc<2.5$ and $\tw>3.5$,
with the edge mode ordering O=$0\ua$, M=$1\ua$ and I=$0\da$. For
$6\leq \tw\leq7$, $\tEc\simeq2.13$ HF converges poorly, 
making it unclear whether there is a direct transition between Phases
B and C, or whether a sliver of Phase A persists between them.  Fig.~\ref{fig1}(b)
shows $\bar{S}_z$ vs. $k\ell$ of the three occupied levels near the
edge at $\tEc=2.3$, $\tw=2.0$ (Phase A). The lines terminate where the
corresponding level crosses $\mu$.  Fig.~\ref{fig1}(b) inset shows the energy
dispersions of the self-consistent HF states for the same
parameters. Fig.~\ref{fig1}(c) shows $\bar{S}_z$ vs. $k\ell$ at the
$A\rightarrow C$ transition ($\tw=4.28$), and Fig.~\ref{fig1}(d) shows the same
in Phase C ($\tw=5.0$).  Figs.~\ref{fig1}(e),~\ref{fig1}(f) show the
expectation value $\bar{S}_{z\mu}$ of the respective levels
of O, M and I that intersect the Fermi energy as a function
of $\tw$ at $\tEc=1.8$ and
2.3. The spin characters of O, M and I show discontinuous jumps,
which indicate $1^{st}$-order transitions in our approximation. The
electron charge density hardly varies through the entire regime, and
shows no sign of charge-driven reconstruction \cite{footnote-charge,Supplemental}.

The emergence of mode-switching is quite robust. The phases and phase
transitions are qualitatively unaffected by including 
LL/spin mixing to higher LLs ($n > 2$).
Phases B and C occur over a very broad range of $\tw$, (Phase
$C$ exists at least up to $\tw=11$). Upon increasing the Zeeman
coupling, the bulk phase boundary between the partially and fully
polarized states moves lower in $\tEc$ and edge Phase C encroaches on
edge Phase B. Furthermore, the lower 
boundary between Phase A and Phase B in Fig.~\ref{fig1}(a) moves upwards.  Reducing
the range of the interaction by increasing $q_{sc}\ell$ moves the 
phase boundaries of edge phases B and C towards 
larger $\tw$. Upon independently varying the strength of the
direct ($E_{cd}$) and exchange ($E_{cx}$) terms, we find that
mode-switching occurs in HF only if $E_{cx}>0.6E_{cd}$, consistent
with our claim that this is an exchange effect \cite{footnote-charge}.

One limitation of HF is that the occupation $n(k)$ of a
single-particle state is either $0$ or $1$ (at $T=0$). To get beyond
this limitation we investigated a class of variational states that do
not conserve particle number and allow continuously varying $0\le
n(k)\le 1$. The simplest such state for the $\nu=1$ spin-polarized
edge is $|\psi\rangle=\prod\limits_{k=1}^{N_s}
(u_k+v_ke^{i\theta_k}c^\dagger_k)|0\rangle$.  Here $u_k,\ v_k$ are
real numbers satisfying $u_k^2+v_k^2=1$, and $n(k)=v_k^2$, $\theta_k$
is a set of phases chosen to minimize the
translation-symmetry-breaking inherent in such states. This class
includes HF states. For $\nu=3$ our variational state is \cite{Supplemental}
\begin{widetext}
\begin{align}
  |\psi \rangle &= \prod\limits_{k} (U_k + V_{0 k} e^{i \theta_{0 k}} c^{\dagger}_{0 k \ua} + V_{1k} e^{i \theta_{1 k}} c^{\dagger}_{0 k \da} c^{\dagger}_{0 k \ua} + V_{2k} e^{i \theta_{2 k}} c^{\dagger}_{1 k \ua} c^{\dagger}_{0 k \ua} + V_{3 k} e^{i \theta_{3 k}} c^{\dagger}_{1 k \ua}  c^{\dagger}_{0 k \da} c^{\dagger}_{0 k \ua}) |0 \rangle .
\end{align}
\end{widetext}
When $E_{cx}<0.4E_{cd}$ this ansatz does produce smoothly varying
$n(k)$ at $\nu=3$, with the variational energy lower than the HF
energy. However, upon increasing $E_{cx}$ we recover
the HF solution, lending further support to the validity of the latter 
(and to the transition being $1^{st}$ order).

%--------------- Fig 2  ----------
\begin{figure*}[t]
 \includegraphics[width=0.3\textwidth]{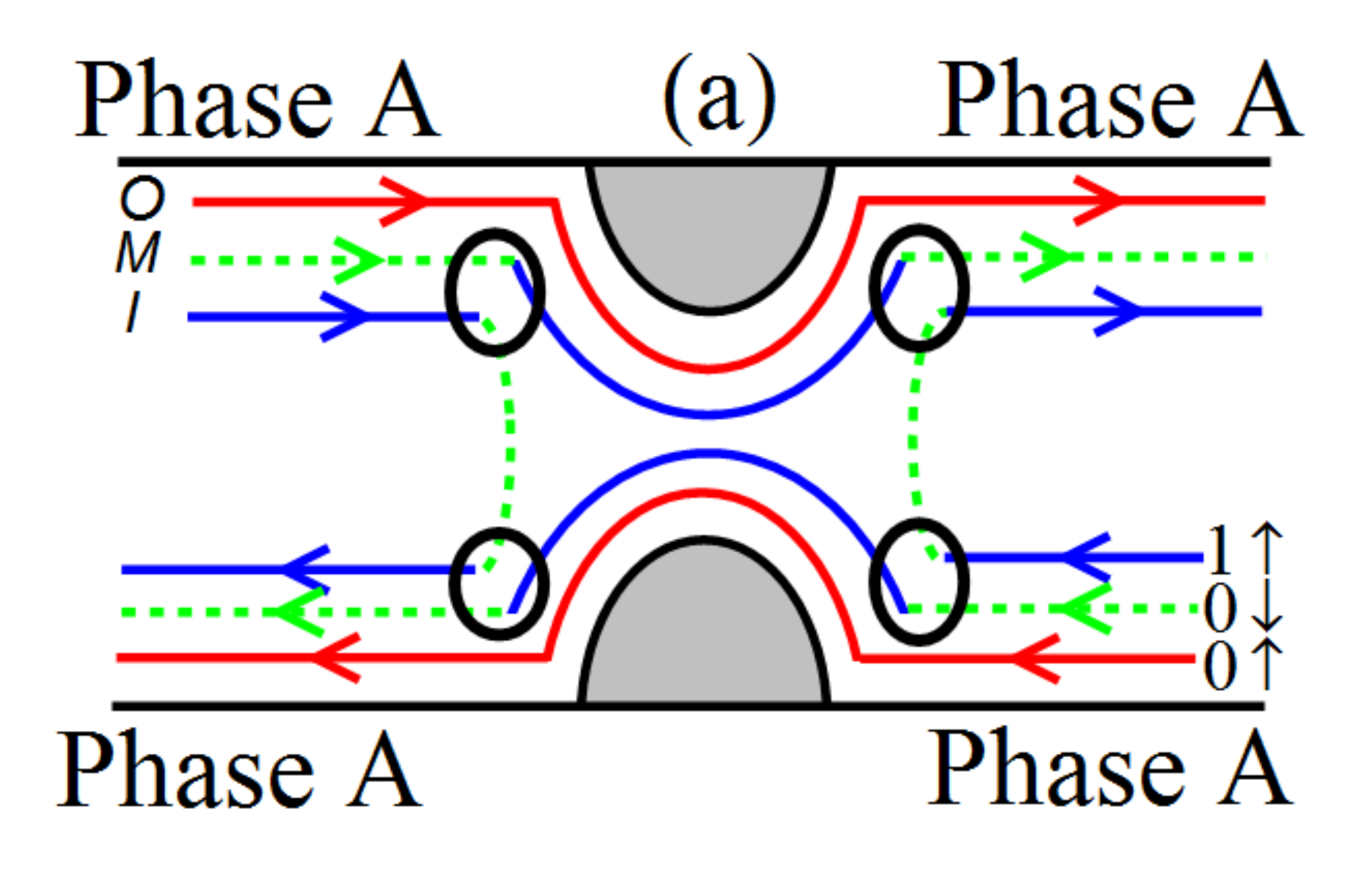}
  \includegraphics[width=0.3\textwidth]{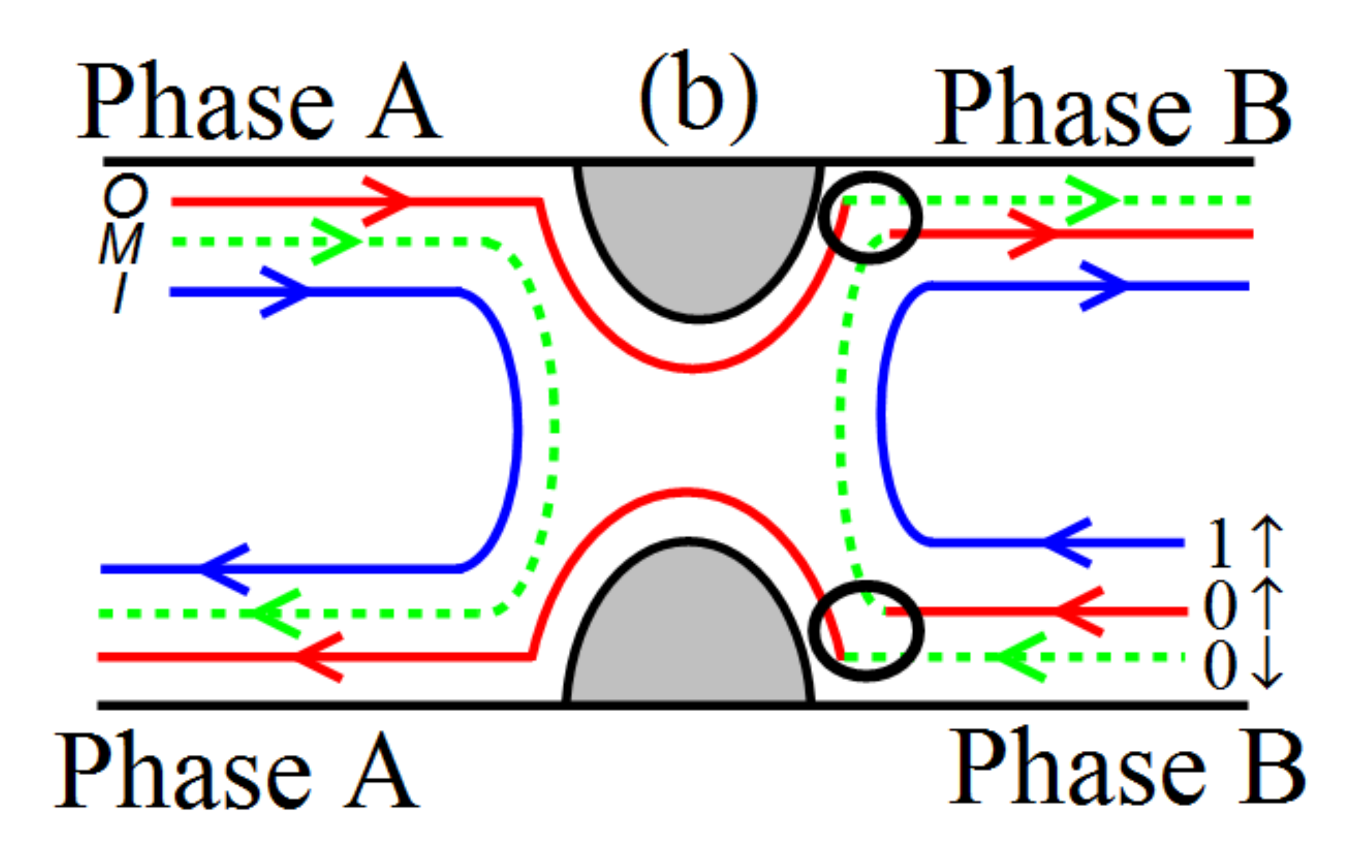}
    \includegraphics[width=0.3\textwidth]{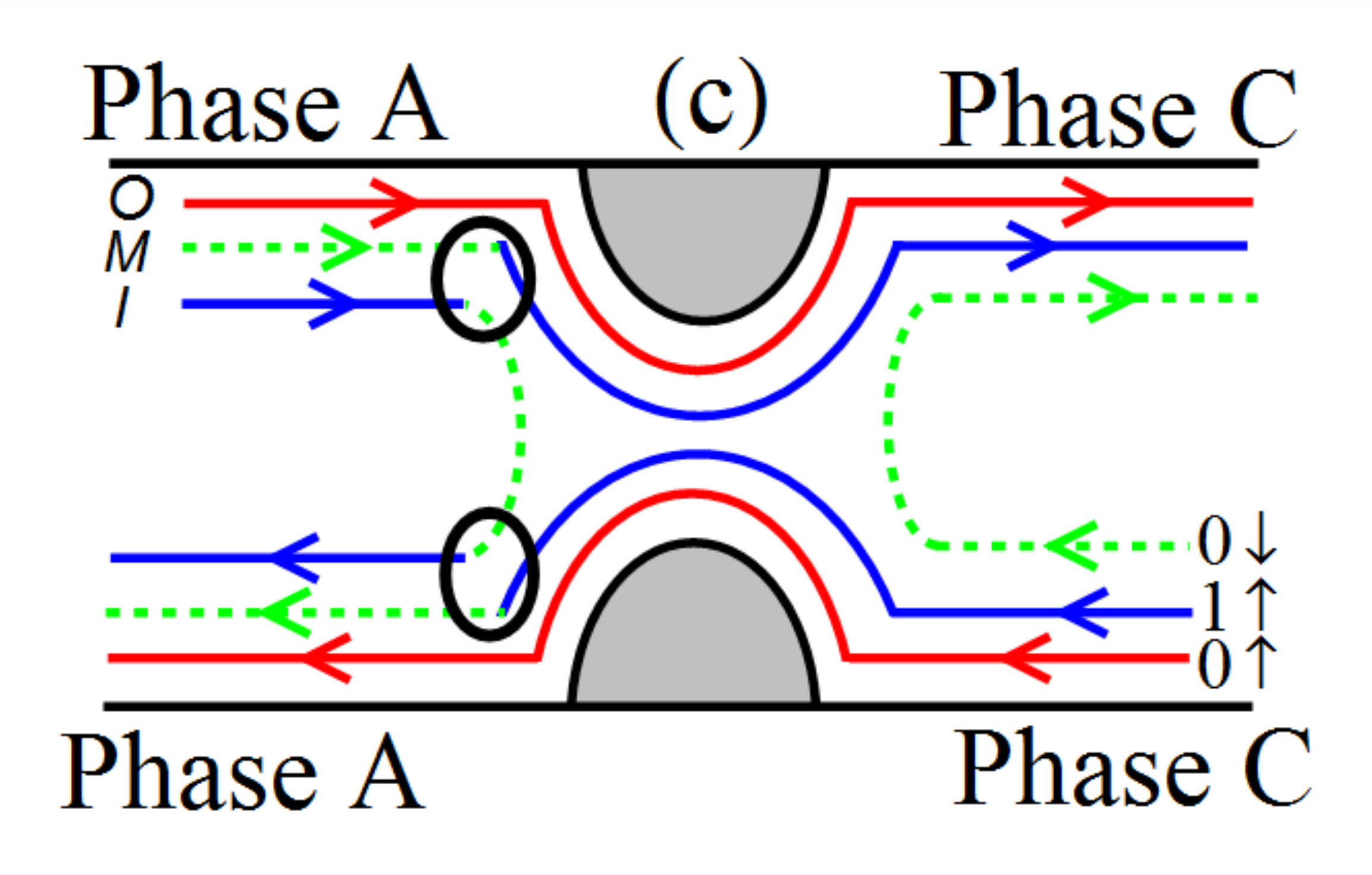}
  \includegraphics[width=0.45\textwidth]{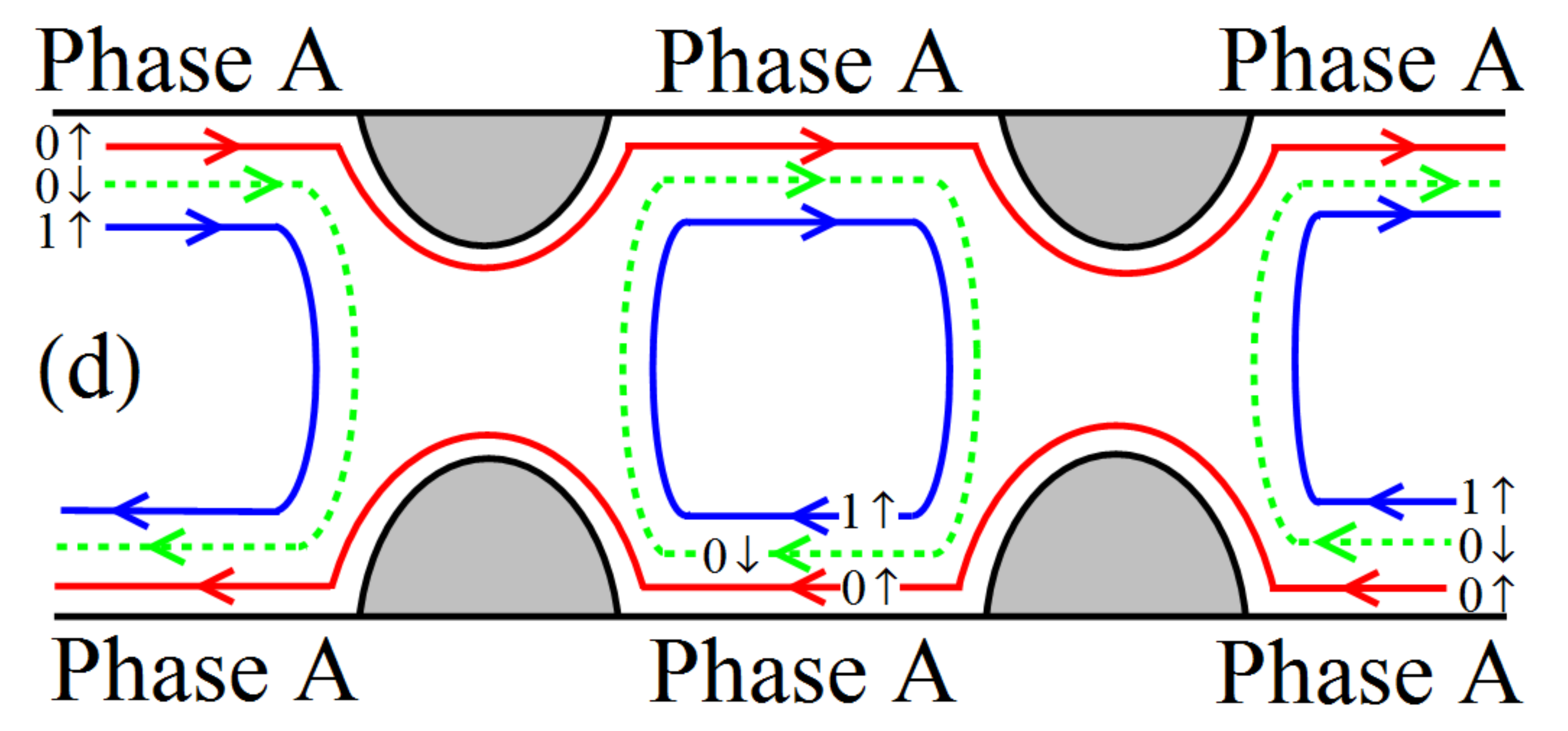}
  \includegraphics[width=0.45\textwidth]{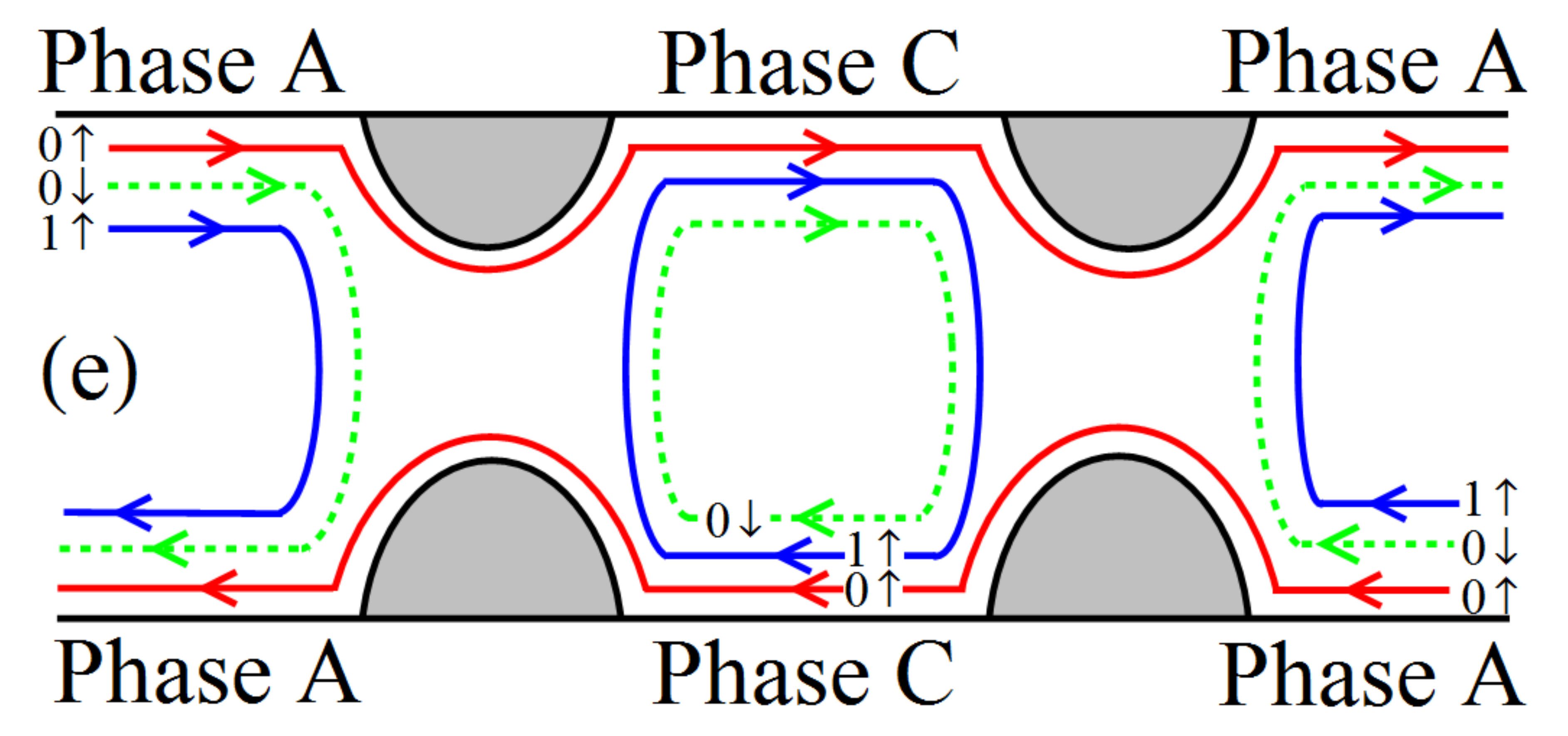}
  \caption{(Color online) Experimental setups to show ``smoking gun'' signatures of
    mode switching. The source (drain) is always on the top left
    (right). Red solid lines depict the $0\ua$ mode, green dashed
    lines the $0\da$ mode and blue solid lines the $1\ua$ mode. The
    edges are labelled on the bottom right of each panel. Spin
    rotations in space are indicated by black circles. \\ (a),(b) and
    (c) Single QPC setups. (a) All the edges are in Phase A, $g_2=2$,
    and $\tEc>2.13$.  The full polarization of the $\nu=2$ QPC region
    forces the M and I modes undergo a spin-rotation upon entering the
    QPC, and an inverse rotation upon exit. The incoming/outgoing
    current is spin
    unpolarized. (b) When the edges to the right of the QPC are in
    Phase B, the current at $g_2 = 1$ reverses spin-polarization from
    $\ua$ to $\da$ at the QPC.  (c) When the edges to the right of the
    QPC are in Phase C, the current (from A to C) at $g_2 = 2$ changes
    from spin-unpolarized to spin-polarized at the QPC. \\ 
    (d) and (e) Two-QPC
    setups at $g_2=1$. (d) When the confining potential in the middle section is sharp on
    both the upper and lower edges (in Phase A), a high quality $g_2=1$
    plateau emerges. (e) When the confining potential in the
    middle section is smooth at both edges (in Phase C), disorder-induced 
    degradation of the conductance plateau due
    to backscattering in the inter-QPC region is expected.} 
 \label{fig2}
	\end{figure*}
%--------------------------------------- 

{\it Experimental signatures.} Before presenting transport signatures
of the switching phenomenon \cite{Supplemental}, we note that whenever
an edge changes from sharp to smooth, spin-mixed edge modes will
undergo avoided crossings with attendant spin rotations along the edge
($x$-direction) \cite{footnote-spatial-crossing,polyakov1,golosov}. Further, the $\nu=2$
state becomes fully polarized at $\tEc\approx 2.13$ for the Coulomb
interaction, in HF. If a QPC is tuned to be at a dimensionless
two-terminal conductance $g_2=2$, the QPC region will be fully
polarized in the regime where Phase C occurs, and unpolarized in the
regime where Phase B occurs.

Our first ``smoking gun'' signature is in spin transport, as
illustrated in Fig.~\ref{fig2}. The system is tuned to be in the $g_2=1$ or $g_2=2$ 
conductance plateau, with the source on the top left, in Fig.~\ref{fig2}. 
Spin rotations in space are indicated by black circles on
the figures, with the modes changing color (and changing from solid to
dashed (i.e., from spin up to spin down) or vice versa).  The current is carried by
the channels O and M (Fig.~\ref{fig2}(a)). For all edges sharp and 
$\tEc>2.13$, there is a nontrivial spin rotation of M as it enters the
QPC region, but it rotates back upon exiting the QPC, so that the
current in the drain (top right) remains unpolarized.  However, when
the right side is in Phase C, (Fig.~\ref{fig2}(c)), the current in the drain is
spin polarized $\ua$. In Fig.~\ref{fig2}(b) we show a QPC tuned to $g_2=1$ in
the regime where Phase B occurs. Recall that Phase B has O=$0\da$,
M=$0\ua$ and I=$1\ua$. The source current (top left) is $\ua$ but the
drain current (top right) is $\da$.

For our next signature, we consider the effect of static, non-magnetic disorder, which
allows tunneling between neighboring chiral modes of the same spin.
In Fig.~\ref{fig2}(d) and~\ref{fig2}(e), the region outside the two QPCs is in Phase
A. We tune the system to the $g_2=1$ plateau. If the inter-QPC region
is in Phase A, the opposite spin polarizations of the two outer
channels $0\ua$,$0\da$ prevent disorder-induced 
tunneling, as shown in Fig.~\ref{fig2}(d). 
On the other hand, when both the top and bottom edges of the inter-QPC region
are in Phase C (Fig.~\ref{fig2}(e)), the two outer channels have same spin 
and disorder-induced tunneling is allowed on both the top and bottom edges.
This leads to backscattering, and hence to a degradation of the quantization of the
conductance plateau. Similar results hold for $g_2=2$ with the
inter-QPC region being either in Phase A (no disorder-induced
degradation) or in Phase B (disorder-induced degradation) \cite{Supplemental}.

{\it Summary and discussion}. We have found spin-exchange driven edge
phases and quantum phase transitions that take place at $\nu=3$ for
low Zeeman energies. Our control parameters are the interaction
strength $\tEc$ and the edge width $\tw$.  We focus on $\tEc \lesssim
2.5$ : a partially polarized bulk state with the LLs
$0\ua$, $0\da$ and $1\ua$ occupied. For small $\tw$ (edge Phase $A$),
the order of the edges follows the bulk order. However, as $\tw$
becomes larger, we find two distinct edge mode-switched phases: For
$1.5 \lesssim \tEc \lesssim 2.13$, Phase $B$ occurs with the edge
ordering O=outermost=$0\da$, M=middle=$0\ua$ and
I=innermost=$1\ua$. For $2.13 \lesssim \tEc \lesssim 2.5$ Phase $C$
occurs with the edge ordering O=$0\ua$, M=$1\ua$ and I=$0\da$. Heuristically, these
phases result from an exchange
attraction between the like-spin edge modes. Employing approximate
analytical methods (the spin unrestricted Hartree-Fock approximation,
and minimization with respect to a particle non-conserving variational
state) we find the transitions to be $1^{st}$-order. We stress that
there is no significant charge rearrangement associated with these
transitions, putting spin-mode-switching in a qualitatively different
category from the extensively investigated phenomena of charge-driven
edge reconstruction. The crucial requirements for the switching
transition to occur are: (i) A partially polarized bulk state. (ii)
Moderate to strong interaction strength $\tEc$. (iii) A smooth edge.
We have also provided experimental signatures of such transitions
in charge and spin transport, relying on experimentally accessible
setups with one or more quantum point contacts.

Our findings have diverse implications, e.g.: (i) Bulk $\nu=1$
supports charged skyrmions \cite{shivaji-skyrmion}, while bulk $\nu=3$
does not \cite{Jain-Wu,Wu-Sondhi}. The $\nu=1$ spinful edge is known
to be unstable to the formation of edge skyrmions
\cite{nu1-edge}. Similar edge spin texture instabilities would likely arise in our
$\nu=3$ system, especially in Phase C, with some similarities to
charge-neutral bilayer graphene \cite{abanin-sid-sondhi}. (ii) Our
results should have direct analogues at $\nu=4$, and more
interestingly, in the QHE in graphene
\cite{Abanin,Brey-Fertig06,Fertig-Brey06}. (iii) Our analysis should
generalize to fractional quantum Hall states such as
$\nu=\frac{3}{7}$, which is the composite fermion analog
\cite{jain-cf} of the $\nu=3$ state.

GM thanks Sid Parameswaran for illuminating discussions, and acknowledges 
the Aspen Center for Physics (NSF Grant No. 1066293) for
its hospitality, the US-Israel
BSF (grants 2012120 and 2016130) and  NSF-DMR 1306897 for support. 
YG thanks Dmitry G. Polyakov for useful discussions, and 
acknowledges  ISF (grant 1349/14), the DFG
(grants RO 2247/8-1, CRC 183), the IMOS
Israel-Russia program, and Minerva for support.

\clearpage

\setcounter{affil}{0}
\renewcommand{\thefigure}{S\arabic{figure}}
\setcounter{figure}{0}
\renewcommand{\theequation}{S\arabic{equation}}
\setcounter{equation}{0}
\renewcommand\thesection{S\arabic{section}}
\setcounter{section}{0}

\title{Supplementary material for \\ Spin Mode-Switching at the Edge of a Quantum Hall System}
\author{Udit Khanna} 
\affiliation{\protect\begin{varwidth}[t]{\linewidth} {\fontsize{9}{10.2}\selectfont {Harish-Chandra Research Institute, Chhatnag Road, Jhunsi, Allahabad 211019, India}}\protect\end{varwidth}}
\affiliation{\protect\begin{varwidth}[t]{\linewidth} {\fontsize{9}{10.2}\selectfont {Homi Bhabha National Institute, Training School Complex, Anushaktinagar, Mumbai, Maharastra 400085, India}}\protect\end{varwidth}}
\author{Ganpathy Murthy}
\affiliation{\protect\begin{varwidth}[t]{\linewidth} {\fontsize{9}{10.2}\selectfont {Department of Physics and Astronomy, University of Kentucky, Lexington KY 40506-0055, USA}}\protect\end{varwidth}}
\author{Sumathi Rao}
\affiliation{\protect\begin{varwidth}[t]{\linewidth} {\fontsize{9}{10.2}\selectfont {Harish-Chandra Research Institute, Chhatnag Road, Jhunsi, Allahabad 211019, India}}\protect\end{varwidth}}
\affiliation{\protect\begin{varwidth}[t]{\linewidth} {\fontsize{9}{10.2}\selectfont {Homi Bhabha National Institute, Training School Complex, Anushaktinagar, Mumbai, Maharastra 400085, India}}\protect\end{varwidth}}
\author{Yuval Gefen} 
\affiliation{\protect\begin{varwidth}[t]{\linewidth} {\fontsize{9}{10.2}\selectfont {Department of Condensed Matter Physics, Weizmann Institute, 76100 Rehovot, Israel}}\protect\end{varwidth}}

\maketitle

In this set of supplemental materials we provide the details of our
theoretical calculations.  In Section I we establish our notation and
present the basic setup. Next, in Section II we present the
Hartree-Fock (HF) approximation both in the bulk and near the edge. In
particular we show the first order bulk transition of the $\nu=3$
quantum Hall state from partially polarized to fully polarized as the
interaction scale increases. In Section II we also present our edge
calculation, where we show the difference between spin-unrestricted HF
and spin-conserving HF, focusing on the region near the spin-mode-switching
transition.  We follow this in Section III with a presentation of the
number non-conserving variational states which go beyond HF. We show
an illustrative example for $\nu = 1$ where we see a smooth
variation of the occupation number and then proceed to the details of
the actual case of interest $\nu = 3$.  Finally, in Section
IV we provide additional transport scenarios, beyond those presented
in the main text, which can be used to test for spin-mode-switching.

\section{I.\,\,\,\,\,\,      Basic Setup}

Consider the integer quantum Hall state on a cylinder. Assuming $x$ to be the periodic direction, 
the wavefunction for single-particle states in Landau gauge $eA_x = -y/\ell^2$, $eA_y = 0$,  is 
\begin{equation}
\Phi_{nk} (x,y) = \frac{e^{i k x}}{\sqrt{L_x}} \frac{e^{-\frac{(y - k \ell^2)^2}{2\ell^2}}}{\sqrt{2^n n! \ell \sqrt{\pi}} } H_n\left(\frac{y - k \ell^2}{\ell} \right)
\end{equation}
\noindent
where $n$ is the Landau level index, $k$ is the guiding center index localized about $Y = k \ell^2$, $\ell = \sqrt{\frac{\hbar}{eB}}$ is the magnetic length and $H_n$ is the $n^{th}$  Hermite polynomial. The Hamiltonian of the system consists of 3 terms: the cyclotron term $H_c$, electron-background attraction   
$H_{bg}$, and electron-electron repulsion $H_{ee}$. 
Using $c_{nks}$ as the destruction operator for single-particle state $\Phi_{nk}$ and spin $s = \ua,\da$, these are 
\begin{align}
H_c &= \sum_{nks} n\hbar \omega_c  c^{\dagger}_{nks} c_{nks} \\
H_{bg} &= -\frac{1}{A} \sum_{\vec{q}} v(\vec{q}) \rho_b(-\vec{q}) \rho_e(\vec{q}) \\
H_{ee} &= \frac{1}{2A} \sum_{\vec{q}} v(\vec{q}) : \rho_e(\vec{q}) \rho_e(-\vec{q}) :,
\end{align}
where $\omega_c = \frac{eB}{m}$ is the cyclotron gap, $A$ is the area of the sample, 
$v(\vec{q})$ is the Fourier transform of the electron-electron interaction, 
$\rho_b(\vec{q})$ is the background density, and 
$\rho_e(\vec{q})$ is the electron density operator, which is
\begin{align}
\rho_e(\vec{q}) = \sum_{\{n\}ks} e^{-i q_y (k + \frac{q_x}{2}) \ell^2} \rho_{n_1 n_2} (\vec{q}) c^{\dagger}_{n_1 ks} c_{n_2 k+q_x s} 
\label{edensity}
\end{align}
where $\{n\}$ denotes the tuple $ (n_1, n_2, \ldots)$.
For $n_1 \geq n_2$ the matrix element $\rho_{n_1 n_2}$ is,
\begin{align}
\rho_{n_1 n_2} (\vec{q}) = \sqrt{\frac{n_2 !}{n_1 !}} \left(\frac{q\ell e^{-i\theta_q}}{\sqrt{2}}\right)^{(n_1 - n_2)} L^{n_1 - n_2}_{n_2} \left(\frac{q^2\ell^2}{2} \right) e^{-\frac{q^2\ell^2}{4}},
\end{align}
where $L_n^m$ is the associated 
Laguerre polynomial and $\rho_{n_2 n_1}(\vec{q}) = [\rho_{n_1 n_2}(-\vec{q})]^{*}$. 

\begin{figure}[t]
\includegraphics[width=0.45\textwidth]{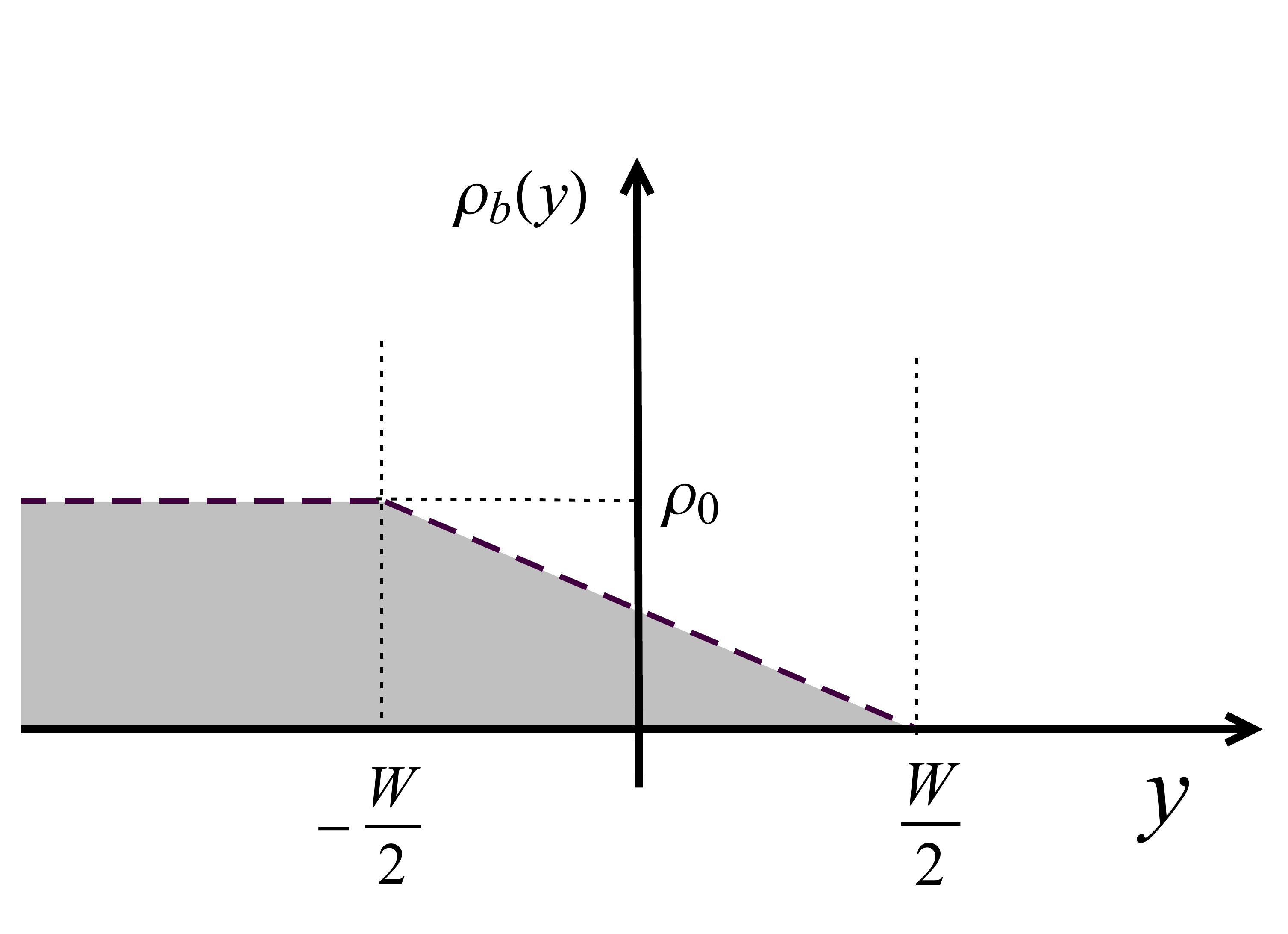}
\caption{(Color online) The background density falls linearly from $\rho_0 = \frac{\nu e}{2\pi \ell^2}$ to 
0 in a distance $W$. }
\label{fig:FigSetup}
\end{figure}

The background charge is assumed to be uniformly distributed in the $x$ direction. Then 
$\rho_b(\vec{q}) = \delta_{q_x,0} \rho_b(q_y)$ and therefore the background term is,
\begin{align}
H_{bg} = -\frac{1}{A} \sum_{\{n\} ks} \sum_{\vec{q}} v(\vec{q}) \Delta_{bg}(\{n\},k,\vec{q}) c^{\dagger}_{n_1 ks} c_{n_2 ks},
\end{align} 
where the background potential is
\begin{align}
\Delta_{bg}(\{n\},k,\vec{q}) = \delta_{q_x,0} \rho_b(-q_y) e^{-i q_y k} \rho_{n_1 n_2}(\vec{q})
\end{align}
Here, the edge is modelled with a background charge density that falls linearly from the 
bulk value $\frac{\nu e}{2\pi \ell^2}$ to 0 over a width of $W$ around $y = 0$ \cite{dempsey-Supp,chamon-wen-Supp} (Fig.~\ref{fig:FigSetup}).
\begin{align}
\rho_{b}(\vec{r}) = \left\{ 
\begin{aligned}
&\frac{\nu e}{2\pi \ell^2} &\text{ for } &y \leq -\frac{W}{2} \\
&\frac{\nu e}{2\pi \ell^2} \frac{W - 2y}{2 W} &\text{ for } &-\frac{W}{2} \leq y \leq \frac{W}{2} \\
&0 &\text{ for } &\frac{W}{2} \leq y
\end{aligned} \right.
\end{align}

%------Fig S1 ------
%------------------

\section{II.\,\,\,\,\,\,   Hartree Fock Calculation}

In the Hartree Fock (HF) approximation, the interaction term $H_{ee}$ is decoupled by 
taking all possible averages and replaced by a one body mean-field 
Hamiltonian $H_{MF}$. Under the assumption that the state has translation invariance in the $x$-direction, the averages have the form,
\begin{align}
\langle c^{\dagger}_{n_1 k_1 s_1} c_{n_2 k_2 s_2} \rangle = 
\delta_{k_1 k_2} \Delta_{n_1 n_2 ; s_1 s_2}(k),
\label{EqS10}
\end{align}
and thus
\begin{align} \nonumber
&H_{MF} = \frac{1}{A} \sum_{\{n, s\}} \sum_{k \vec{q}} v(\vec{q}) 
\left[
\begin{aligned} &V_H(\{n,s\},k,\vec{q}) \\ &- V_F(\{n,s\},k,\vec{q})
\end{aligned} \right]
 c^{\dagger}_{n_1 k s_1} c_{n_2 k s_2} \\ \label{EqH}
&V_H(\{n,s \},k_1,\vec{q}) = \sum_{\{m\} k_2 \sigma} \left[
\begin{aligned} & \delta_{s_1 s_2} \delta_{q_x,0} e^{-i q_y (k_1 - k_2) \ell^2} \\
&\times \Delta_{m_1 m_2; \sigma \sigma}(k_2) \\ 
&\times \rho_{n_1 n_2} (\vec{q}) \rho_{m_1 m_2} (-\vec{q}) \end{aligned} \right] \\ \label{EqF}
&V_F(\{n,s\},k,\vec{q}) = \sum_{\{m\}} \left[
\begin{aligned}  &\Delta_{m_1 m_2; s_2 s_1}(k + q_x) \\ 
&\times \rho_{n_1 m_2} (\vec{q}) \rho_{m_1 n_2} (-\vec{q})  \end{aligned} \right],
\end{align} 
where the Hartree and Fock potentials ($V_{H/F}$) have to be
computed self-consistently.  The Hartree potential arises from the
classical density-density interaction and therefore is same for both spin
states. The spin-dependent Fock potential arises due to the exchange
of electrons and promotes ferromagnetic behaviour.

{\bf Bulk Solution:} In the bulk, the electron charge density is fully
translation invariant, implying that the matrix elements $\Delta_{n_1
  n_2 ; s_1 s_2}$ are independent of the guiding center label
$k$.  In the presence of an infinitesimal Zeeman field, the spins of
all the single-particle states will be either parallel or antiparallel to the quantization axis, implying that 
$\Delta$ is  diagonal in spin
labels. Furthermore, no LL-mixing occurs in translation invariant HF
states. Therefore $\Delta$ reduces to a diagonal matrix in both spin and
Landau level indices, in which the diagonal elements are the occupations of the single-particle levels. Thus,  the HF potentials depend only on the
occupations of the single-particle levels.  In the bulk, the Hartree
potential cancels the background potential exactly due to charge
neutrality.  The Fock potentials reduce to
\begin{align*}
  V_F(n,s,\vec{q}) = \sum_{m} n_f(m;s) \rho_{nm}(\vec{q}) \rho_{mn}(-\vec{q}).
\end{align*}
The energy (per particle) of a translation-invariant bulk HF state
with occupations $n_f(m;s)$ is
\begin{align*}
  E[{n_f(m;s)}] = \sum_{\{m\},s} \frac{n_f(m_1;s)}{\nu} \left[ 
  \begin{aligned}
   &\delta_{m_1,m_2} m_1 \hbar \omega_c \\
   &- E_{ex}(m_1,m_2) n_f(m_2;s) \end{aligned} \right],   
\end{align*}
where the exchange energy  is 
\begin{align*}
E_{ex}(m_1,m_2) = \frac{1}{2} \int \frac{d^2 q}{(2\pi)^2} v(\vec{q}) \rho_{m_1 m_2} (\vec{q}) \rho_{m_2 m_1} (-\vec{q}).  
\end{align*}
These integrals can be calculated analytically for certain special
forms of $v(\vec{q})$, such as the Coulomb interaction.  At zero
temperature, the occupation of a single-particle level is either 1 or
0. Given the occupations of a HF state, its energy can be calculated
easily. We drop the Hartree term in the bulk, since it contributes
equally to all states.

At filling factor $\nu = 3$, the two possible translation-invariant
ground states are the partially polarized state 
$|\psi_1 \rangle = |0\ua,0\da,1\ua \rangle$ 
and the fully polarized
state $|\psi_2 \rangle = |0\ua,1\ua,2\ua \rangle$. We assume
a screened Coulomb interaction of the form
\begin{align*}
v(\vec{q}) = \frac{2\pi E_c}{q + q_{sc}}, 
\end{align*}
for which the energy can be expressed in terms of the Error function. 
$E_c = \tilde{E_c} \hbar \omega_c = \frac{e^2}{4 \pi \epsilon \ell}$ is the 
interaction scale as defined in the main text. 
For a long range Coulomb interaction ($q_{sc} = 0$), 
the expression simplifies to - 
\begin{align*}
E[|\psi_1\rangle] &= \frac{1}{3} \hbar \omega_c - \frac{5}{8} E_c \sqrt{\frac{\pi}{2}} \\
E[|\psi_2\rangle] &= \hbar \omega_c - \frac{107}{128} E_c \sqrt{\frac{\pi}{2}}
\end{align*}
Comparing the two energies, we see that there is a $1^{st}$ order
\cite{Jungwirth1-Supp,Jungwirth2-Supp} bulk transition from $|\psi_1 \rangle$ to
$|\psi_2 \rangle$ at $\tilde{E_c} \approx 2.52$.  At finite
values of $q_{sc}$, the exchange interaction is weaker and the
transition occurs at a larger value of $\tilde{E_c}$. In the numerical
calculations to follow, we have used $q_{sc}\ell = 0.01$ for which the
transition occurs at $\tilde{E_c} \approx 2.53$.  Similar
considerations lead to a bulk $1^{st}$ order transition of the $\nu=2$
state from unpolarized to fully polarized at $\tilde{E_c}=2.13$.
%------ Fig S2 ------
\begin{figure*}[t]
\includegraphics[width=0.33\textwidth]{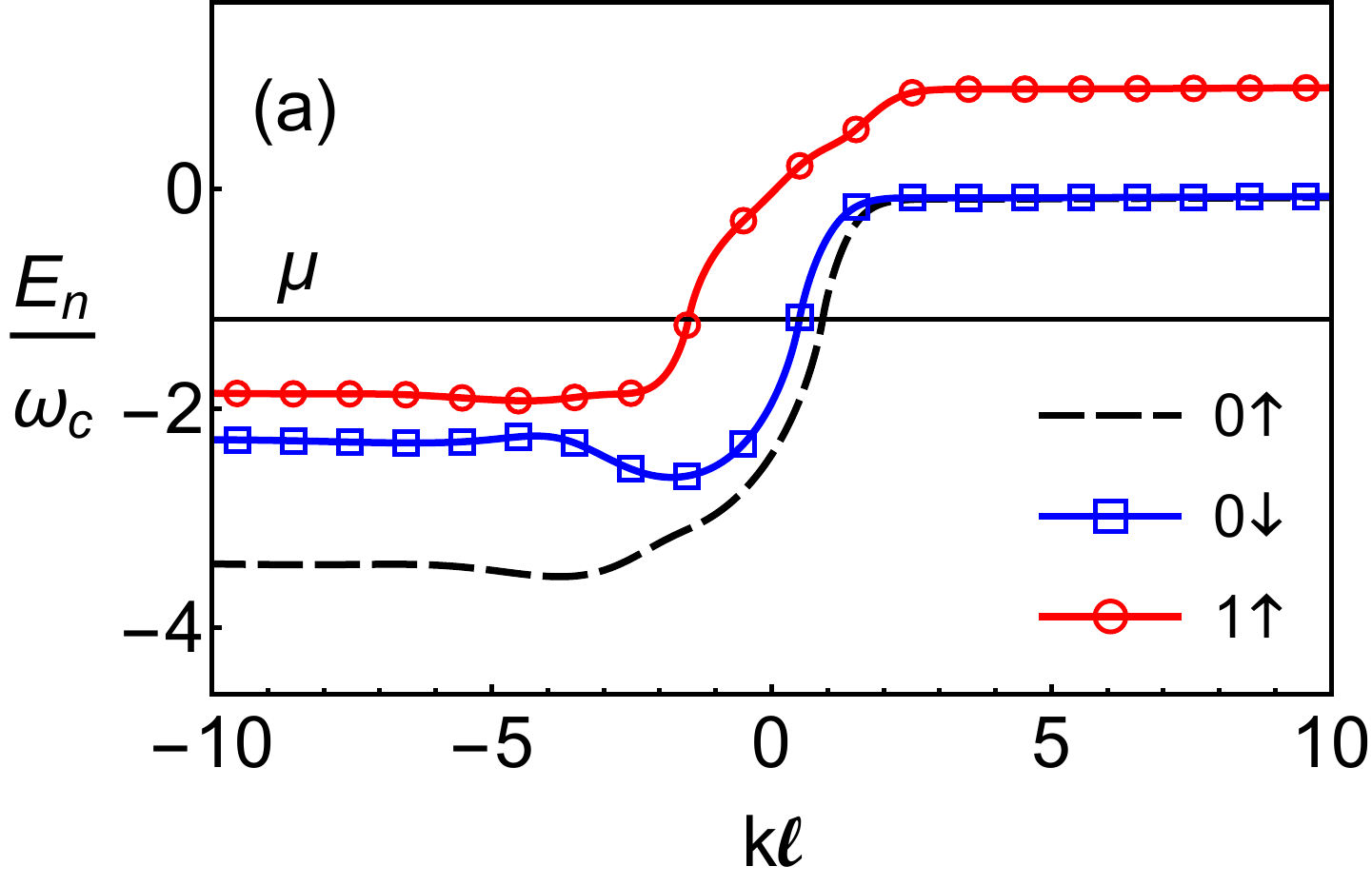}
\includegraphics[width=0.33\textwidth]{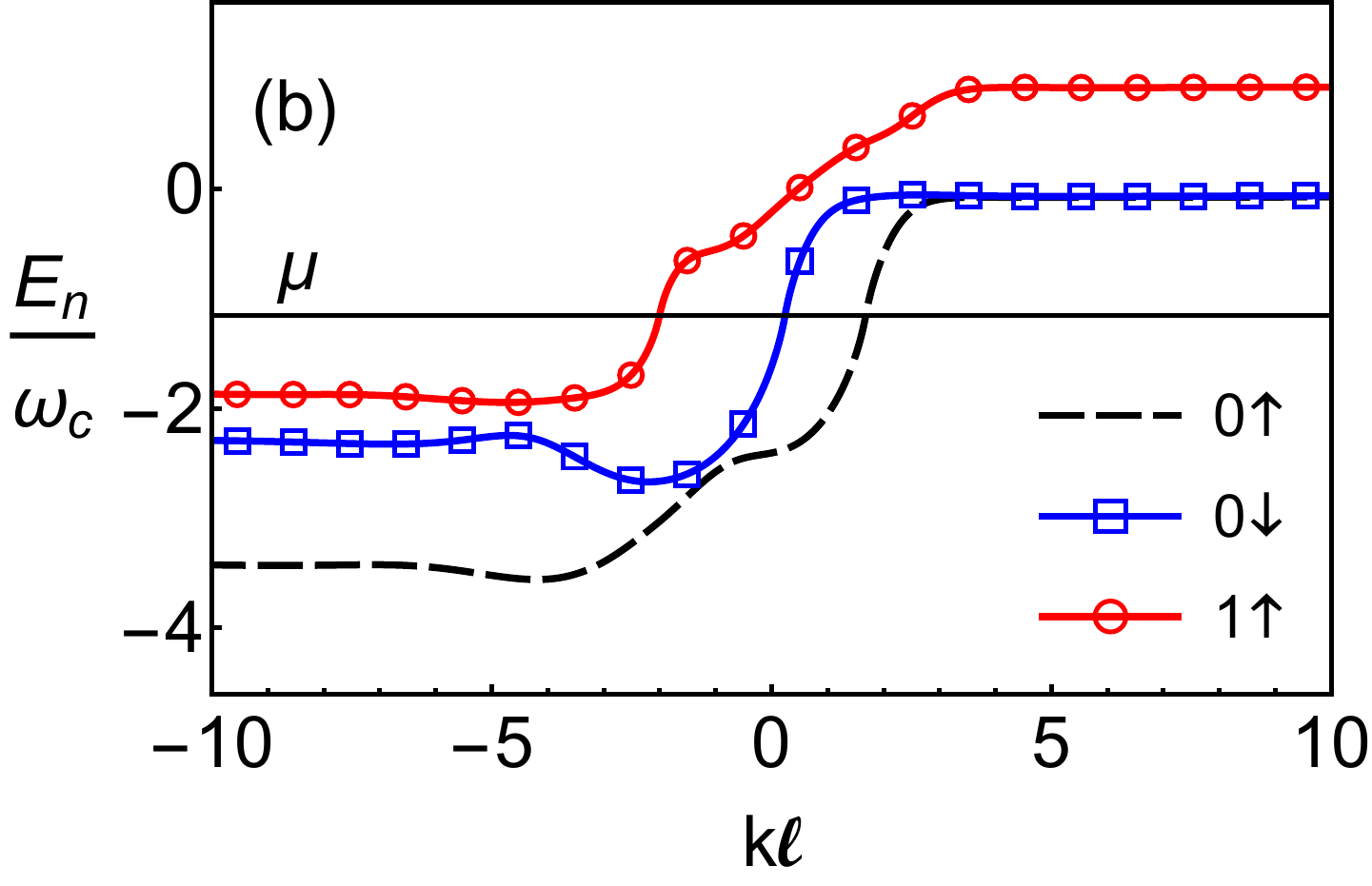}
\includegraphics[width=0.33\textwidth]{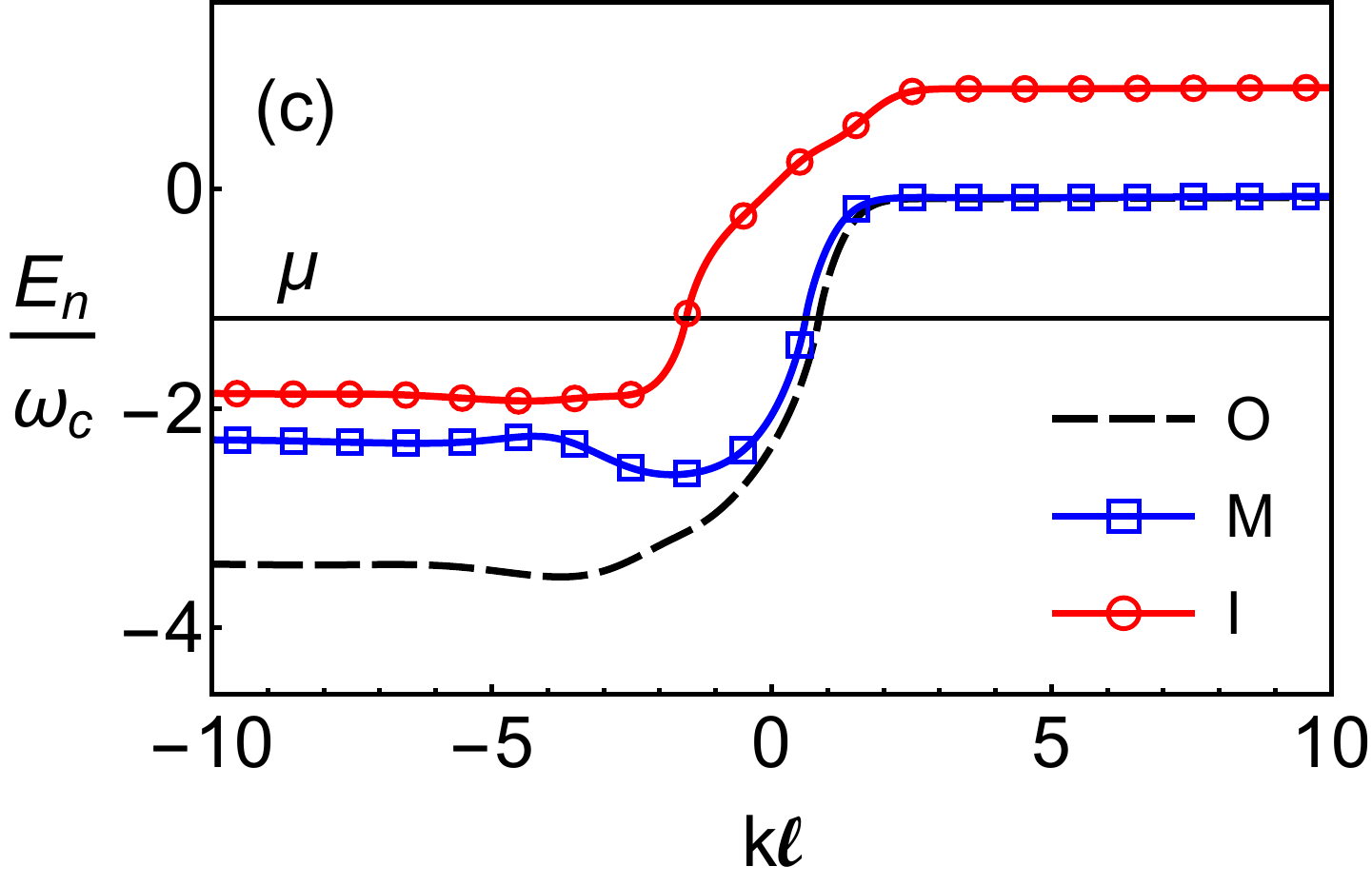}
\includegraphics[width=0.33\textwidth]{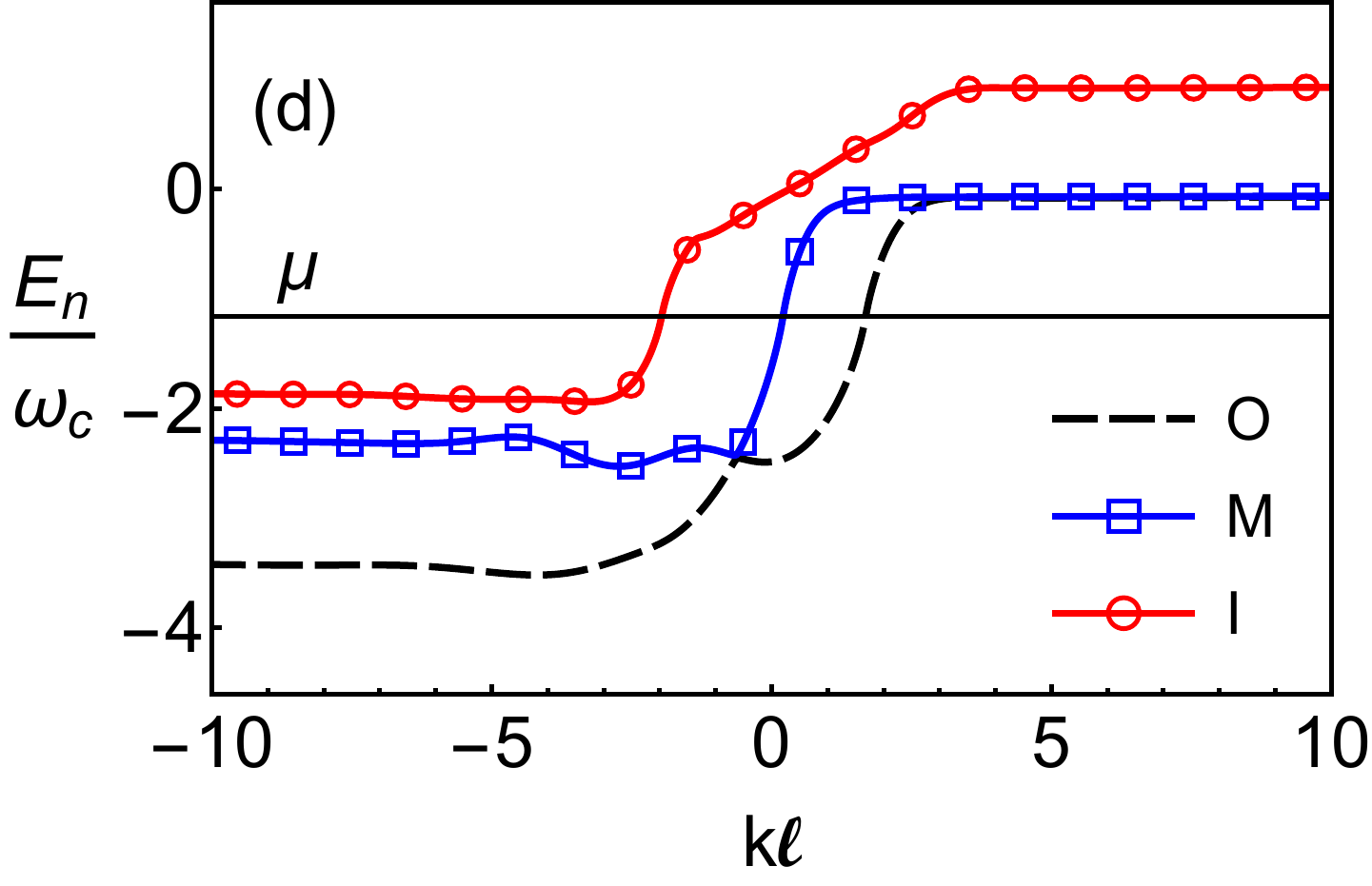}
\caption{(Color online) Comparison of the self-consistent single
  particle energy level dispersion in RHF (spin-restricted HF, cf. Eq.~\ref{EqS10} and~\ref{EqS14})
  and UHF (spin-unrestricted HF, i.e. without the restriction in Eq.~\ref{EqS14}) at $\tilde{E_c} = 1.8$. 
  In RHF we can label an energy dispersion with the LL-index it
  had in the bulk and its spin. In UHF, however, spin is not
  conserved, so we label the modes in the order they appear at the
  edge from outside in, as $i$ = O (outermost), M (middle), and I
  (innermost). (a) Self-consistent single-particle energy dispersions
  vs. $k\ell$ in RHF for $ W = 2.0 \ell$ (Phase A). The order of the edges,
  from outermost to innermost, is the same as the bulk order of
  energies from lowest to highest. (b) Single-particle energy
  dispersions vs. $k\ell$ for $W = 4.0 \ell$ in RHF. The true ground
  state is  Phase B, but RHF is unable to reach the true ground
  state, so the figure shows the dispersions in a metastable Phase
  A. (c) Single-particle energy dispersions vs. $k\ell$ for $ W = 2.0
  \ell $ in UHF. Since the true ground state is that of Phase A,
  almost no spin-rotations are required, and the dispersions are very
  close to those of RHF.  (d) Single-particle energy dispersions
  vs. $k\ell$ at $ W = 4.0 \ell$ in UHF. Now the system is in Phase
  B. Note that the apparent touching of the O and M modes near
  $k\ell=0$ is actually an avoided crossing where the spin character
  of the single particle levels changes (see Fig.~\ref{fig:FigSz1.8}(b) below). }
\label{fig:FigEnLv1.8}
\end{figure*}
%-------------------
%------ Fig S3 ------
\begin{figure*}[t]
\includegraphics[width=0.32\textwidth]{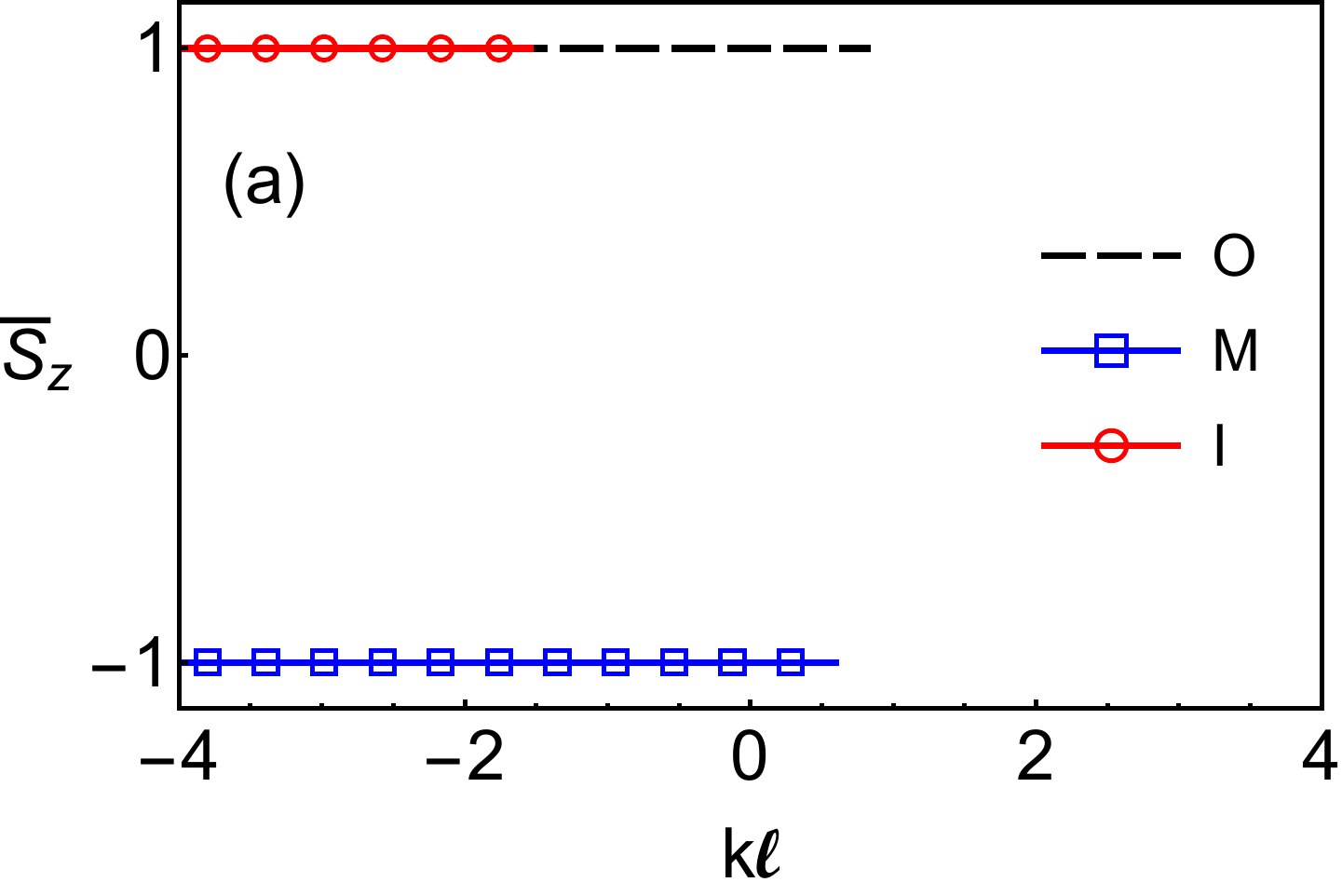}
\includegraphics[width=0.32\textwidth]{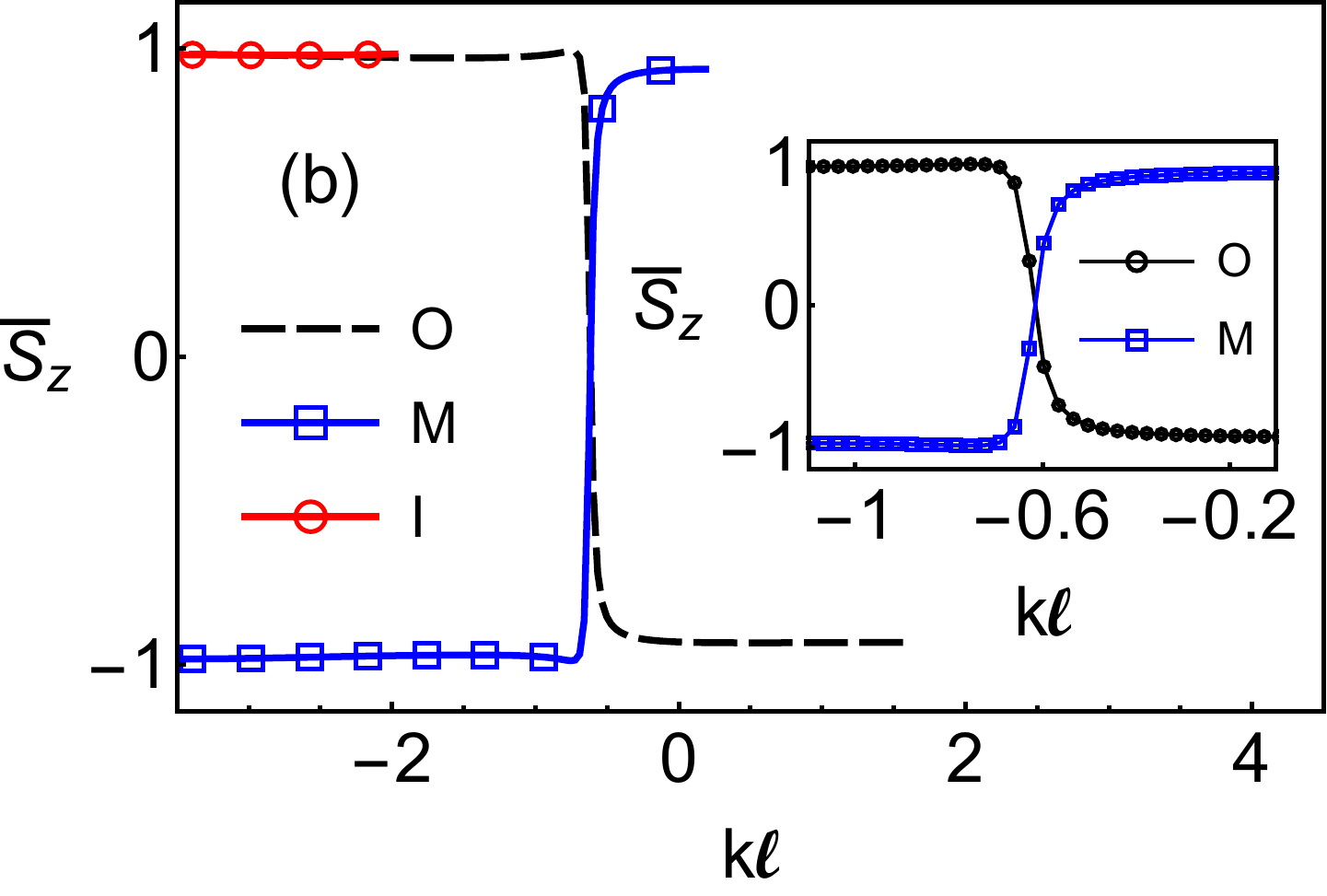}
\includegraphics[width=0.32\textwidth]{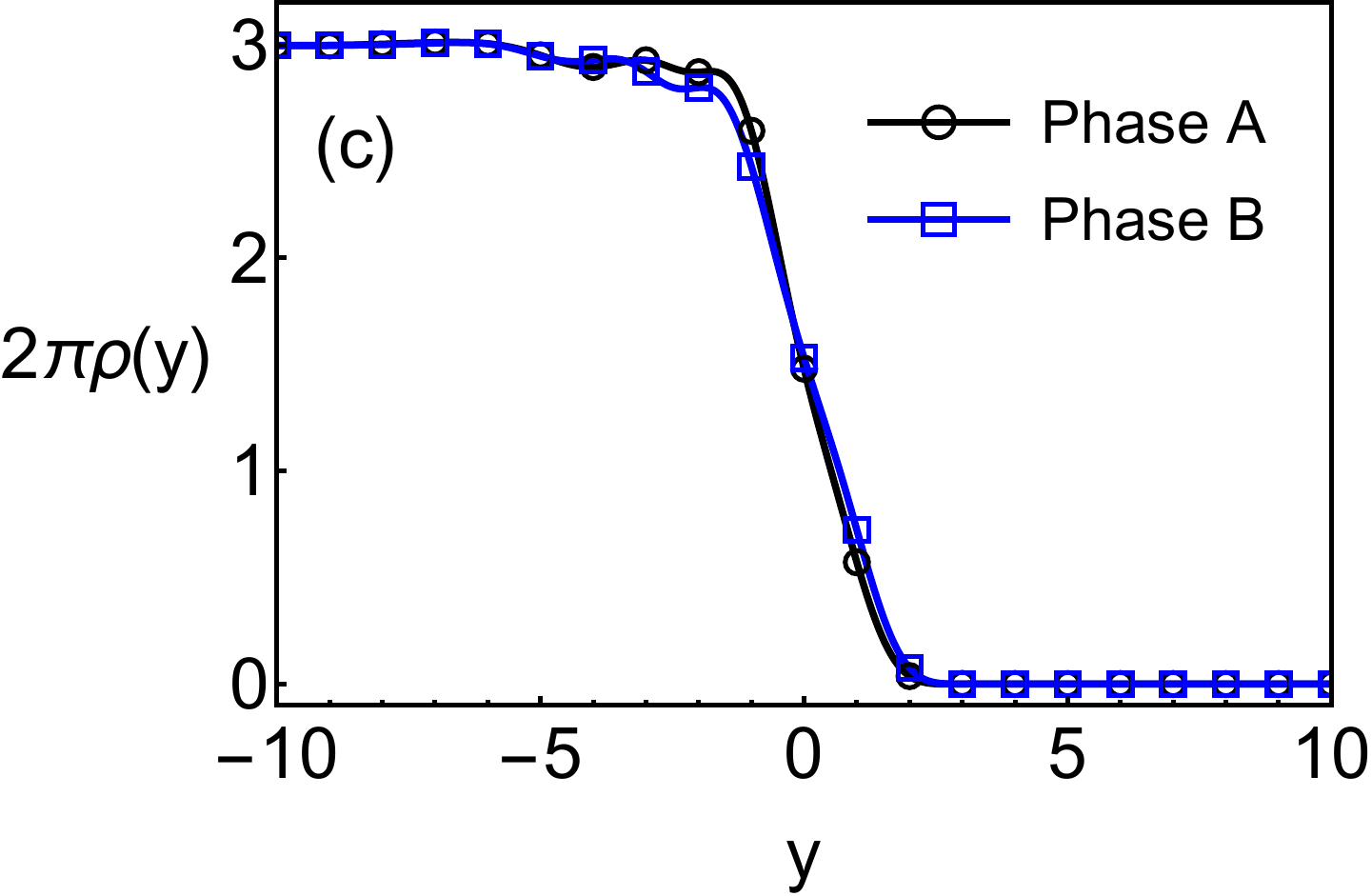}
\caption{(Color online) (a) $\bar{S}_z(i,k)$ of single particle energy
  levels in UHF vs. $k\ell$ at $\tilde{E_c} = 1.8$ and $ W = 2.0 \ell$
  (Phase A). There are no spin-rotations.  (b) $\bar{S}_z(i,k)$ in UHF
  vs. $k\ell$ at $\tilde{E_c} = 1.8$ and $ W = 4.0 \ell$ (Phase
  B). Note that $\bar{S}_z(i,k)$ for $i=$ O and M vary continuously
  near $k\ell=0$, precisely where the corresponding energy dispersions
  (Fig.~\ref{fig:FigEnLv1.8}(d)) appear to touch. The inset  makes the smooth
  variation of $\bar{S}_z(i,k)$ unambiguous. This is evidence that the
  apparent touching is actually an avoided crossing. At an actual
  crossing, $\bar{S}_z(i,k)$ would have changed abruptly.  (c) The
  average of the electronic density $\rho(y)$ vs. $y$ just before and
  after the mode-switching transition in Phase A (at $ W = 3.25 \ell$,
  $\tEc = 1.8$) and in Phase B (at $W = 3.75 \ell$, $\tEc = 1.8$). It
  is seen that there is very little variation of the electron charge
  density across the transition, which is the basis of our conclusion
  that the mode-switching transition is not driven by charge effects.
}
\label{fig:FigSz1.8}
\end{figure*}
%-------------------
%------ Fig S4 ------
\begin{figure*}[ht]
\includegraphics[width=0.33\textwidth]{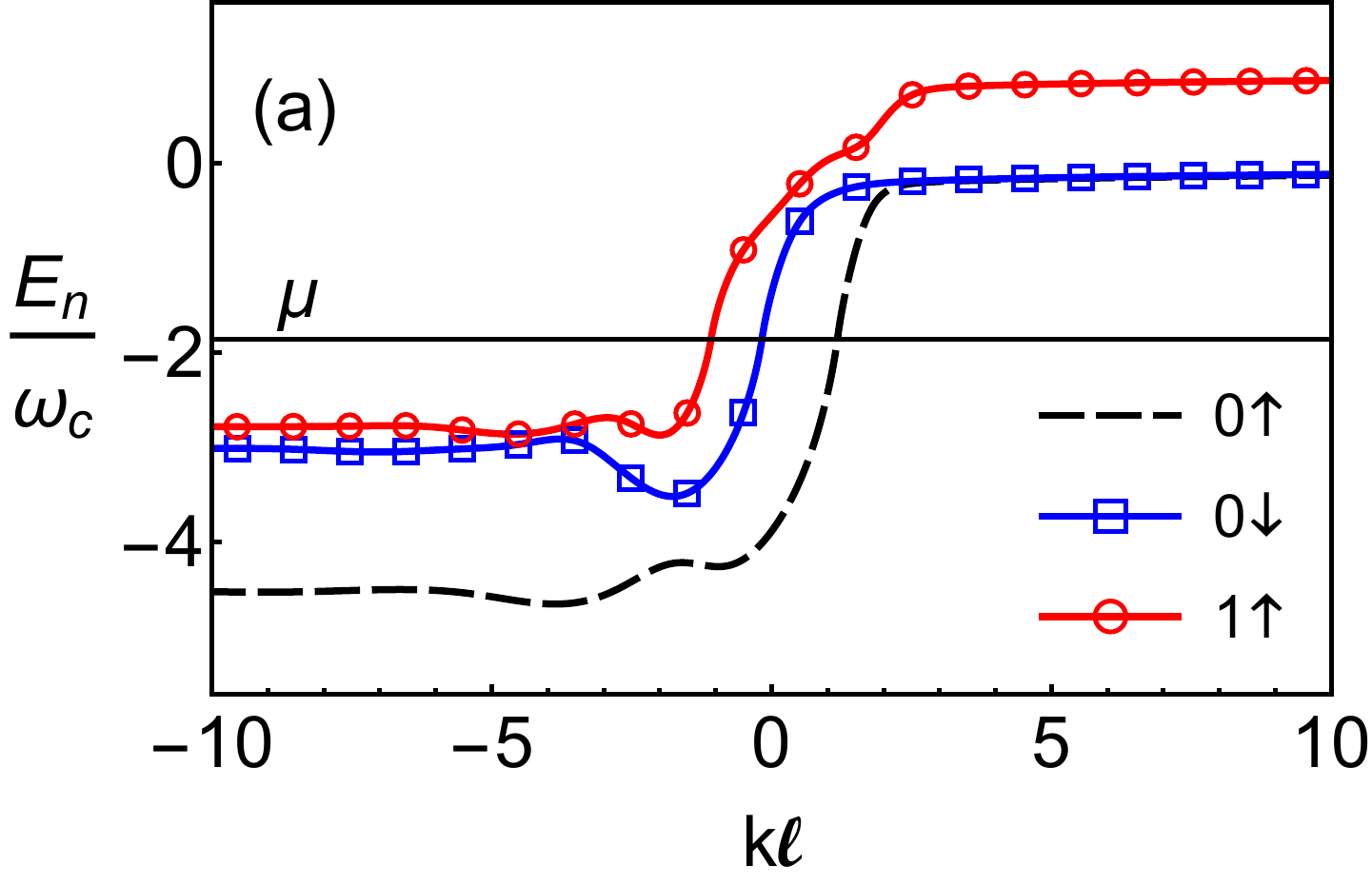}
\includegraphics[width=0.33\textwidth]{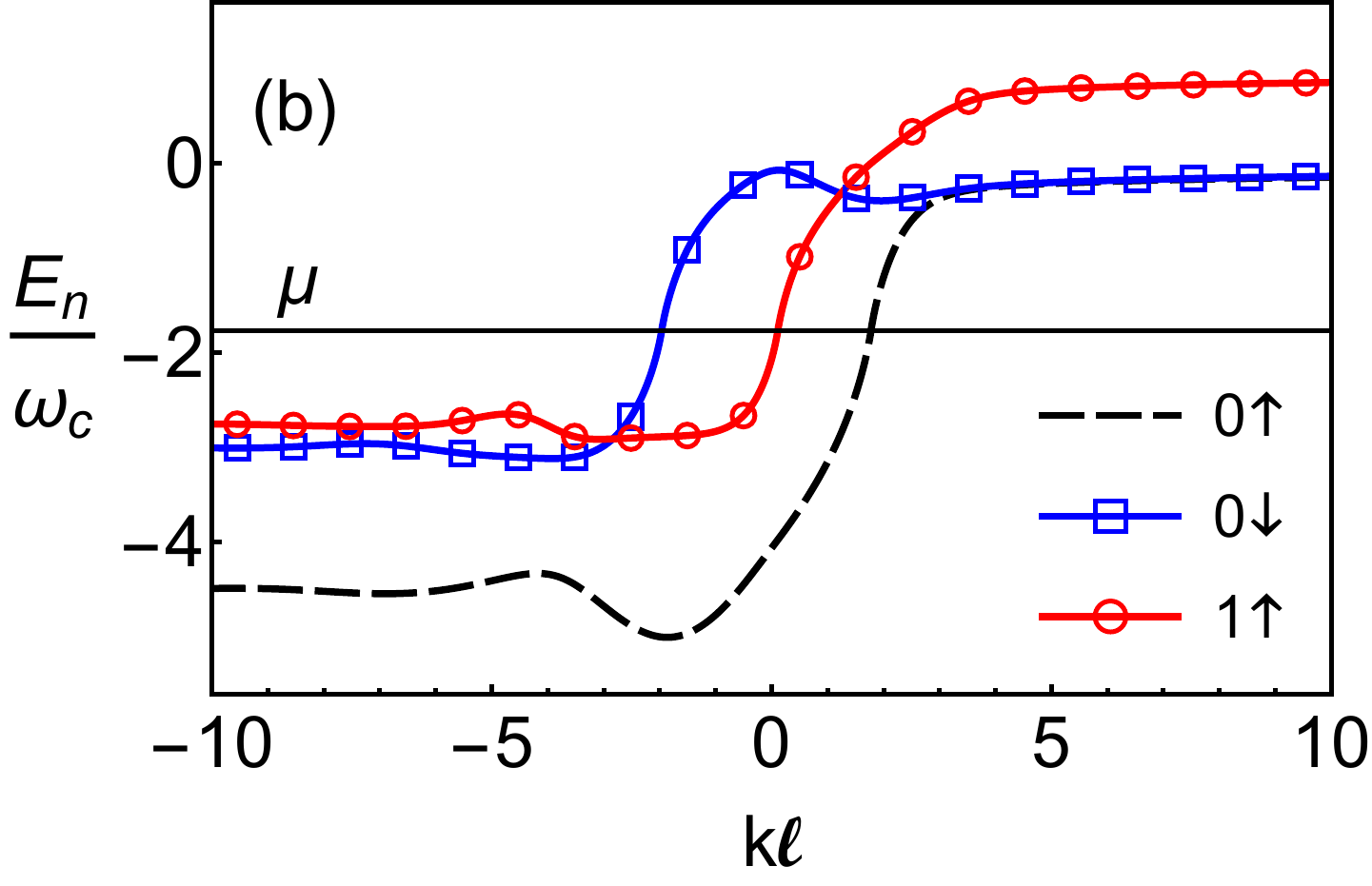}
\includegraphics[width=0.33\textwidth]{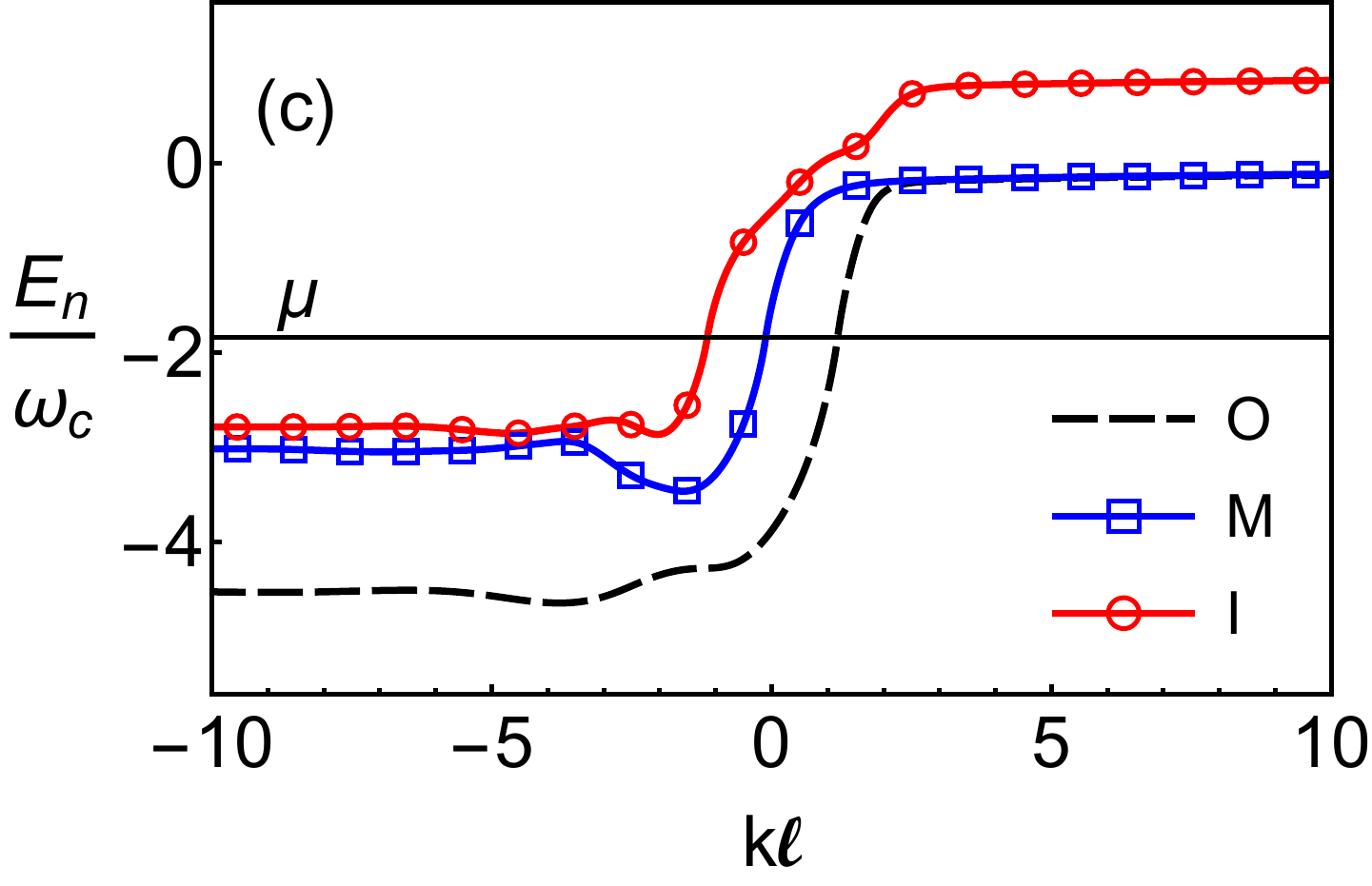}
\includegraphics[width=0.33\textwidth]{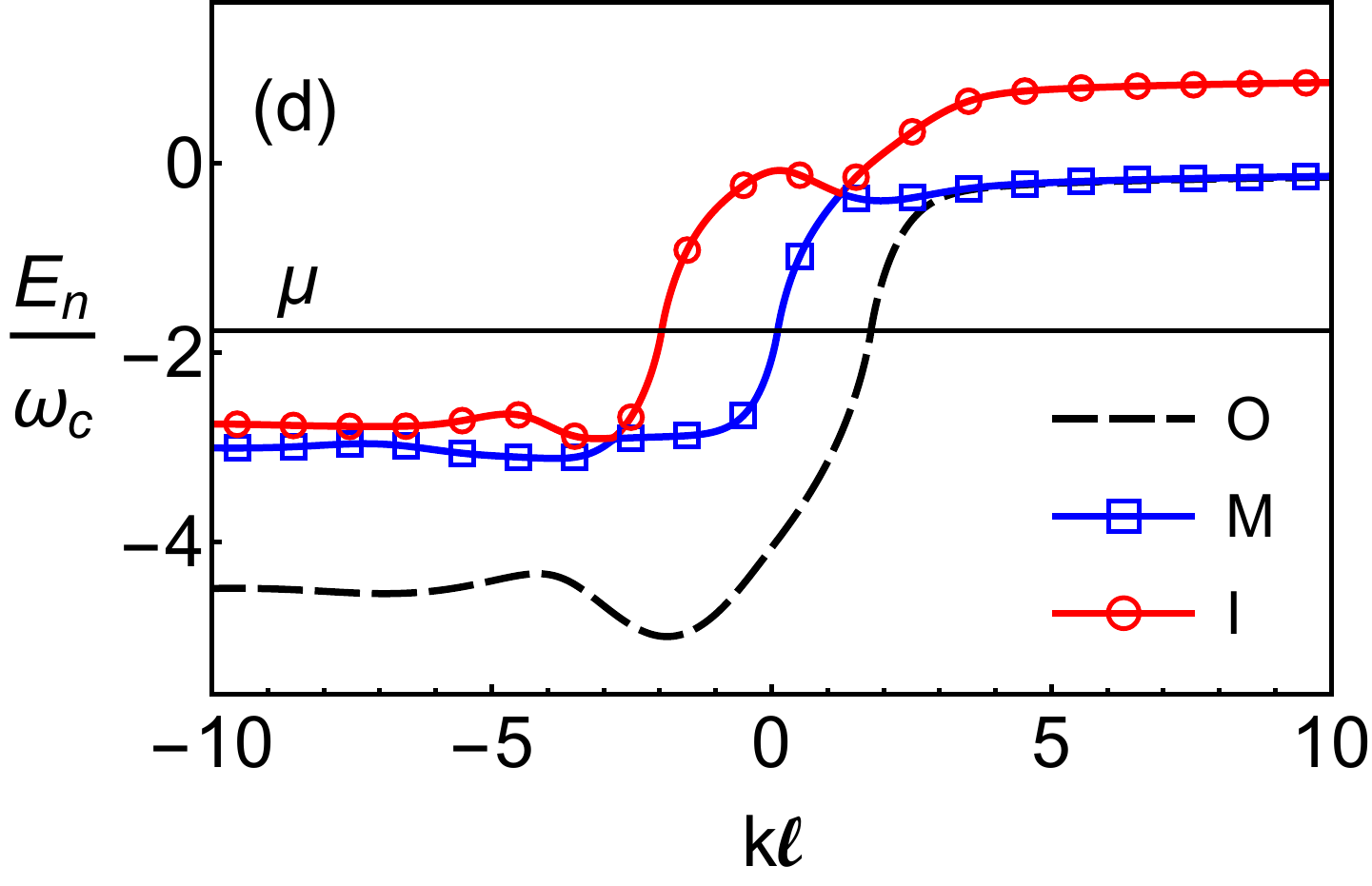}
\caption{(Color online) Comparison of the self-consistent energy
  dispersions in RHF and UHF at $\tilde{E_c} = 2.3$.  (a) Energy
  dispersions vs. $k\ell$ at $ W = 2.0 \ell$ in RHF. As always in
  Phase A, there are no level crossings. (b) Energy dispersions
  vs. $k\ell$ at $ W = 4.5 \ell$ in RHF (Phase C). RHF is able to
  access Phase C, and one sees the $0\da$ and $1\ua$ levels cross
  below $\mu$, so that the ordering at the edge is, from the outermost
  in, $0\ua$, $1\ua$, and $0\da$. (c) Energy dispersions vs. $k\ell$
  at $ W = 2.0 \ell $ in UHF (Phase A). There are no level
  crossings. (d) Energy dispersions vs. $k\ell$ at $ W = 4.5 \ell$ in
  UHF (Phase C). Note that the apparent touching of M and I near
  $k\ell=-3$ is actually an avoided crossing, as evidenced by
  $\bar{S}_z(i,k)$ (see Fig.~\ref{fig:FigSz2.3}(b) below). }
\label{fig:FigEnLv2.3}
\end{figure*}
%-------------------
%------ Fig S5 ------
\begin{figure*}[t]
\includegraphics[width=0.32\textwidth]{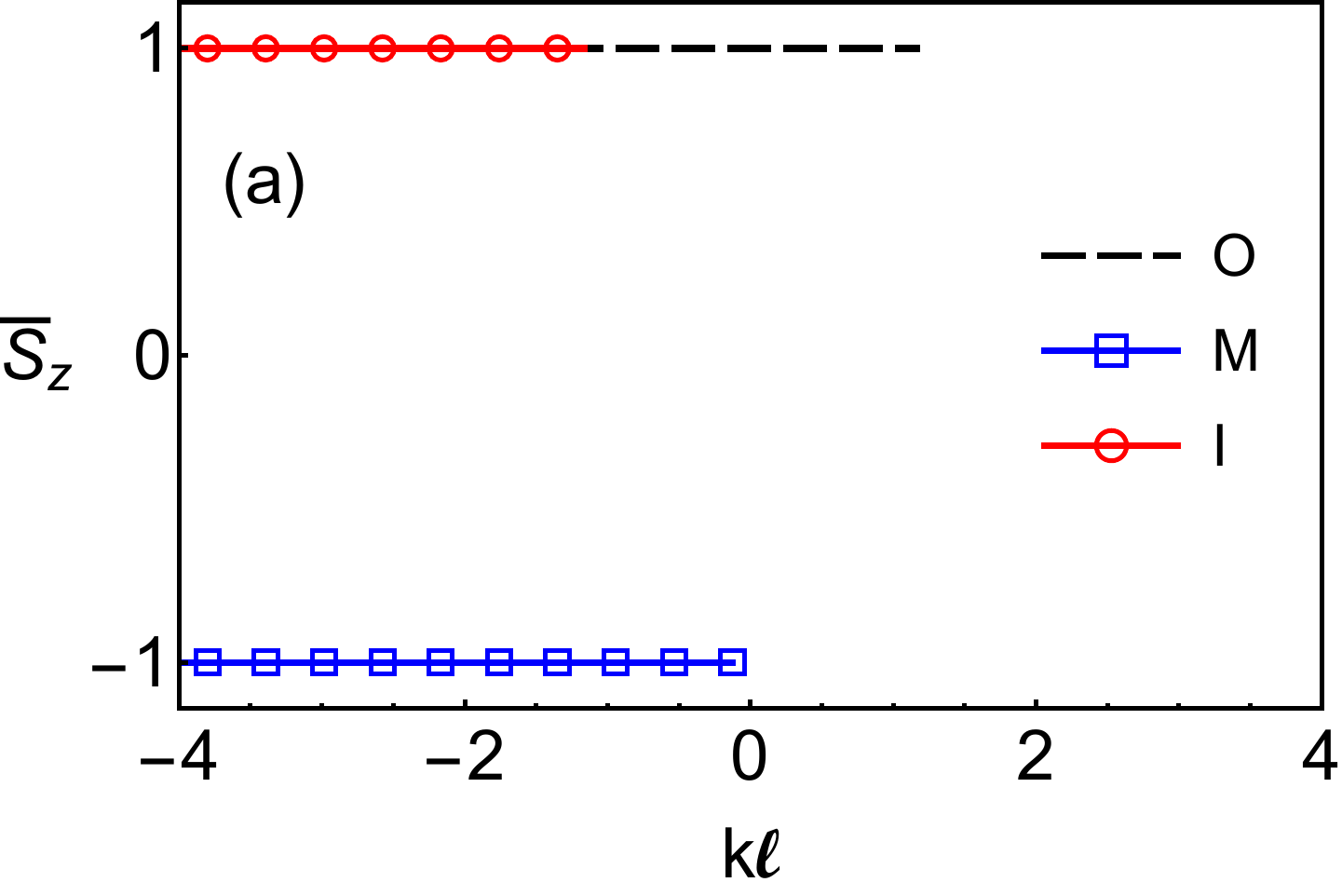}
\includegraphics[width=0.32\textwidth]{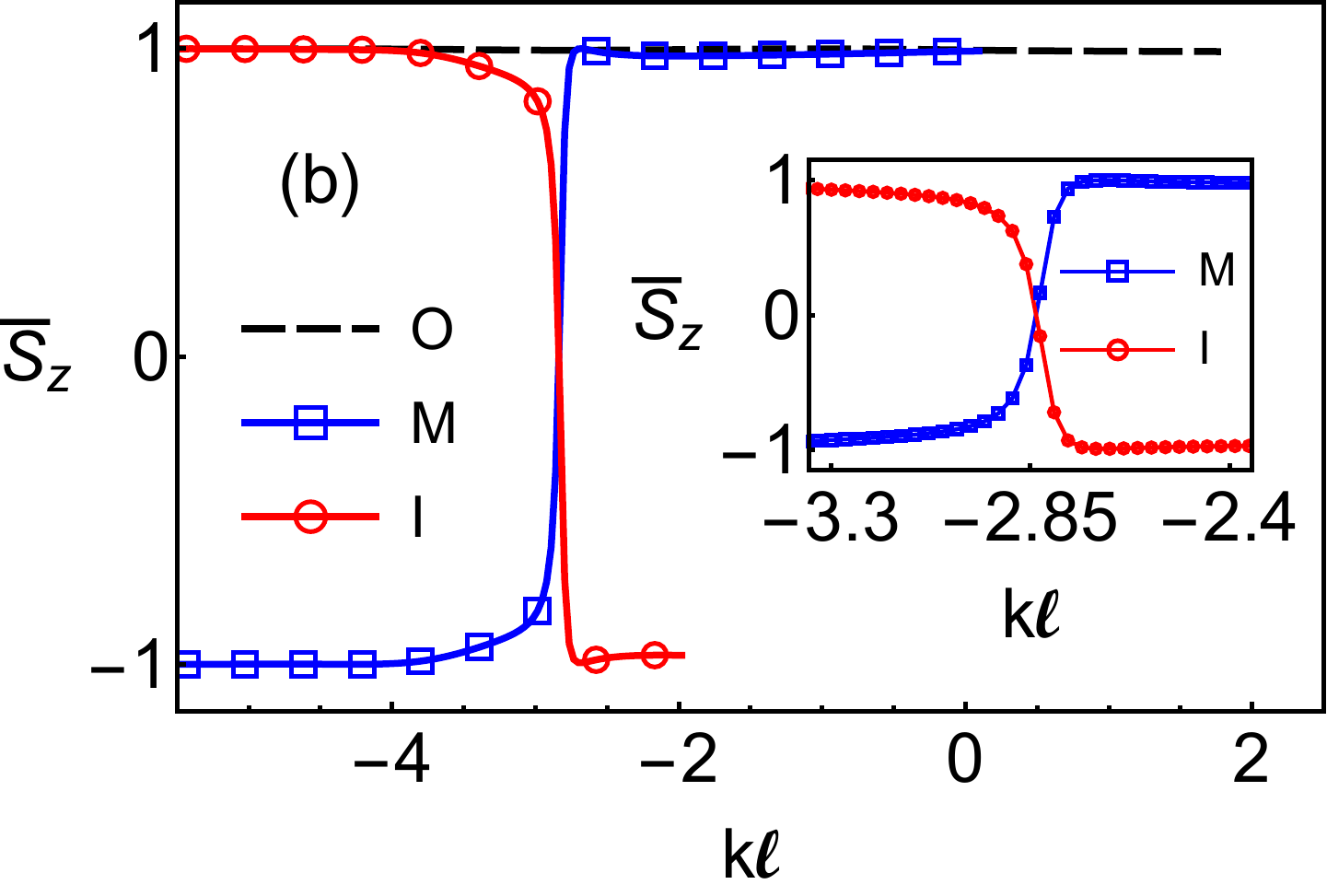}
\includegraphics[width=0.32\textwidth]{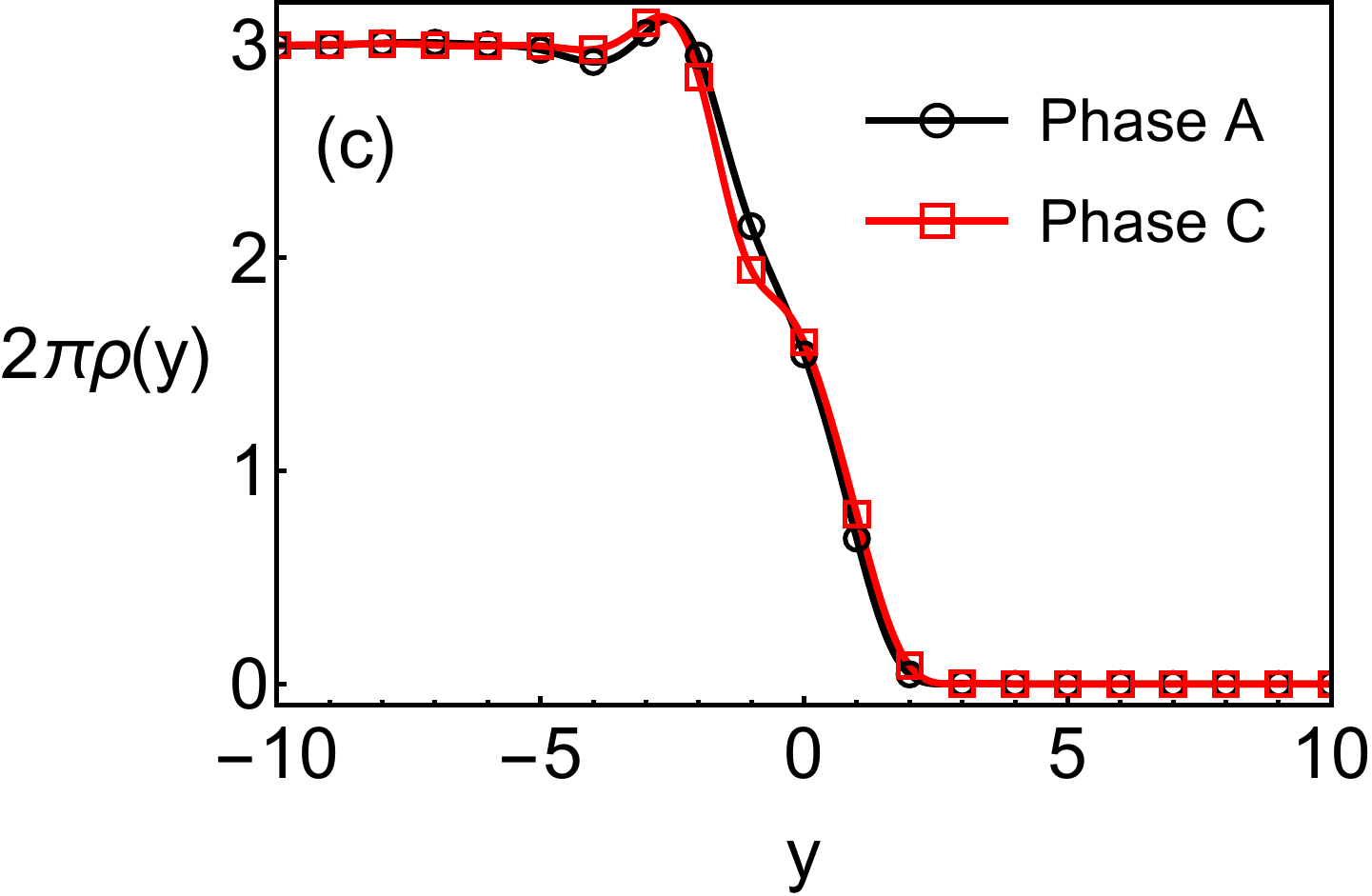}
\caption{(Color online) (a) and (b) $\bar{S}_z(i,k)$
  vs. $k\ell$ of single particle energy levels at $\tilde{E_c} = 2.3$
  in UHF.  (a) At $ W = 2.0 \ell$ the system is in Phase A, and
  $\bar{S}_z(i,k)$ are independent of $k$. (b) At $ W = 4.5 \ell$ the
  system is in Phase C. Note that the spin character in (b) changes
  smoothly near the guiding centers where the energy levels M and I
  show an avoided crossing (Fig.~\ref{fig:FigEnLv2.3}(d)). The inset makes this smooth
  variation clear. (c) The average of the electronic charge density
  $\rho(y)$ vs. $y$ just before and after the mode-switching
  transition in Phase A (at $ W = 4.0 \ell$, $\tEc = 2.3$) and in
  Phase C (at $W = 4.5 \ell$, $\tEc = 2.3$). Once again, we see that
  there is hardly any variation across the transition, supporting our
  conclusion that the spin-mode-switching transition is not driven by
  charge effects, but rather primarily by spin-exchange.  }
\label{fig:FigSz2.3}
\end{figure*}
%-------------------
%------ Fig R2 ------
\begin{figure*}[t]
\includegraphics[width=0.329\textwidth]{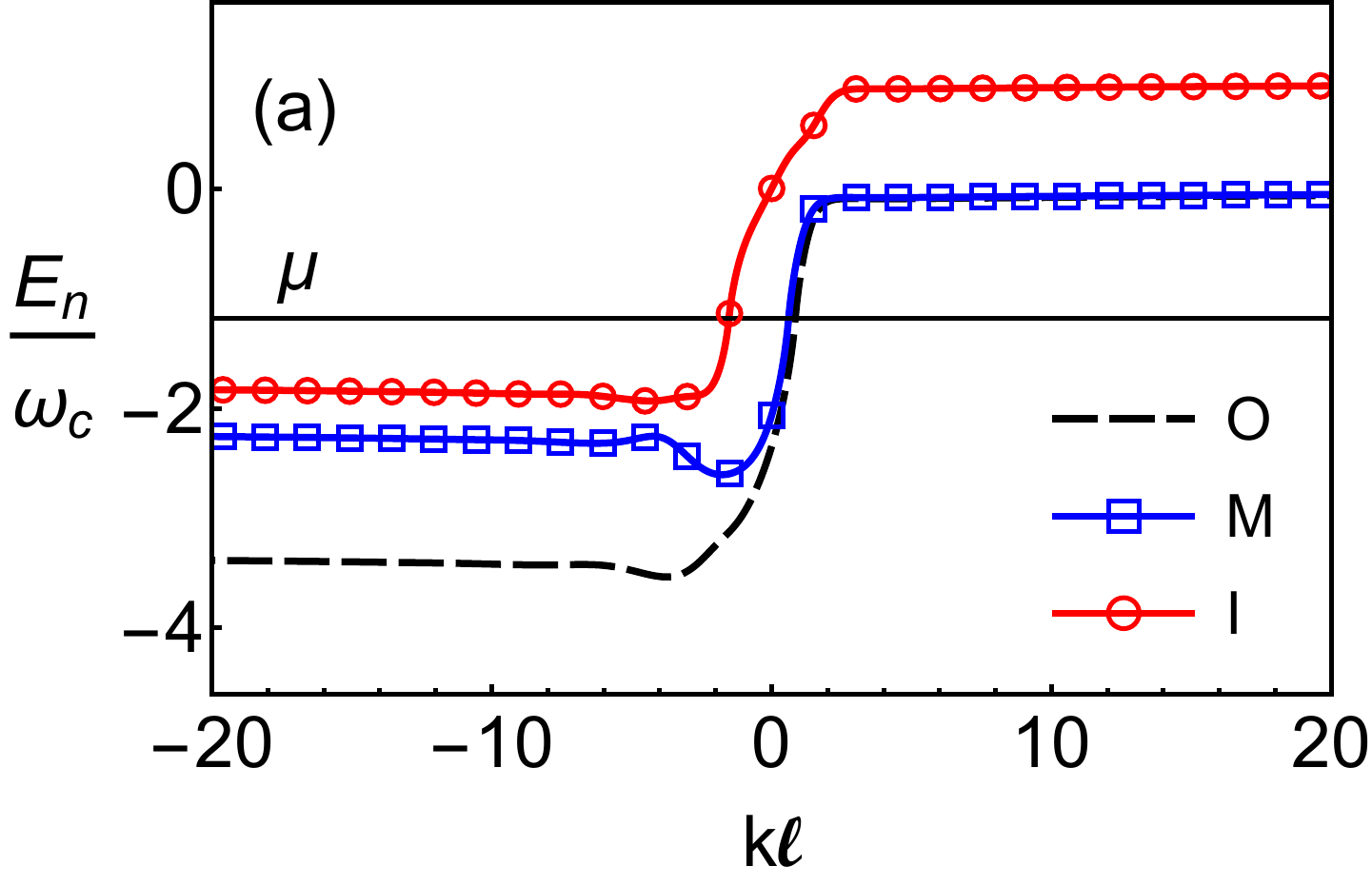}
\includegraphics[width=0.329\textwidth]{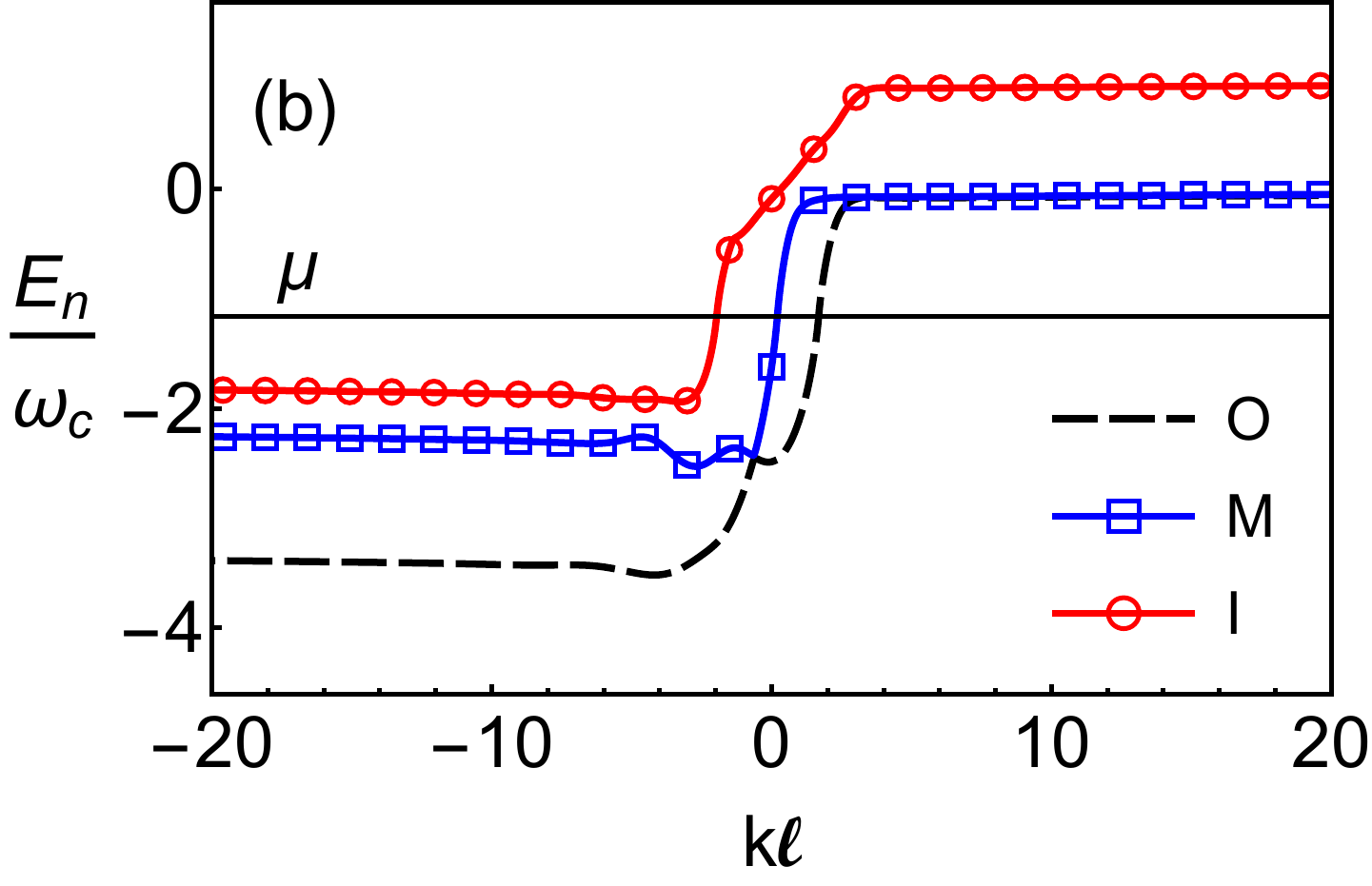}
\includegraphics[width=0.329\textwidth]{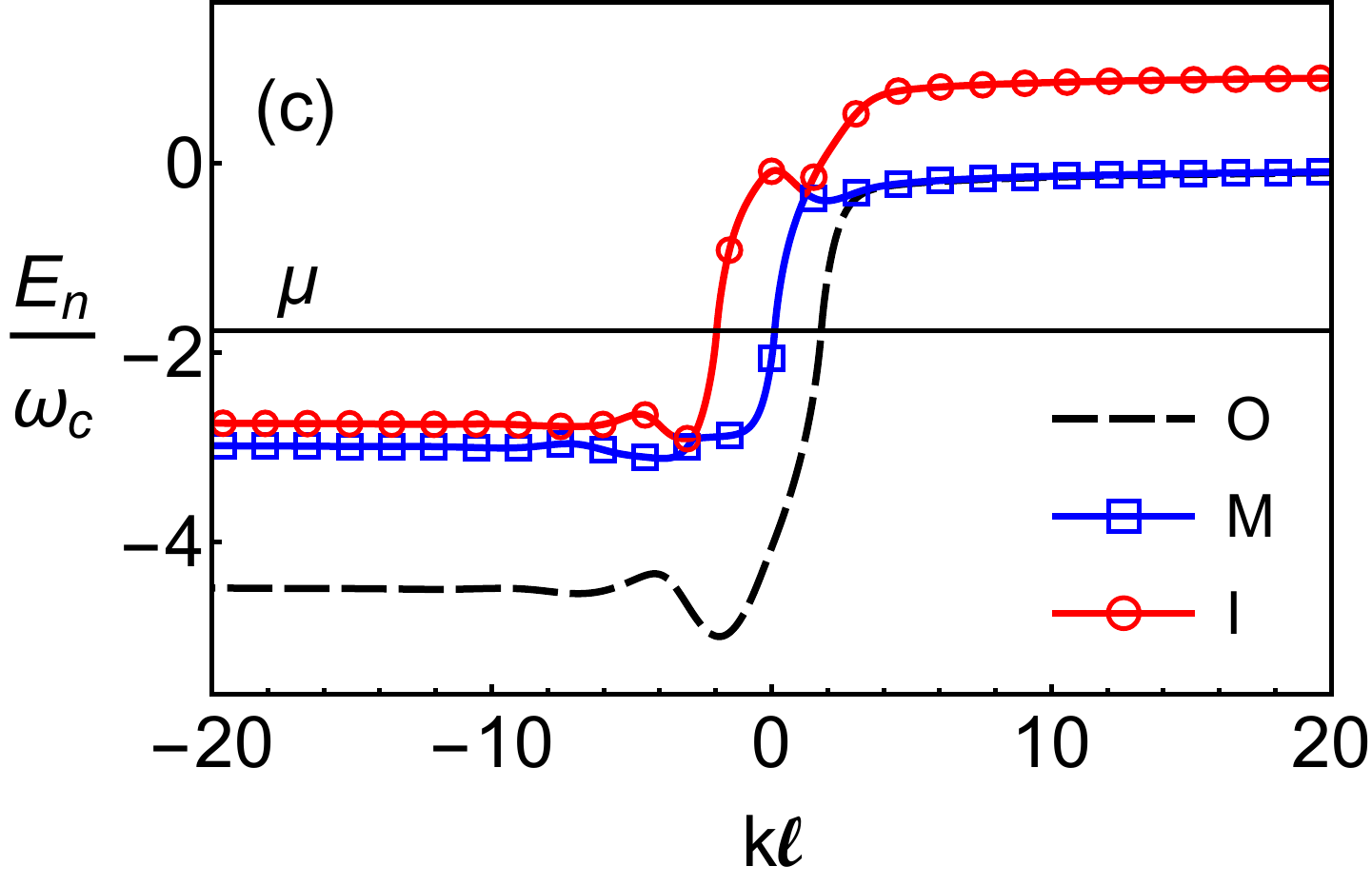}
\caption{(Color online) The self-consistent single
  particle energy level dispersion in UHF (spin-unrestricted HF, 
  i.e. without the restriction in Eq.~\ref{EqS14}) in the full 
  range of active guiding centers. The dispersion neatly converges 
  to the expected value on both the boundaries for all the values 
  of $W$ considered in this work. 
  (a) At $ W = 2.0 \ell$, $\tEc = 1.8$ (Phase A). 
  (b) At $ W = 4.0 \ell$, $\tEc = 1.8$ (Phase B). 
  (c) At $ W = 4.75 \ell$, $\tEc = 2.3$ (Phase C). } 
\label{fig:FigEnLvFull}
\end{figure*}
%-------------------
%------ Fig 6 ------
\begin{center}
\begin{figure*}[t]
\includegraphics[width=0.32\textwidth]{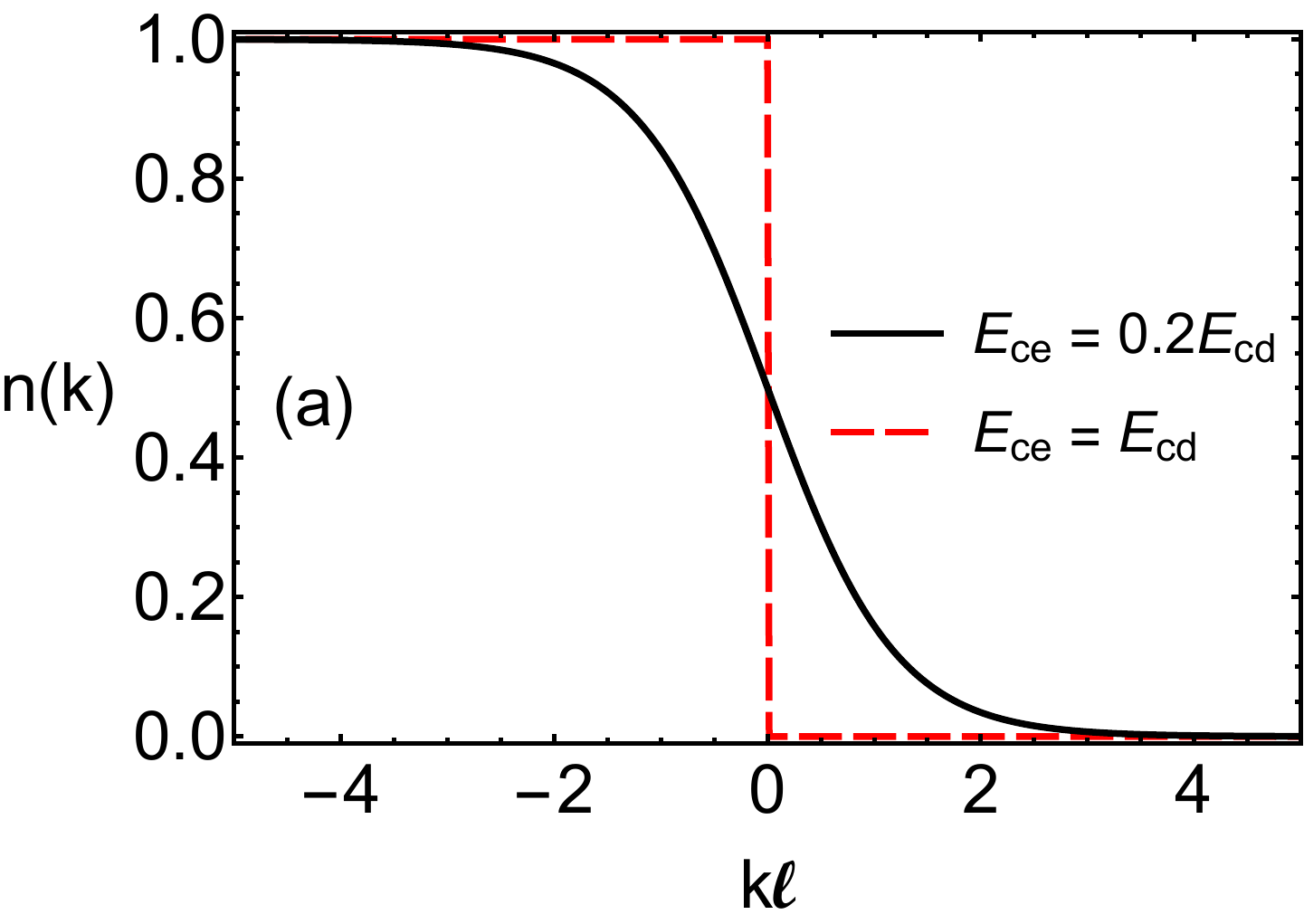}
\includegraphics[width=0.32\textwidth]{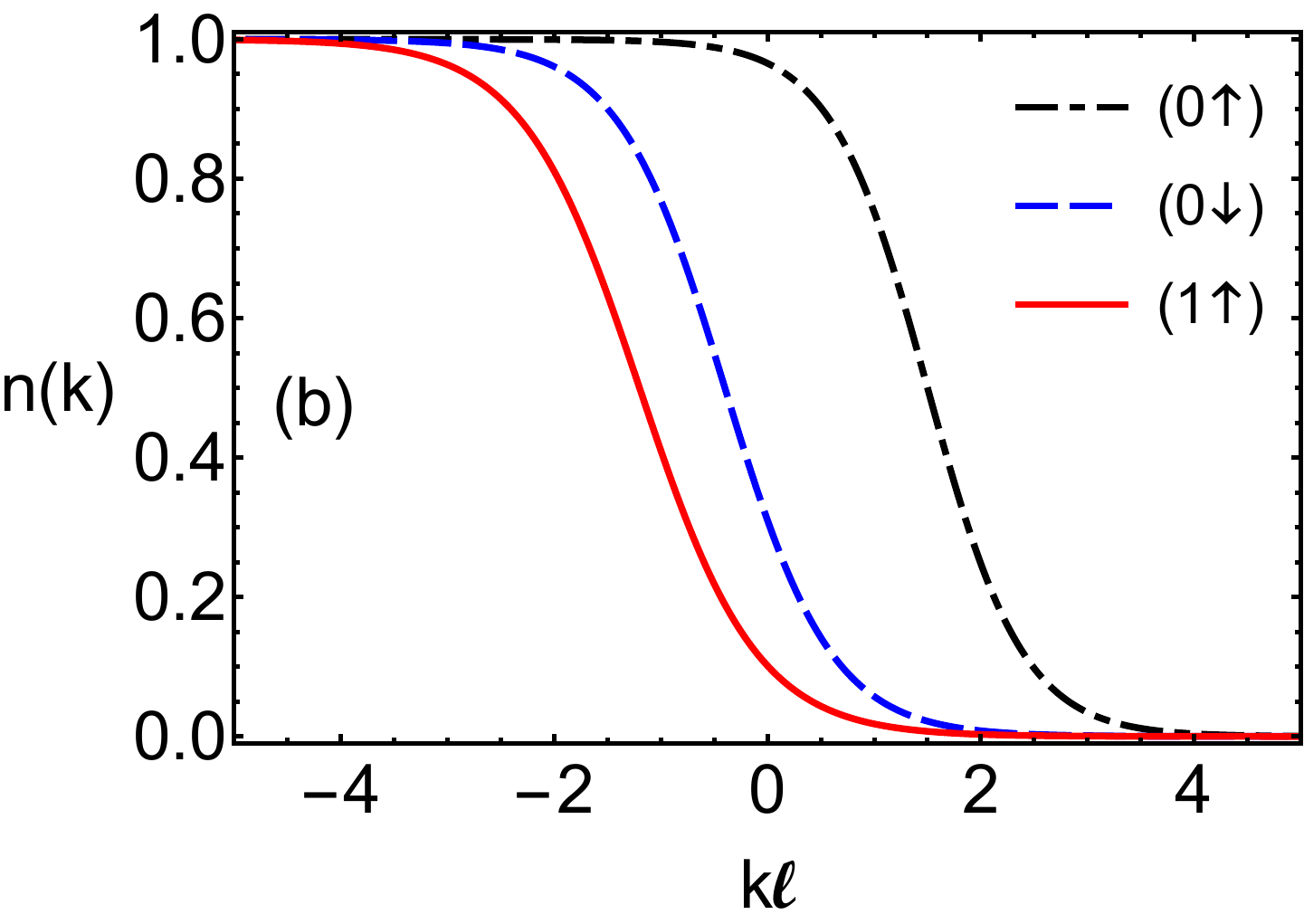}
\caption{(Color online) Ground state occupations in the variational state (a) at $\nu=1$ and (b) at $\nu=3$ (at $E_{cx} = 0.2E_{cd}$).
  At $E_{cx} = E_{cd}$, the variational ansatz reproduces the HF state.}
\label{fig:FigVariational}
\end{figure*}
\end{center}
%------------------
%------ Fig 7 ------
\begin{figure*}[ht]
\includegraphics[width=0.23\textwidth]{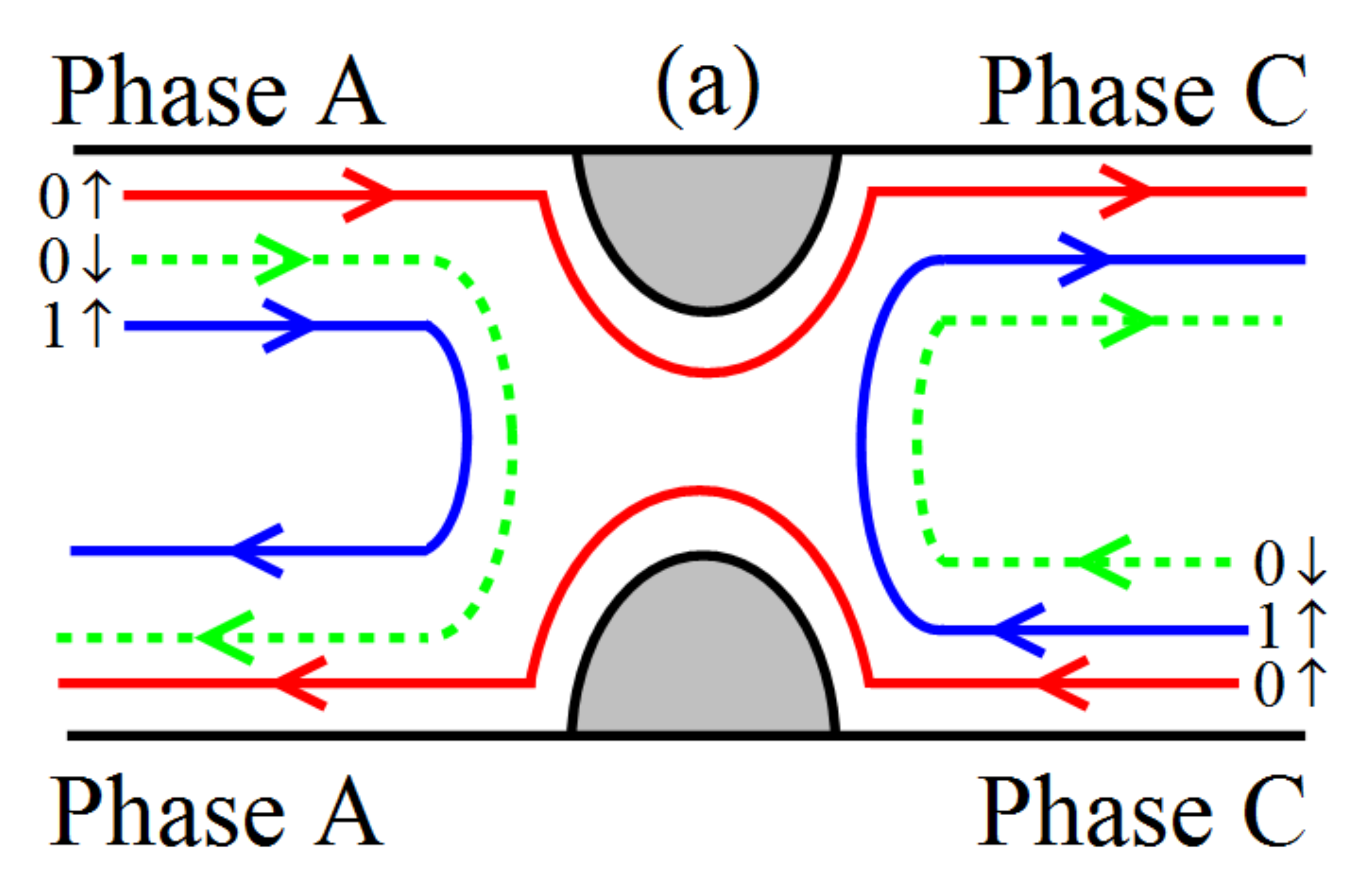}
\includegraphics[width=0.23\textwidth]{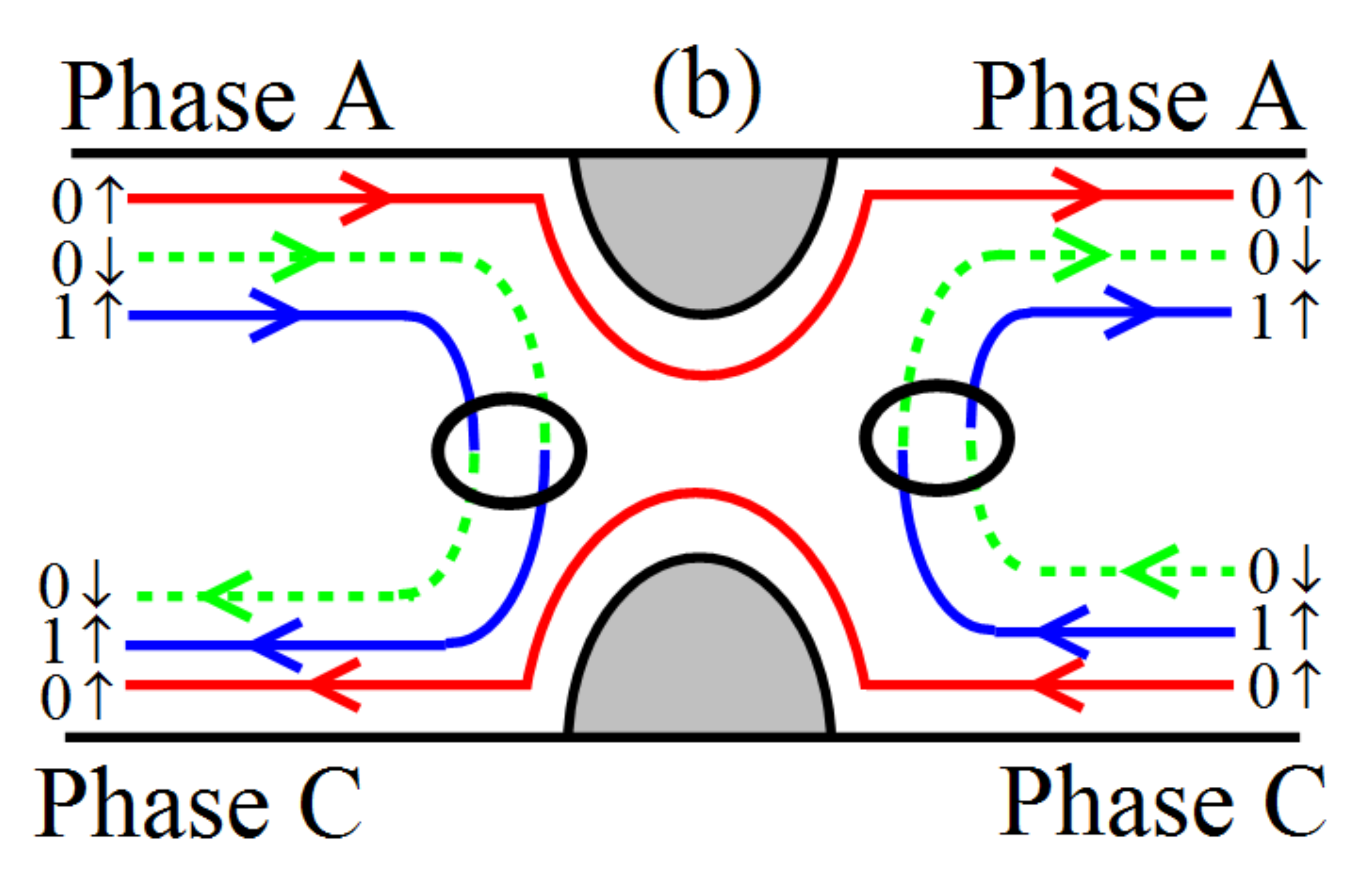}
\includegraphics[width=0.23\textwidth]{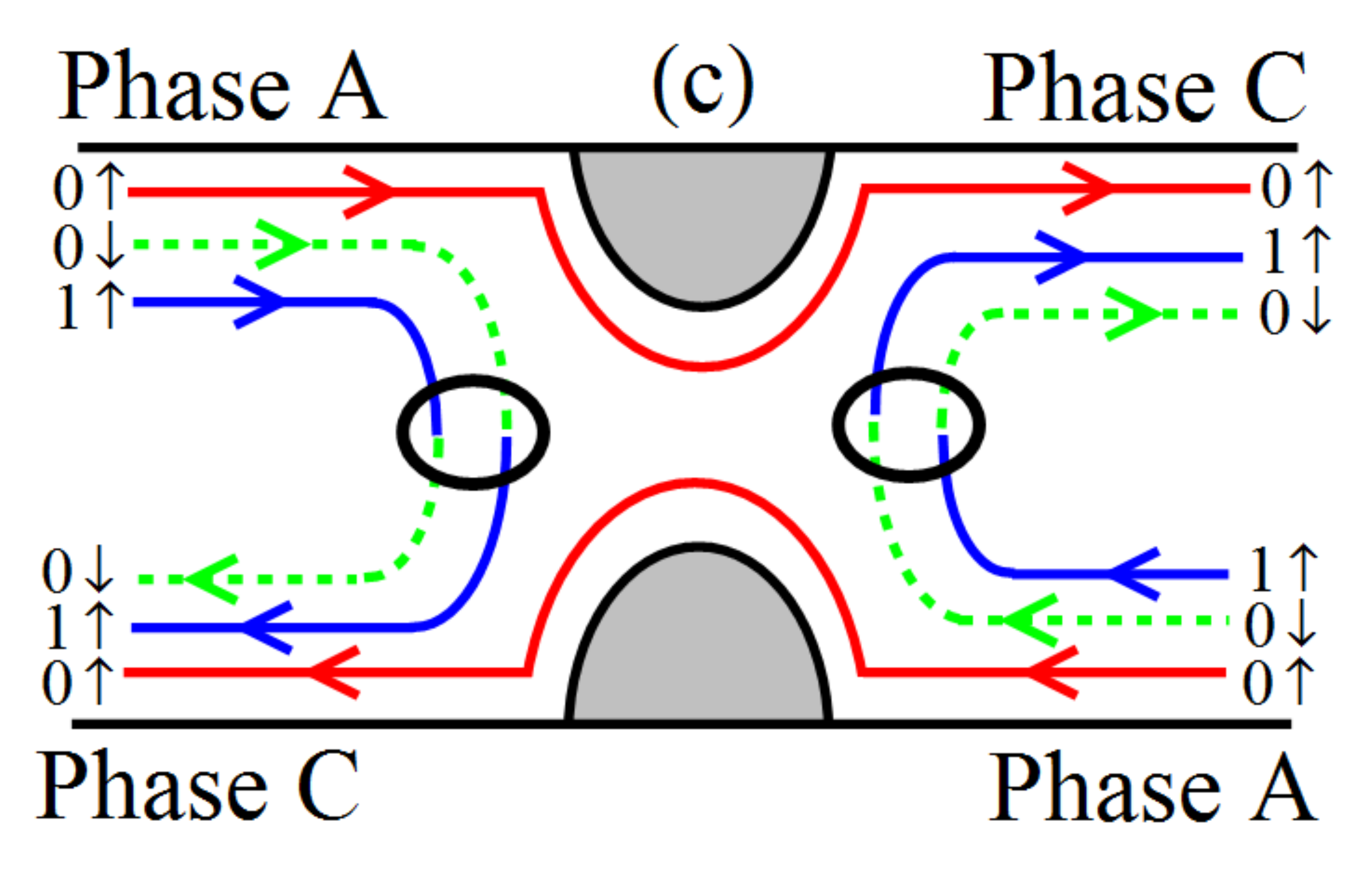}
\includegraphics[width=0.23\textwidth]{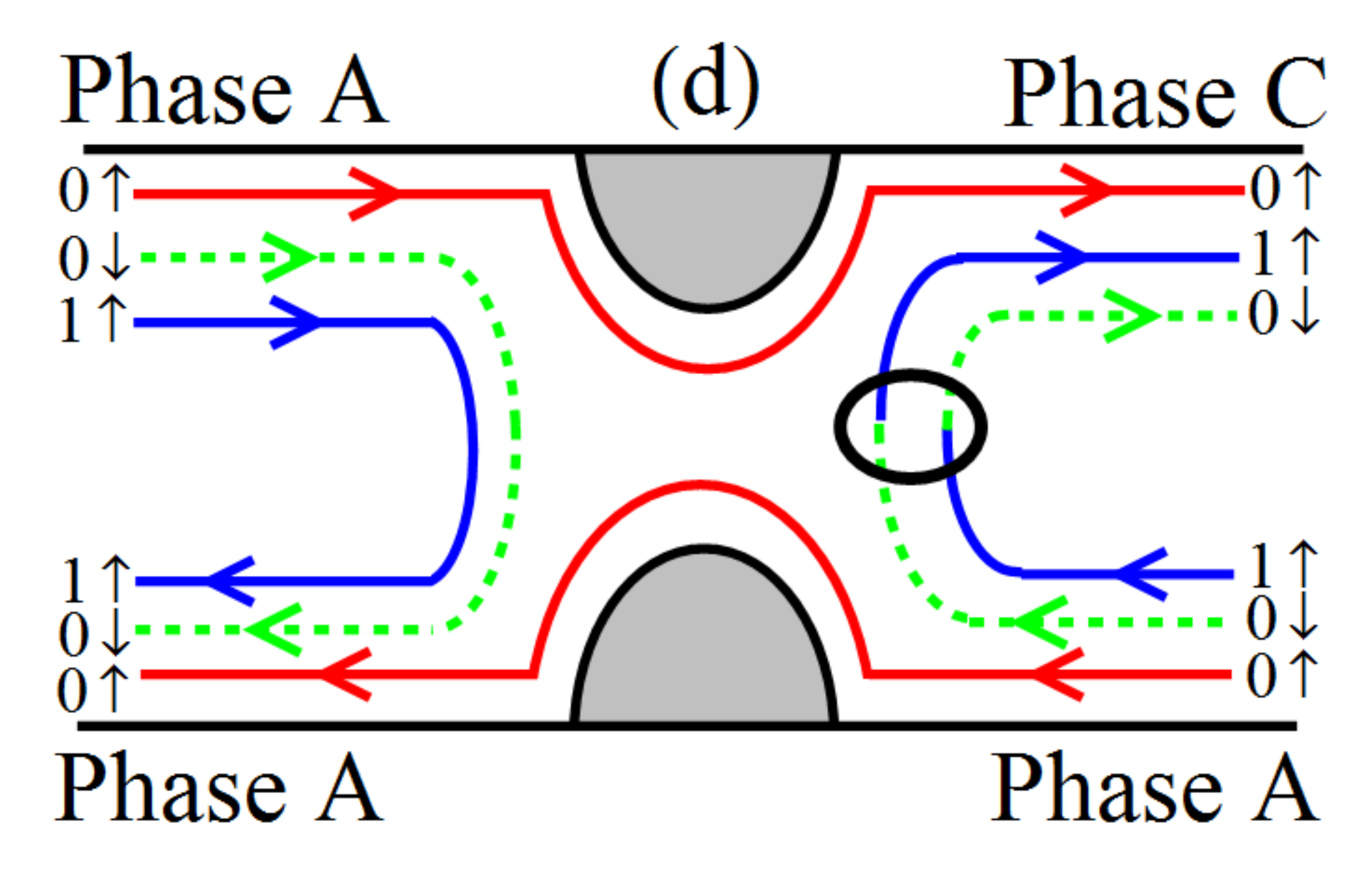}
\caption{(Color online) A single QPC tuned at the $g_2 = 1$ plateau,
  connecting regions with different confining potentials. Here
  $\tEc>2.13$ and hence the smooth edges are in Phase C. Spin
  rotations along the edge take place in the regions marked by black
  circles. The red (solid), green (dashed) and blue (solid) lines
  denote the ($0\ua$), ($0\da$) and ($1\ua$) modes
  respectively.  Disorder-induced tunneling between neighbouring
  same-spin voltage biased edge modes is implied.  (a,b) If the source
  is at the top left, neighboring modes in Phase A are unable to
  tunnel into each other, being of opposite spin. Thus, no
  backscattering is expected. However, if the source is at the bottom
  right in Phase C, disorder induced tunneling can degrade the
  current, and reduce the quality of the conductance plateau. Thus,
  the left-right symmetry of the conductance plateau is broken.  (c,d)
  Regardless of whether the source is at the top right or the bottom
  left, the source current is always in Phase A, which does not allow
  disorder-induced tunneling between neighboring modes. Thus, the
  left-right symmetry of the conductance plateau is not broken. }
\label{fig:FigS7}
\end{figure*}
%------------------
%------ Fig 8 ------
\begin{figure*}[ht]
\includegraphics[width=0.23\textwidth]{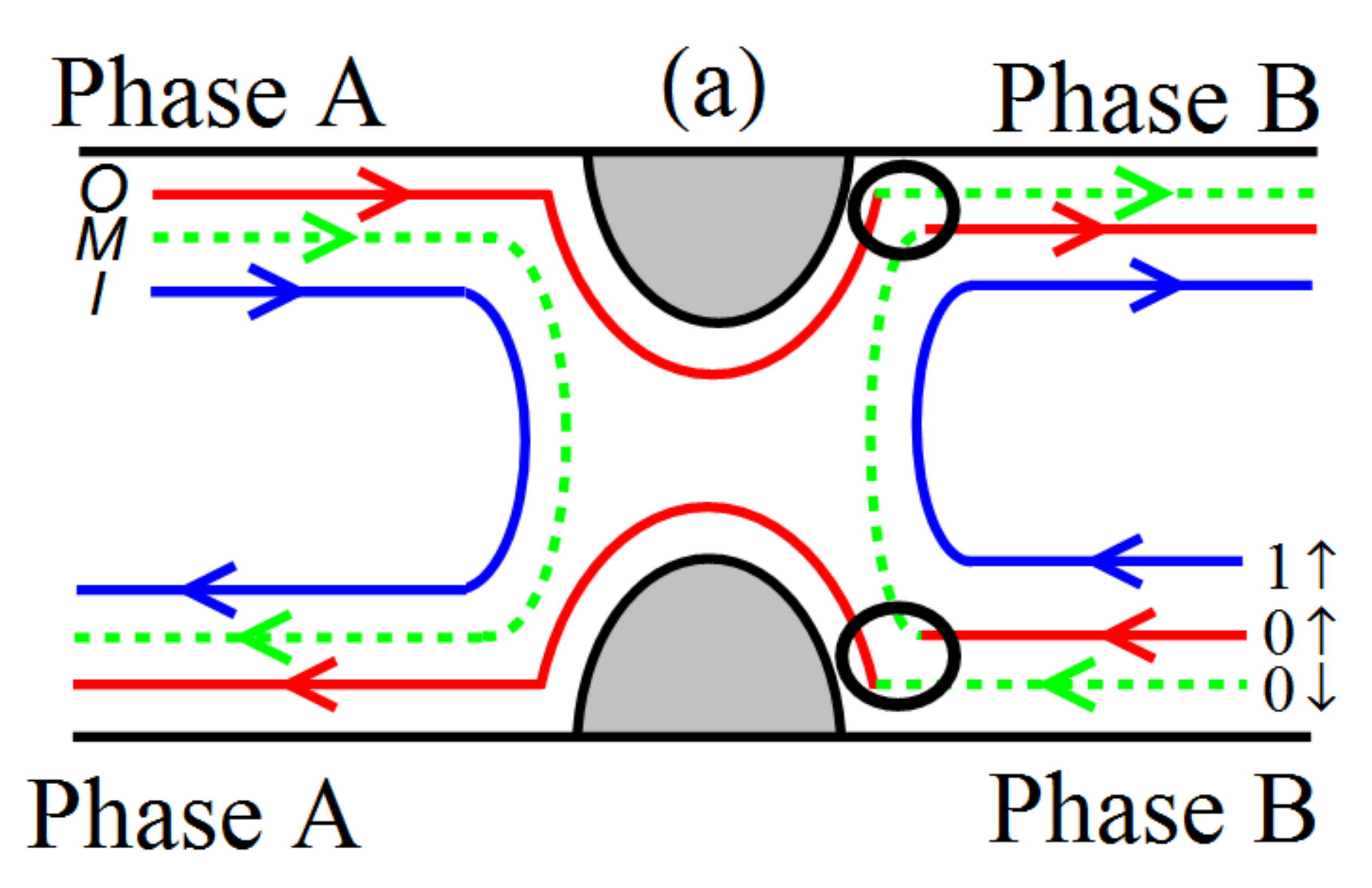}
\includegraphics[width=0.23\textwidth]{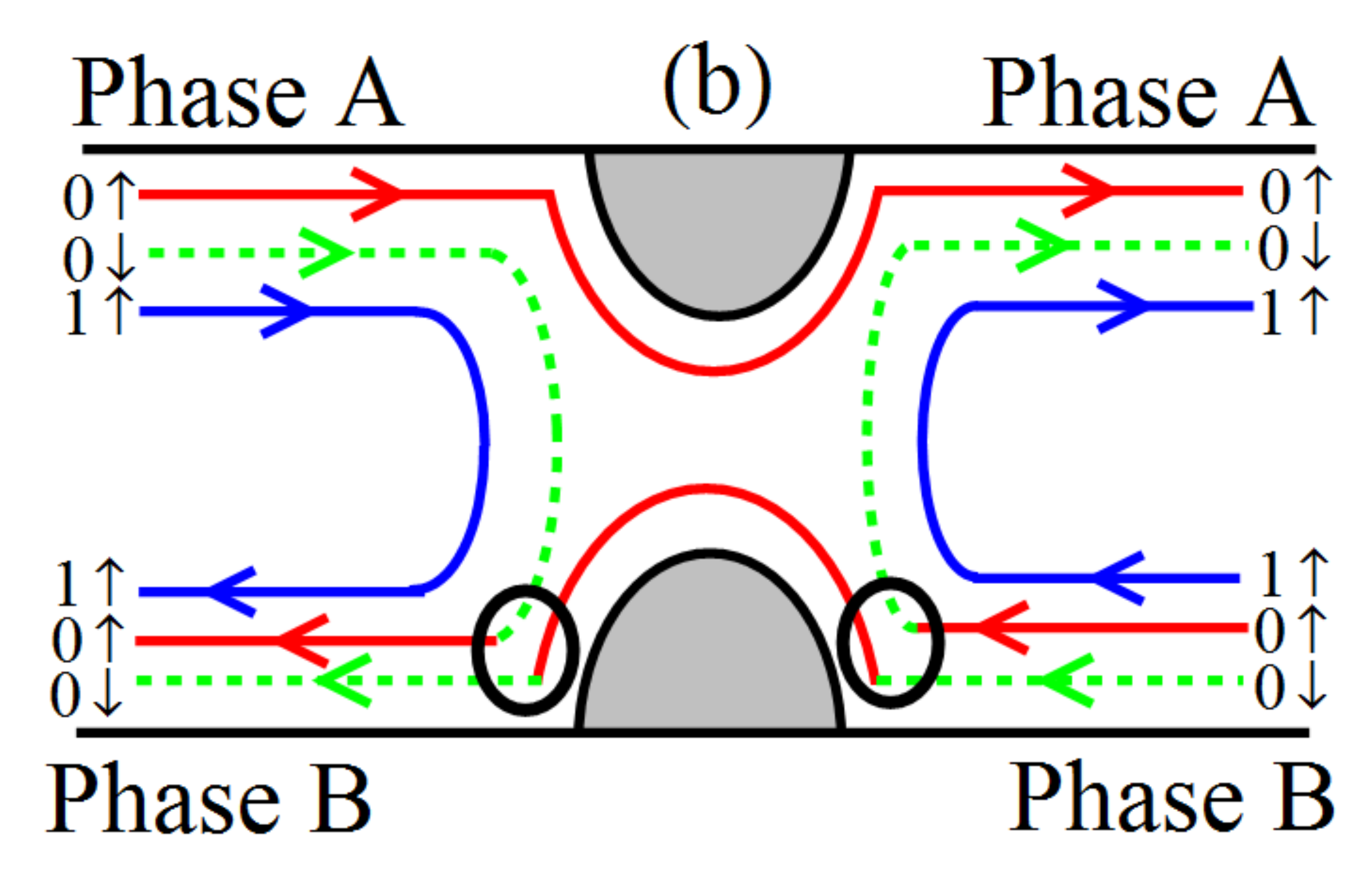}
\includegraphics[width=0.23\textwidth]{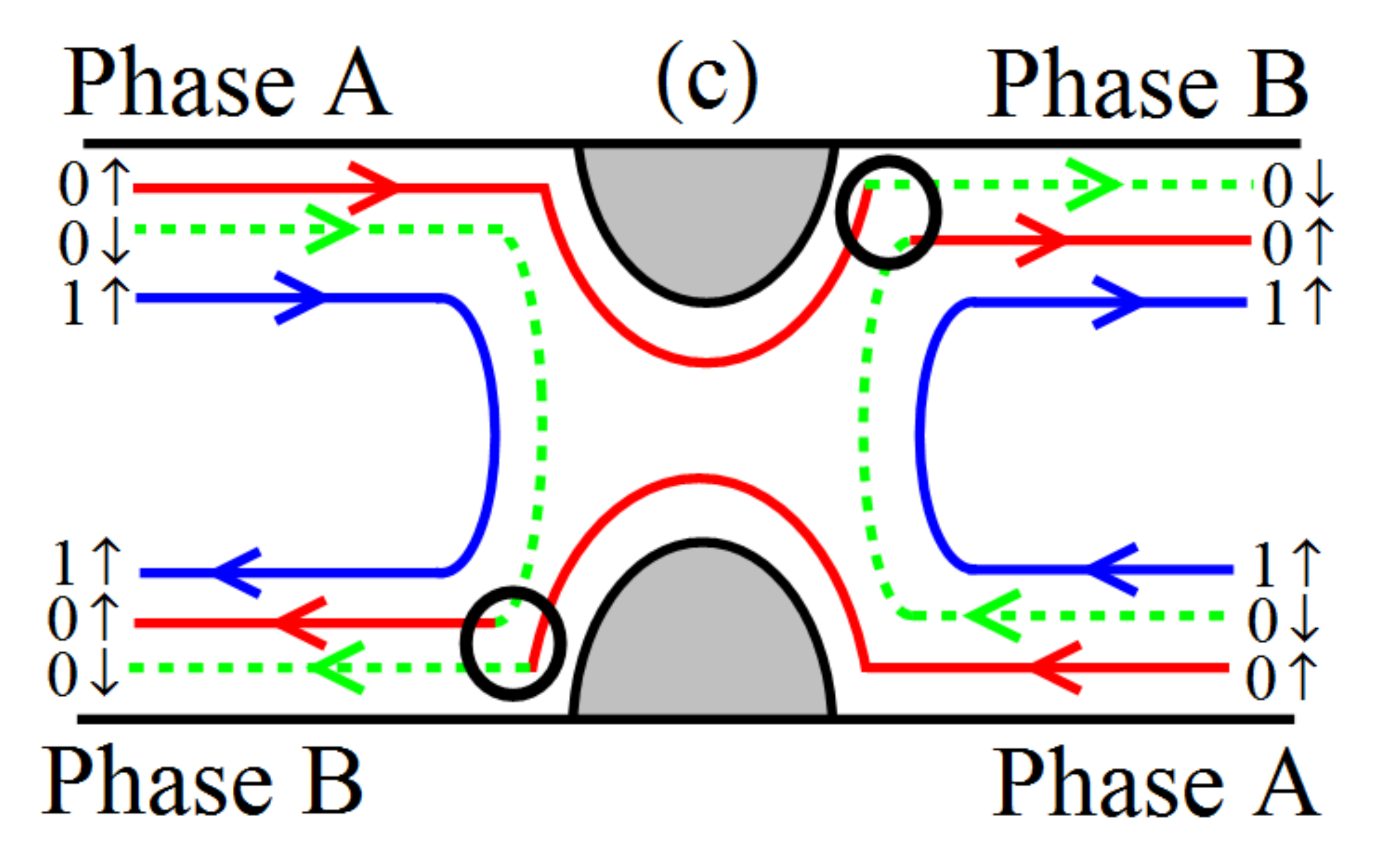}
\includegraphics[width=0.23\textwidth]{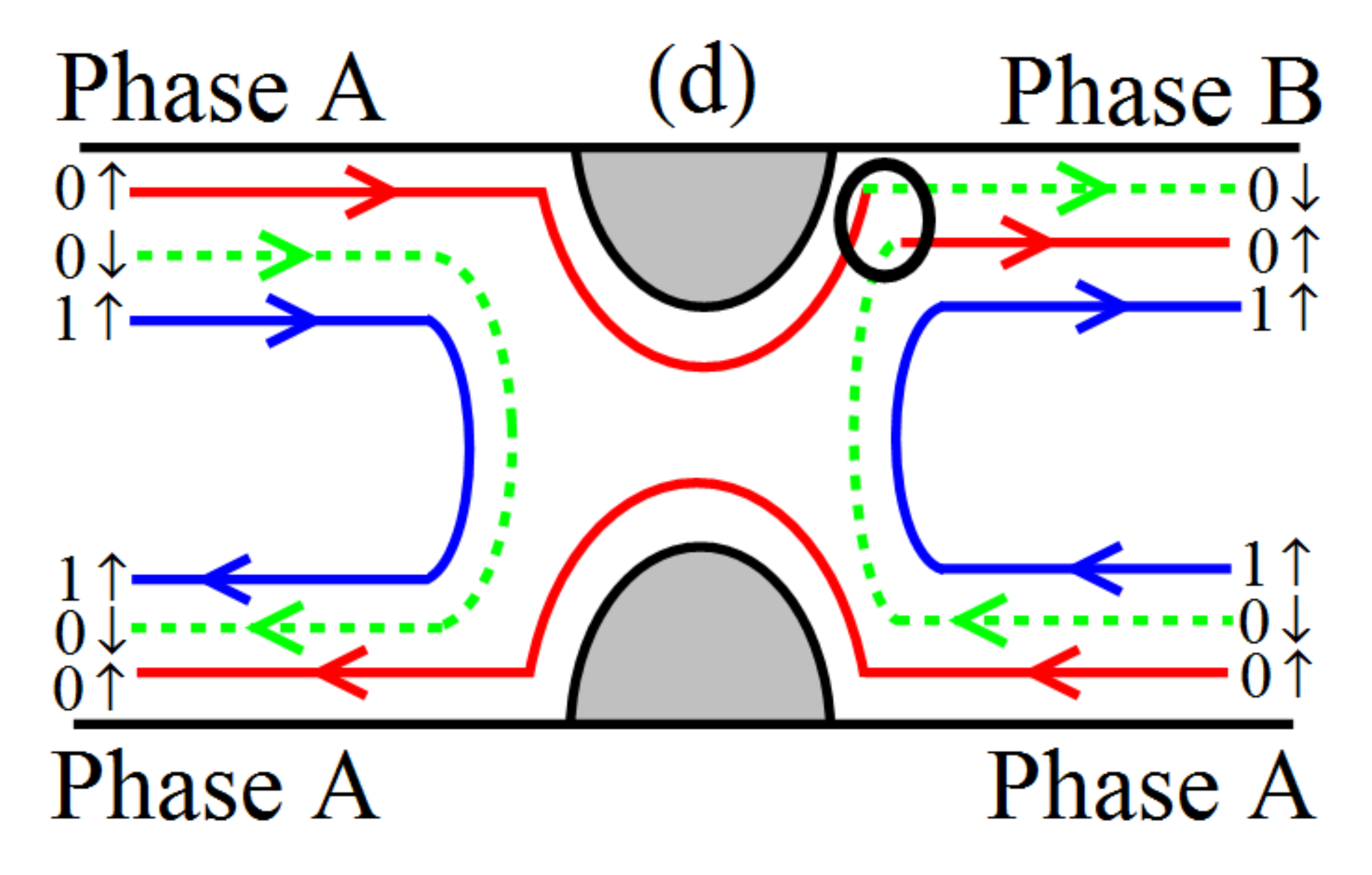}
\caption{(Color online) A single QPC tuned at the $g_2 = 1$ plateau,
  connecting regions with different confining potentials. Here
  $\tEc<2.13$, hence the smooth edges are in Phase B.  We employ the
  same notation as in Fig.~\ref{fig:FigS7}.  It is clear that in both
  Phases A and B, the two outer edges are of opposite spin. Therefore,
  disorder-induced tunneling is ineffective in degrading the
  conductance plateau at $g_2=1$, regardless of the locations of the source and
  drain. }
\label{fig:FigS8}
\end{figure*}
%------------------
%------ Fig 9 ------
\begin{figure*}[th]
\includegraphics[width=0.23\textwidth]{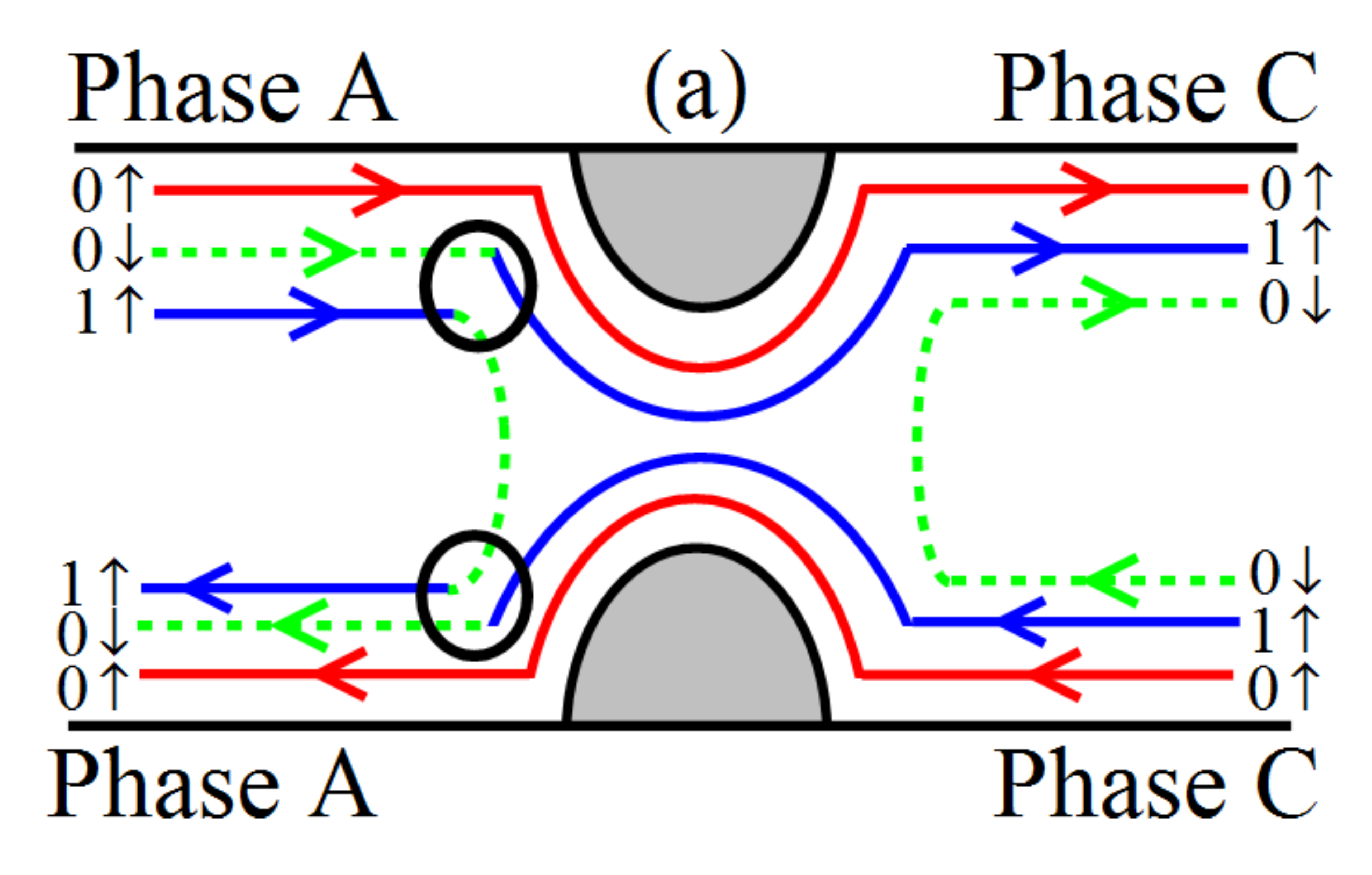}
\includegraphics[width=0.23\textwidth]{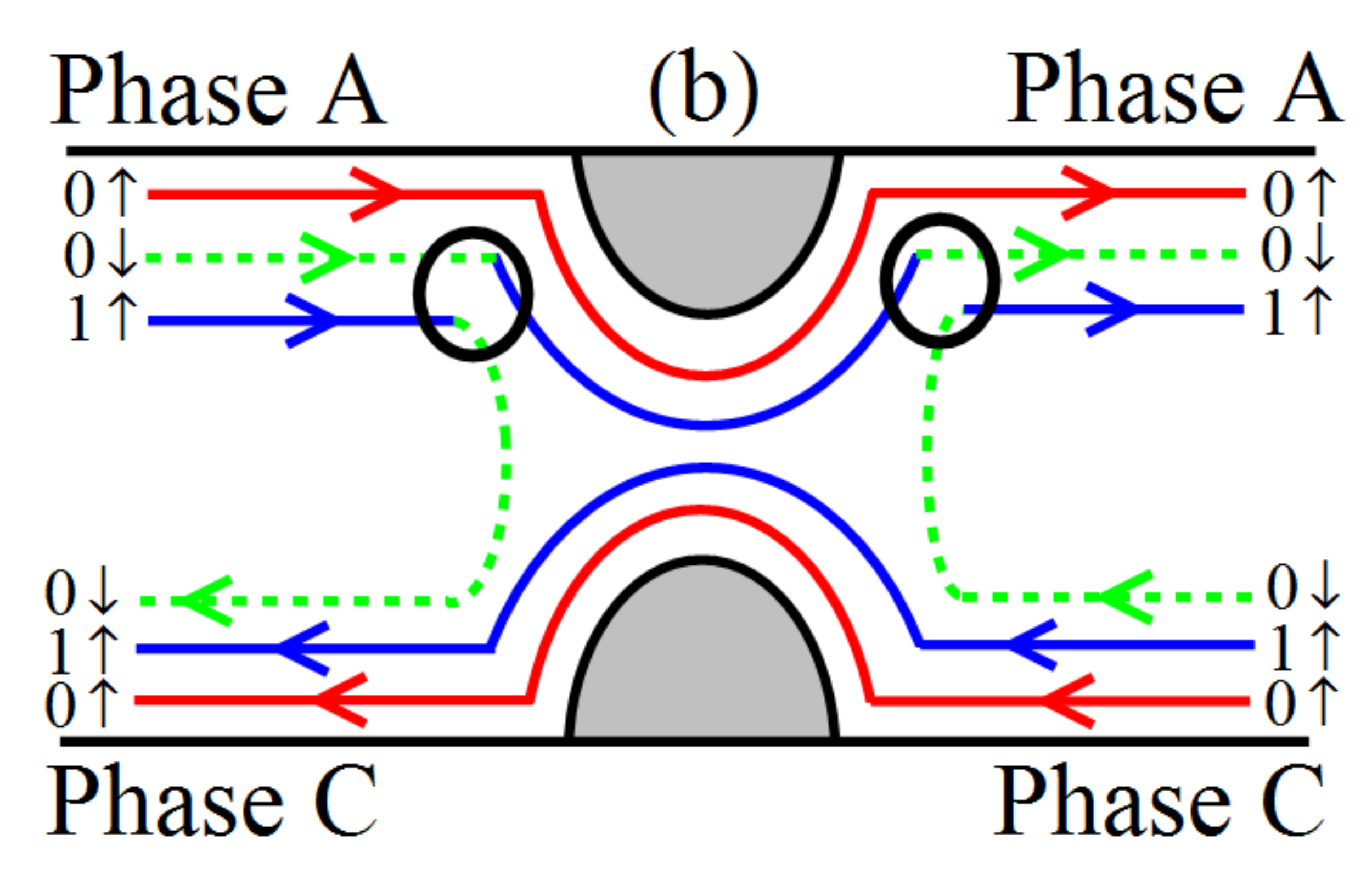}
\includegraphics[width=0.23\textwidth]{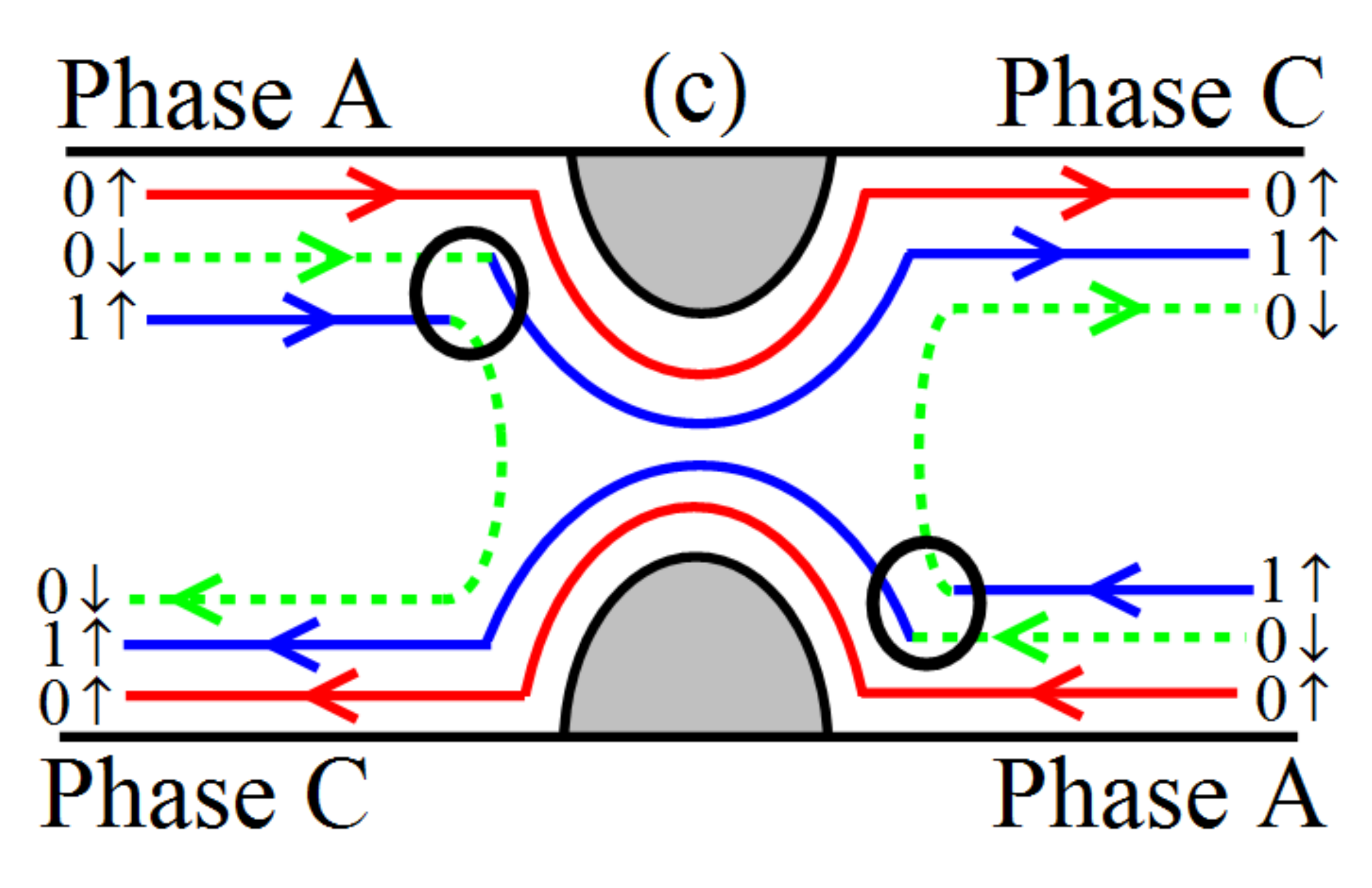}
\includegraphics[width=0.23\textwidth]{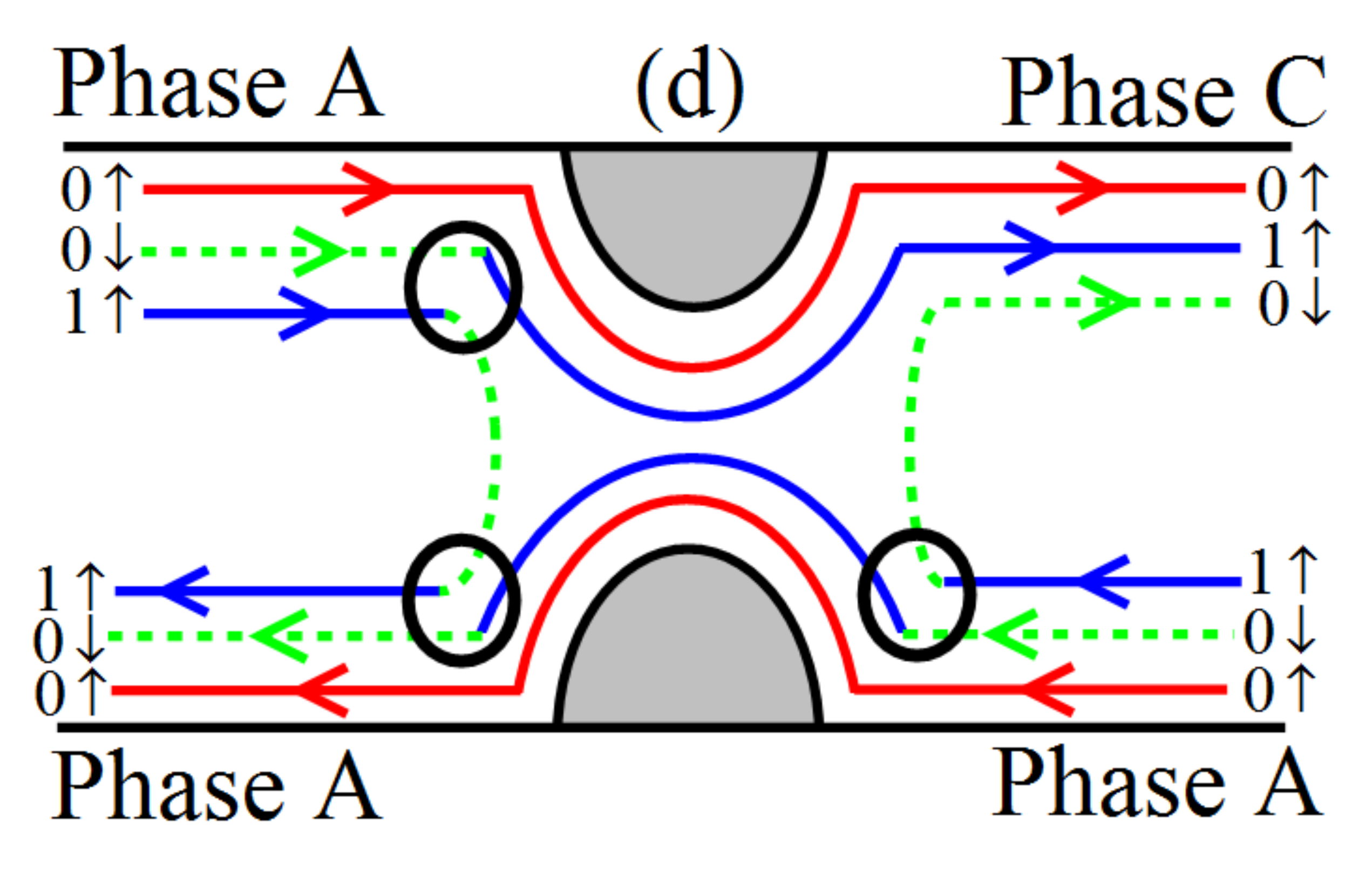}
\caption{(Color online) A single QPC tuned at the $g_2 = 2$ plateau,
  connecting regions with different confining potentials. In this
  figure $\tEc>2.13$, hence the smooth edges are in Phase C, and the
  $\nu=2$ QPC region is fully polarized.  The two outer edge modes
  carry the current. If the source is at the top left (Phase A), no
  disorder-induced tunneling can take place between any pair of
  neighboring modes to the left of the QPC. If the source is at the
  bottom right (Phase C), the innermost mode has the opposite spin of
  the two outer modes, and therefore disorder-induced tunneling is
  ineffective. Thus, the two-terminal conductance plateau at $g_2=2$
  is not undermined by disorder-induced backscattering.}
\label{fig:FigS9}
\end{figure*}
%------------------
%------ Fig 10 ------
\begin{figure*}[ht]
\includegraphics[width=0.23\textwidth]{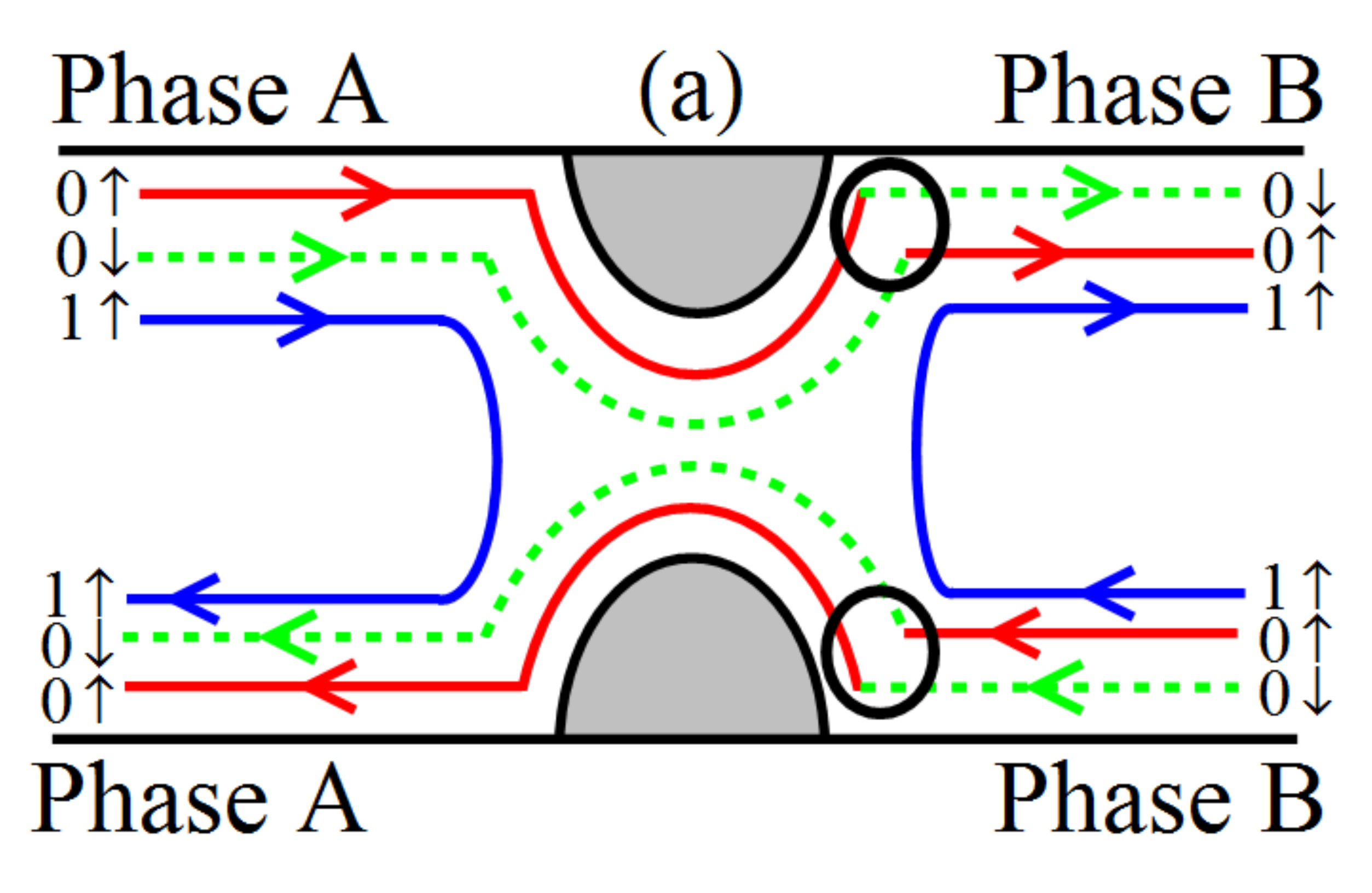}
\includegraphics[width=0.23\textwidth]{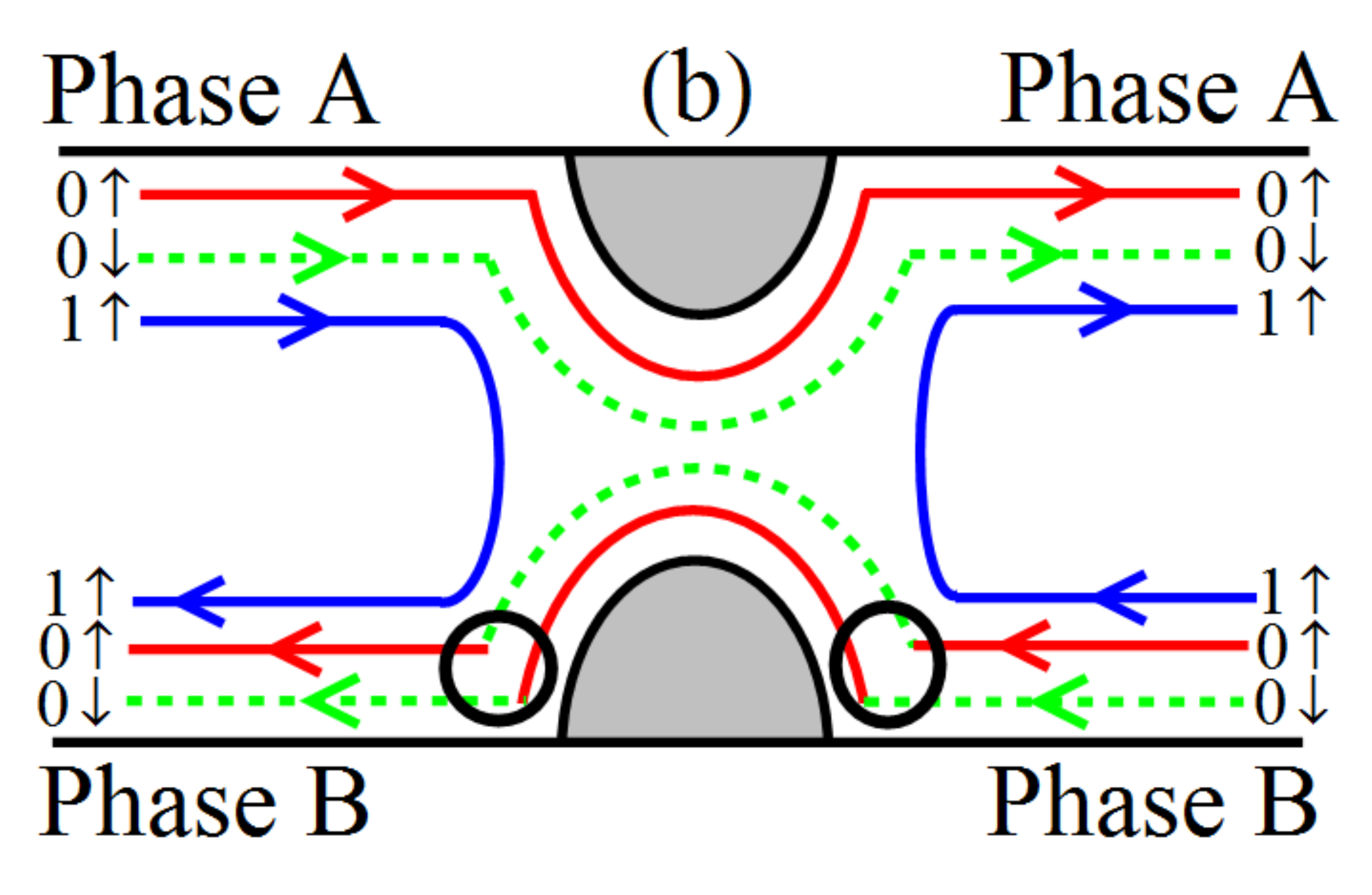}
\includegraphics[width=0.23\textwidth]{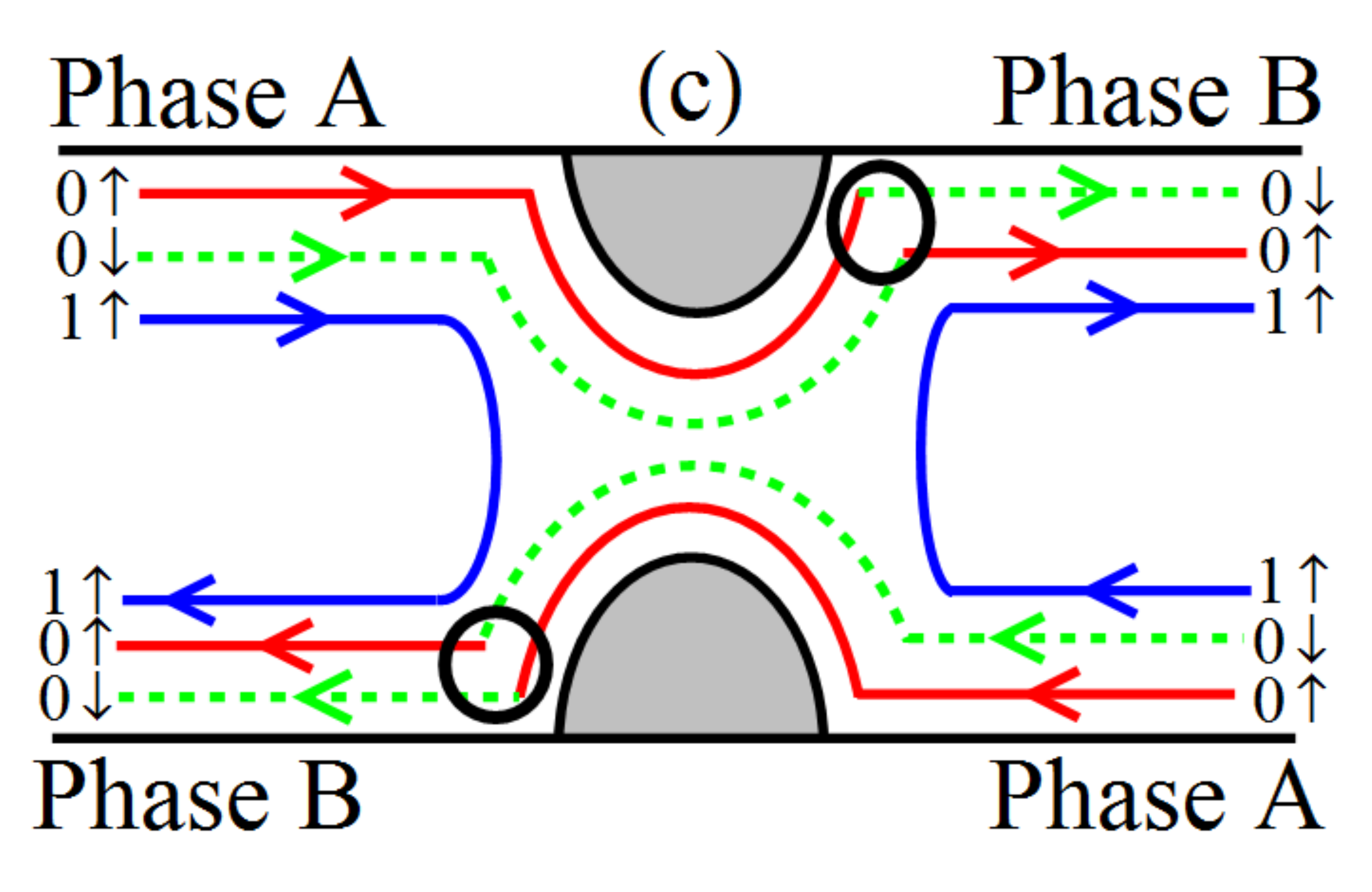}
\includegraphics[width=0.23\textwidth]{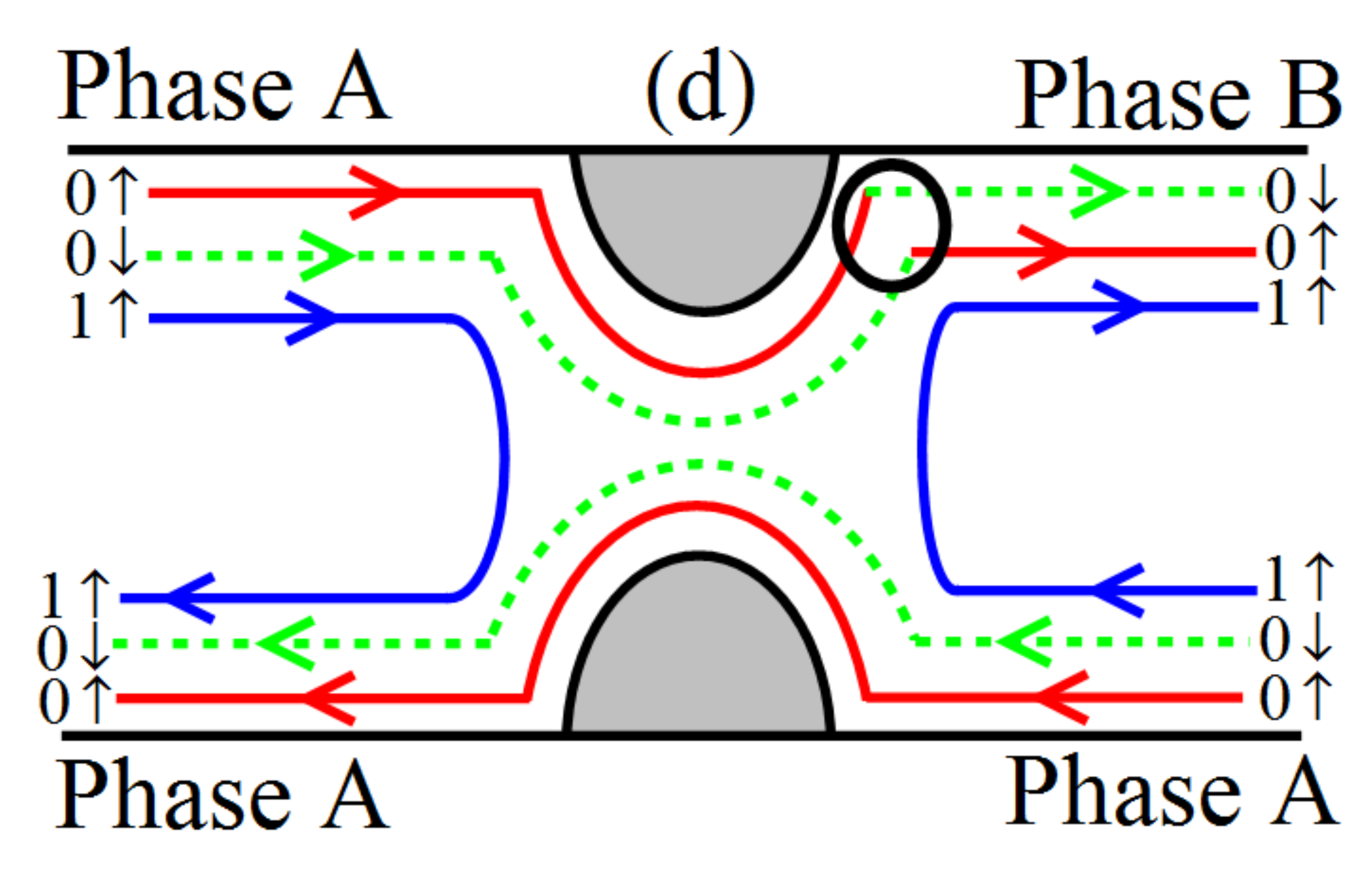}
\caption{(Color online) A single QPC tuned at the $g_2 = 2$ plateau,
  connecting regions with different confining potentials. Here
  $\tEc<2.13$, hence the smooth edges are in Phase B, and the $\nu=2$
  QPC region is unpolarized.  If the source is at the top left (Phase
  A), disorder-induced tunneling cannot degrade the current. If the
  source is at the bottom left, then in (a), (b), Phase B does allow
  disorder-induced tunneling to degrade the source current, whereas in
  (c), (d), Phase A does not. Thus, the setups of (a) and (b) will
  have a left-right asymmetry of the $g_2=2$ conductance plateau,
  while the setups of (c), (d) will not.  }
\label{fig:FigS10}
\end{figure*}
%------------------

{\bf Edge Solution:} We will assume translation-invariance along the
edge ($x$-direction) so that equation~\ref{EqS10} holds and
$\Delta_{n_1n_2;s_1 s_2}(k)$ remains diagonal in $k$.
However, close to the edge $\Delta$ does depend on $k$. Furthermore,
Landau levels with different $n$ will generically hybridize in order
to lower the overall energy, leading to LL-mixing.  We compute the
ground state in both the (spin-)restricted HF (RHF), which assumes
that $\Delta$ remains diagonal in spin indices, and in
(spin-)unrestricted HF (UHF) which relaxes this assumption and allows
spin-mixing. We describe below the numerical method used to compute
the ground state and then compare the results at the edge for
restricted and unrestricted HF.
 
After the HF averages are taken, the HF Hamiltonian becomes diagonal
in the guiding center label $k$.  We choose $L_x=200\ell$, which leads
to roughly 32 guiding centers per magnetic length $\ell$. We truncate
the Hilbert space and restrict the number of Landau levels (per spin)
to $N$ (where $N=3$ for all our results).  Two kinds of $k$ labels
enter the calculation: ``frozen'' and active. The frozen $k$ labels
occur in the range $-35<k\ell<-20$. The occupations in these values of $k$
are fixed to be those of the partially polarized bulk $\nu=3$ state
throughout the calculation. The active levels occur for $-20\leq
k\ell\leq 20$, and for these levels we allow all values of $\Delta$
consistent with translation symmetry along the edge.  In computing the
$V_{H/F}$ potentials for the active states it is important to include
the contribution of the frozen states. The presence of these frozen
states serves to impose the proper bulk boundary condition for the
active states.

The self-consistent ground state can then be found by the following
iterative procedure.  For a given set of matrix elements 
$\Delta_{n_1 n_2;s_1 s_2}(k)$ (such that the total number of electrons
is $N_e$ consistent with charge neutrality, at temperature $T$), the
HF potentials are computed through equations~(\ref{EqH}) 
and~(\ref{EqF}). Diagonalizing the $2N \times 2N$ Hamiltonian at each
guiding center gives the new single-particle states with energy
$\epsilon_{nks}$.  Next, the new chemical potential is computed
by imposing the charge neutrality condition
\begin{align}
\sum_{n,k,s} n_f(\epsilon_{nks}, \mu, T) = N_e,
\end{align}
where $n_f(\epsilon,\mu,T) = 1/(e^{\frac{(\epsilon-\mu)}{T}}+1)$ is
the Fermi-Dirac distribution.  This chemical potential is then used to
compute the new matrix elements $\Delta_{n_1 n_2; s_1 s_2}(k)$. 
The cycle is repeated until self-consistency is
achieved. In order to remove any dependence on the initial averages,
the self-consistent solution is used as the seed for a new round of
iterations in which the temperature is gradually increased and then
decreased to zero.

In RHF, we assume 
\begin{align} 
\Delta_{n_1 n_2;s_1 s_2} (k) = \delta_{ s_1 s_2 } \Delta_{n_1 n_2}(k,s_1). 
\label{EqS14}
\end{align}
Since spin is conserved in this approximation, we can label the final
set of hybridized single-particle levels by spin.  We use the
non-interacting ground state as the starting seed for the iteration in
this case. The self-consistent single-particle energy levels
vs. $k\ell$ at $\tilde{E_c}=1.8$ and two different values of $W$ are
shown in Fig.~\ref{fig:FigEnLv1.8}(a),(b).  Note that at $\tilde{E_c}
= 1.8$ (Fig.~\ref{fig:FigEnLv1.8}(a) and (b)) the order of the levels,
from the outermost in, at the chemical potential $\mu$ follows the
bulk order or increasing energies. This is expected in Phase A, but is
unexpected in Phase B. The reason for this result is that RHF is not
able to find Phase B, so the RHF result depicts a
metastable version of Phase A.

Now we turn to Fig.~\ref{fig:FigEnLv2.3}(a),(b) where we show the RHF
results at $\tilde{E_c} = 2.3$ for two different values of $W$. Once
again, at $W=2\ell$ (Fig.~\ref{fig:FigEnLv2.3}(a)) we are in Phase A, and no level
crossings are expected, and indeed none are seen.  However, at
$W=4.5\ell$ (Fig.~\ref{fig:FigEnLv2.3}(b)) there is a crossing of the $0\da$ and $1\ua$
levels below $\mu$. This leads to the order of the edge modes being,
from the outside in, $0\ua$, $1\ua$ and $0\da$, thus showing a
spin-mode-switching transition from phase A to C. In the absence of
spin-mixing, the $0\da$ and $1\ua$ levels can be degenerate, and as
shown in Fig.~\ref{fig:FigEnLv2.3}(b), cross each other below and/or 
above the chemical potential. 

Now we allow spin-mixing to occur in UHF. Operationally, we generate a
spin-mixed seed by rotating the spins of the various single-particle
levels in the RHF ground state in a small region close to $k\ell=0$.
The single-particle levels can no longer be labelled by spin, so we
index them by $i=$ O (outermost), M (middle), and I
(innermost).  The energy level dispersions vs. $k$ for
$\tilde{E}_c=1.8$ for two values of $W$ are shown in
Fig.~\ref{fig:FigEnLv1.8}(c),(d), and for $\tilde{E}_c=2.3$ for two
values of $W$ are shown in \ref{fig:FigEnLv2.3}(c),(d). 

As in the main paper, to elucidate the physics of the
spin-mode-switched phases, we compute ${\bar S}_z(i,k)$, the average
of the operator $S_z$ in the single-particle state labelled by $i=$ O,
M, I, at position $k\ell$. The importance of ${\bar S}_z(i,k)$ is that
its value at the position where it crosses the chemical potential,
$\bar{S}_{z\mu}(i)$, determines the spin of the chiral edge
mode. ${\bar S}_z(i,k)$ vs. $k\ell$ for occupied levels $i$ are shown
in Fig.~\ref{fig:FigSz1.8}(a),(b) and \ref{fig:FigSz2.3}(a),(b),
corresponding to the level dispersions of
Fig.~\ref{fig:FigEnLv1.8}(c),(d) and \ref{fig:FigEnLv2.3}(c),(d)
respectively.

As seen in Fig.~\ref{fig:FigSz1.8}(a) and \ref{fig:FigSz2.3}(a), in
Phase A ($W = 2.0 \ell$), $\bar{S}_z(i,k)$ are almost independent of
$k$. In particular, at the point where the single-particle levels
cross the chemical potential, the innermost (I) and outermost (O)
levels have $\bar{S}_z\approx 1$ while the middle level (M) has
$\bar{S}_z\approx -1$.  Fig.~\ref{fig:FigSz1.8}(b) and
\ref{fig:FigSz2.3}(b) show the $\bar{S}_z$ after the mode-switching
transition in Phase B ($W = 4.0 \ell$) and C ($W = 4.5 \ell$)
respectively. Clearly, the ${\bar S}_z(i,k)$ of two of the single
particle levels (O \& M in Phase B and I \& M in Phase C) vary
smoothly as they go through the avoided crossing. The insets show an
expanded view near the location where most of the changes in ${\bar
  S}_z(i,k)$ occur. This is, in fact, the best evidence that the
apparent touching of levels seen in UHF in 
Fig.~\ref{fig:FigEnLv1.8}(d) and~\ref{fig:FigEnLv2.3}(d) are
actually avoided crossings. Had they been actual crossings, 
${\bar S}_z(i,k)$ would have changed discontinuously as functions of $k$.

To summarize, in UHF, levels undergo avoided crossing, and the ${\bar S}_z(i,k)$
change smoothly as a function of $k\ell$. However, since the avoided
crossings occur below $\mu$, the values of ${\bar S}_z(i,k)$ at the
$\mu$-crossing, ${\bar S}_{z\mu}(i)$, do change discontinuously
through the transition as shown in Fig.~1(e) and Fig.~1(f) of the main
text.

Finally, let us look at Fig.~\ref{fig:FigSz1.8}(c) and \ref{fig:FigSz2.3}(c). 
These show the average of the electronic charge density close to the spin-mode-switching
transition, but on either side of it. In Fig.~\ref{fig:FigSz1.8}(c), we focus on the
Phase A $\to$ Phase B transition. It is seen that the electronic
charge density hardly changes across the transition. The same is true
of the Phase A $\to$ Phase C transition in Fig.~\ref{fig:FigSz2.3}(c). This is the
basis of our conclusion in the main text that the spin-mode-switching
transitions are not primarily driven by charge effects, but rather
primarily by spin-exchange effects.

Fig.~\ref{fig:FigEnLvFull}(a)-(c) show the single particle energy levels
in the full range of active guiding centers ($-20 < k \ell < 20$) 
in Phases A, B and C. We note that the occupations and dispersions converge
to the expected values in the bulk (on the left) and in the vacuum (on the
right) in all three cases shown here and for all values of $W$ considered 
in this work.

\section{III.\,\,\,\,\,\,   Variational Calculation}

One limitation of HF is that the ground state can only be a Slater
determinant of some set of single-particle states. Thus, at $T=0$ the
occupation of any single-particle state is either $0$ or $1$. To
overcome this limitation, we consider a class of variational states
that allow the occupations to lie between $0$ and $1$ even at $T=0$.
These states do not conserve particle number but we choose parameters
so that the average number maintains charge neutrality.  To be
specific, for the edge of $\nu=3$ we consider states of the form
\begin{widetext}
\begin{align}
  |\psi \rangle &= \prod\limits_{k} (U_k + V_{0 k} e^{i \theta_{0 k}} c^{\dagger}_{0 k \ua} + V_{1k} e^{i \theta_{1 k}} c^{\dagger}_{0 k \da} c^{\dagger}_{0 k \ua} + V_{2k} e^{i \theta_{2 k}} c^{\dagger}_{1 k \ua} c^{\dagger}_{0 k \ua} + V_{3 k} e^{i \theta_{3 k}} c^{\dagger}_{1 k \ua}  c^{\dagger}_{0 k \da} c^{\dagger}_{0 k \ua}) |0 \rangle .
\end{align}
\end{widetext}
\noindent
where $U_k$, $V_{i k}$ and $\theta_{i k}$ are real
numbers. Normalizing $|\psi \rangle$ imposes the condition $U_k^2 +
\sum_i V_{i k}^2 = 1$ for each guiding center $k$, leaving us with 
8 free parameters for each guiding center ($V_{ik}$ and
$\theta_{ik}$).  Note that the HF state without LL-mixing is
a member of this class of states. The average occupation of each LL is,
\begin{align}
  \langle c^{\dagger}_{0 k \ua} c_{0 k \ua} \rangle &= 
  \sum_{i=0}^{4} V_{i k}^2 \\
  \langle c^{\dagger}_{0 k \da} c_{0 k \da} \rangle &= V_{1k}^2 + V_{3k}^2 \\
  \langle c^{\dagger}_{1 k \ua} c_{1 k \ua} \rangle &= V_{2k}^2 + V_{3k}^2 
\end{align}
We also note that $|\psi \rangle$ does not allow the states of $0\ua$ level to mix with 
those of $0\da$ or $1\ua$ but allows a mixing of the latter two.
\begin{align}
  \langle c^{\dagger}_{0 k \da} c_{0 k \ua} \rangle &= 0 \\
  \langle c^{\dagger}_{1 k \ua} c_{0 k \ua} \rangle &= 0 \\ 
  \langle c^{\dagger}_{1 k \ua} c_{0 k \da} \rangle &= V_{2k} V_{1k} e^{i (\theta_{1k} - \theta_{2k} ) } 
\end{align}
Furthermore since the particle number is not conserved we have,
\begin{align}
  \langle c^{\dagger}_{0 k \ua} \rangle &= U_k V_{0k} e^{-i \theta_{0k}} \\
  \langle c^{\dagger}_{0 k \da} \rangle &= 
  V_{0k} V_{1k} e^{i (\theta_{0k} - \theta_{1k})} - V_{2k} V_{3k} e^{i (\theta_{2k} - \theta_{3k})} \\
  \langle c^{\dagger}_{1 k \ua} \rangle &= 
  V_{0k} V_{2k} e^{i (\theta_{0k} - \theta_{2k})} + V_{1k} V_{3k} e^{i (\theta_{1k} - \theta_{3k})} 
\end{align}

Although $|\psi \rangle$ is a product state and does not couple
different guiding centers directly, the non-conservation of particle
number produces non-zero averages for operators of the form
$O_{n_1n_2;s_1s_2}(k,q_x)=c^{\dagger}_{n_1 k s_1} c_{n_2 k+q_x s_2}$ for $q_x \neq 0$. 

Our variational states therefore inherently violate translation
invariance in the $x$-direction (along the edge).  While one cannot
avoid this, one may choose states where the {\it electron density} is
(almost) uniform.  Recalling that the electron density
$\rho_e(q_x,q_y)$ is a superposition of the operators
$O_{n_1n_2;ss}(k,q_x)$ (see Eq.~\ref{edensity}), we can arrange for
values of $\theta_k$ to make $\langle\psi|
\rho_e(q_x,q_y)|\psi\rangle$ vanish. We achieve this by imposing
the conditions $\theta_{1k} + \theta_{2k} = 0 = \theta_{0k} +
\theta_{3k}$ at each $k$ and $ \theta_{0k} = k^2/(N_p \ell^2) =
\frac{1}{2} \theta_{1k} $ where $N_p$ is a large prime number.  There
remain 4 variational parameters ($V_{ik}$) for each guiding center.
To find the ground state, we need to minimise the energy functional
$\mathcal{F}[\{V_{ik}\}]$,
 
\begin{align*}
\mathcal{F}[\{V_{ik}\}] = \langle \psi | H | \psi \rangle + \lambda (\langle \psi | N | \psi \rangle - N_e)^2
\end{align*}
where we have added a Lagrange multiplier $\lambda$ to fix the average particle number to be 
$N_e$. 

Unfortunately, with a large number of guiding centers, this is still
computationally prohibitive.  To make further progress, we assume
that spin is a good quantum number so that $0\da$ and $1\ua$ 
do not mix and that the occupations in the Landau levels
vary smoothly, except, possibly, for a discontinuous jump at the
chemical potential. The matrix elements are assumed to have the
functional form

\begin{widetext}
\begin{align}
 \langle c^{\dagger}_{1 k \ua} c_{0 k \da} \rangle &= 0 \label{ofd} \\
  \langle c^{\dagger}_{0 k \ua} c_{0 k \ua} \rangle &= \begin{cases}
      \frac{1}{2}\left[1 - \tanh \left(\alpha_1(k - K_{1} - \beta_1) \right) \right] & \text{if } k \leq K_{1} \\ 
      \frac{1}{2}\left[1 - \tanh \left(\alpha_1(k - K_{1} + \beta_1) \right) \right] & \text{if } k > K_{1}
    \end{cases} \\
  \langle c^{\dagger}_{0 k \da} c_{0 k \da} \rangle &= \begin{cases}
  \frac{1}{2}\left[1 - \tanh \left(\alpha_2(k - K_{2} - \beta_2) \right) \right] & \text{if } k \leq K_{2} \\ 
  \frac{1}{2}\left[1 - \tanh \left(\alpha_2(k - K_{2} + \beta_2) \right) \right] & \text{if } k > K_{2}
\end{cases} \\
  \langle c^{\dagger}_{1 k \ua} c_{1 k \ua} \rangle &= \begin{cases}
      \frac{1}{2}\left[1 - \tanh \left(\alpha_3(k - K_{3} - \beta_3) \right) \right] & \text{if } k \leq K_{3} \\ 
      \frac{1}{2}\left[1 - \tanh \left(\alpha_3(k - K_{3} + \beta_3) \right) \right] & \text{if } k > K_{3}
    \end{cases},
\end{align}
\end{widetext}

\noindent
where $\alpha_i, \beta_i$ and $K_{i} (i = 1,2,3)$ are 9 variational
parameters.  Here the bulk state is always $|0\ua,0\da,1\ua \rangle$ 
but the order of edge modes can change. The
variational parameters $K_{o i}$ denote the positions where the Landau
levels cross the chemical potential, the parameters $\beta_i$
characterize the discontinuity in the occupation of level $i$ at
$\mu$, while $\alpha_i$ allow a smooth relaxation back to $0$ or $1$
away from the discontinuity.

\noindent
To satisfy the constraint in Eq.~(\ref{ofd}) we must have,
\begin{align*}
  V_{2k} &= 0   \text{ if } K_{3} < K_{2}\\
  V_{1k} &= 0   \text{ if } K_{3} > K_{2}
\end{align*}
Then we can replace the remaining $V_{ik}$ in the functional $\mathcal{F}$ with the 
functional forms defined above. The new functional $\mathcal{F}[\{\alpha,\beta,K\}]$ 
is then minimised using the method of steepest descent.

In the absence of exchange interaction, the energy functional is
dominated by the classical Coulomb interaction between electron
density and background density. Therefore, if we vary the strengths of
the direct ($E_{cd}$) and exchange ($E_{cx}$) terms independently and
set $E_{cx} = 0$, the occupations can be expected to change very
smoothly from 1 to 0 so as to make the electron density completely
cancel the background density.  Indeed we find that in this regime,
our variational calculation gives $\beta_i = 0$, implying a continuous
variation of occupations, as shown in
Fig.~\ref{fig:FigVariational}. As the exchange interaction is
increased in strength, the smoothly varying $n(k)$ revert back to the
HF solution with $n(k)=0, 1$. As explained in the main text,
mode-switching is an exchange-driven effect and only occurs if the
exchange interaction is larger than a certain minimum fraction of the
direct interaction. We find that in the regime where HF shows
spin-mode-switching, the variational calculation always gives the HF state
as the ground state, whereas in the regime where the occupations vary
smoothly, there is no spin-mode-switching.

%------ Fig 11 ------
\begin{figure*}[ht]
\includegraphics[width=0.33\textwidth]{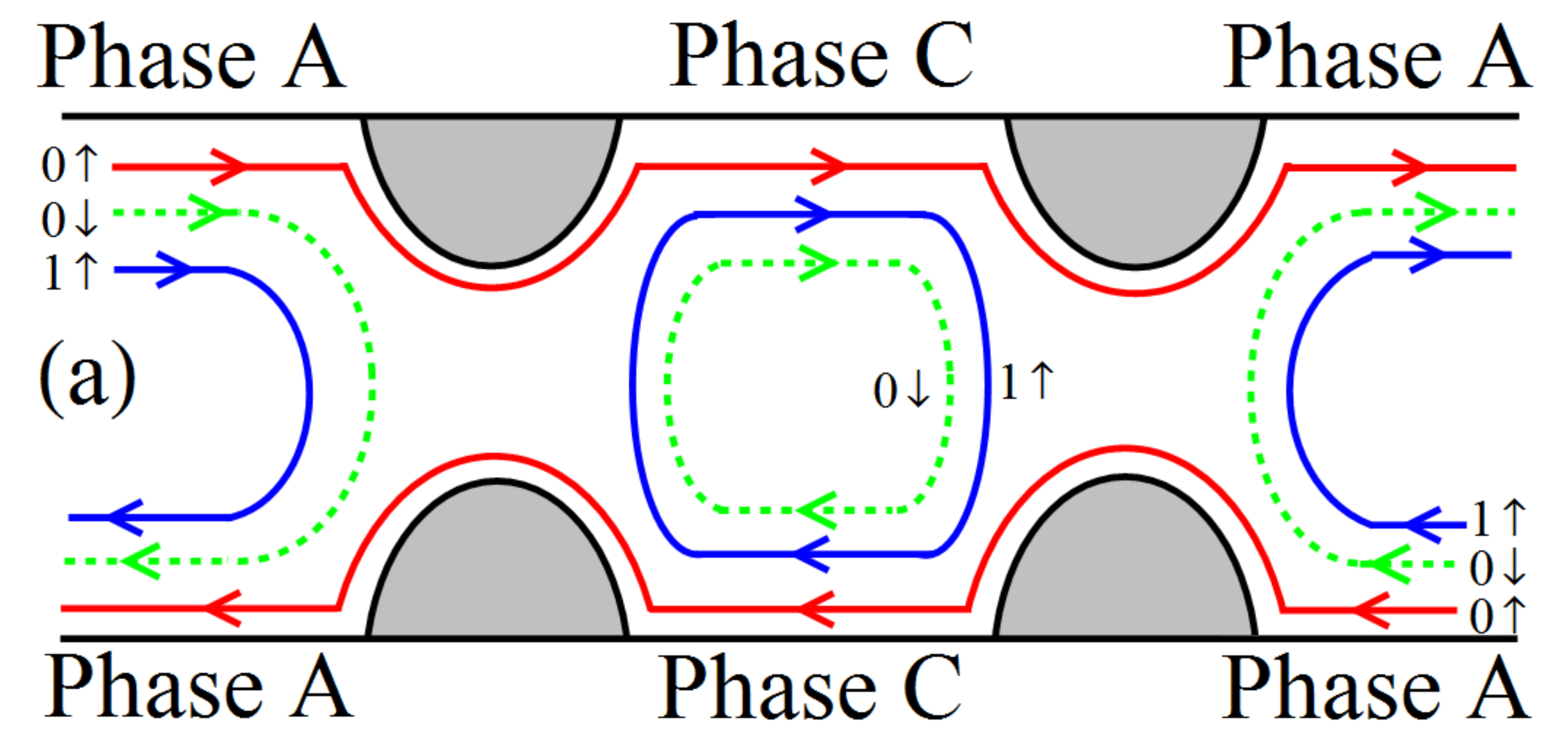}
\includegraphics[width=0.33\textwidth]{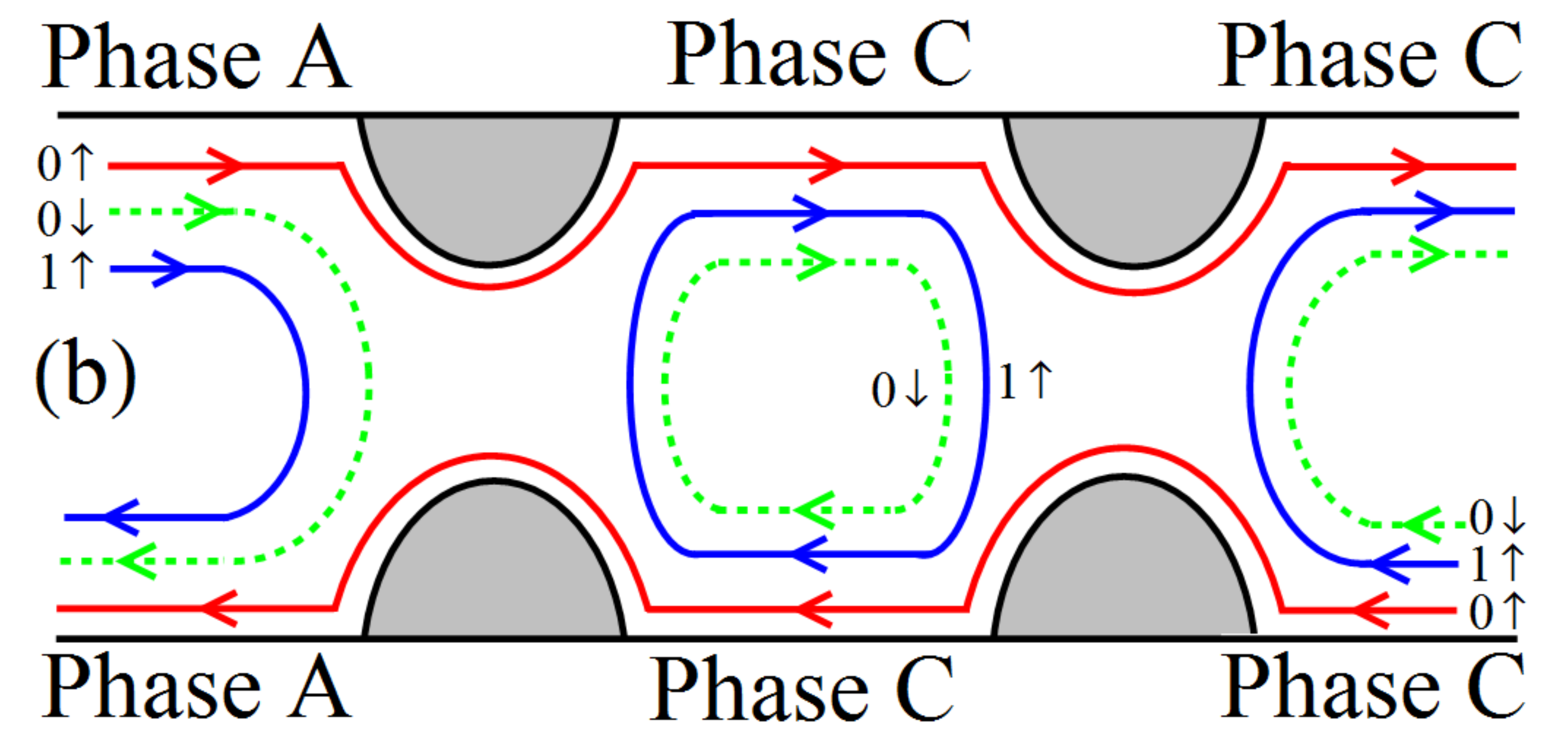}
\includegraphics[width=0.33\textwidth]{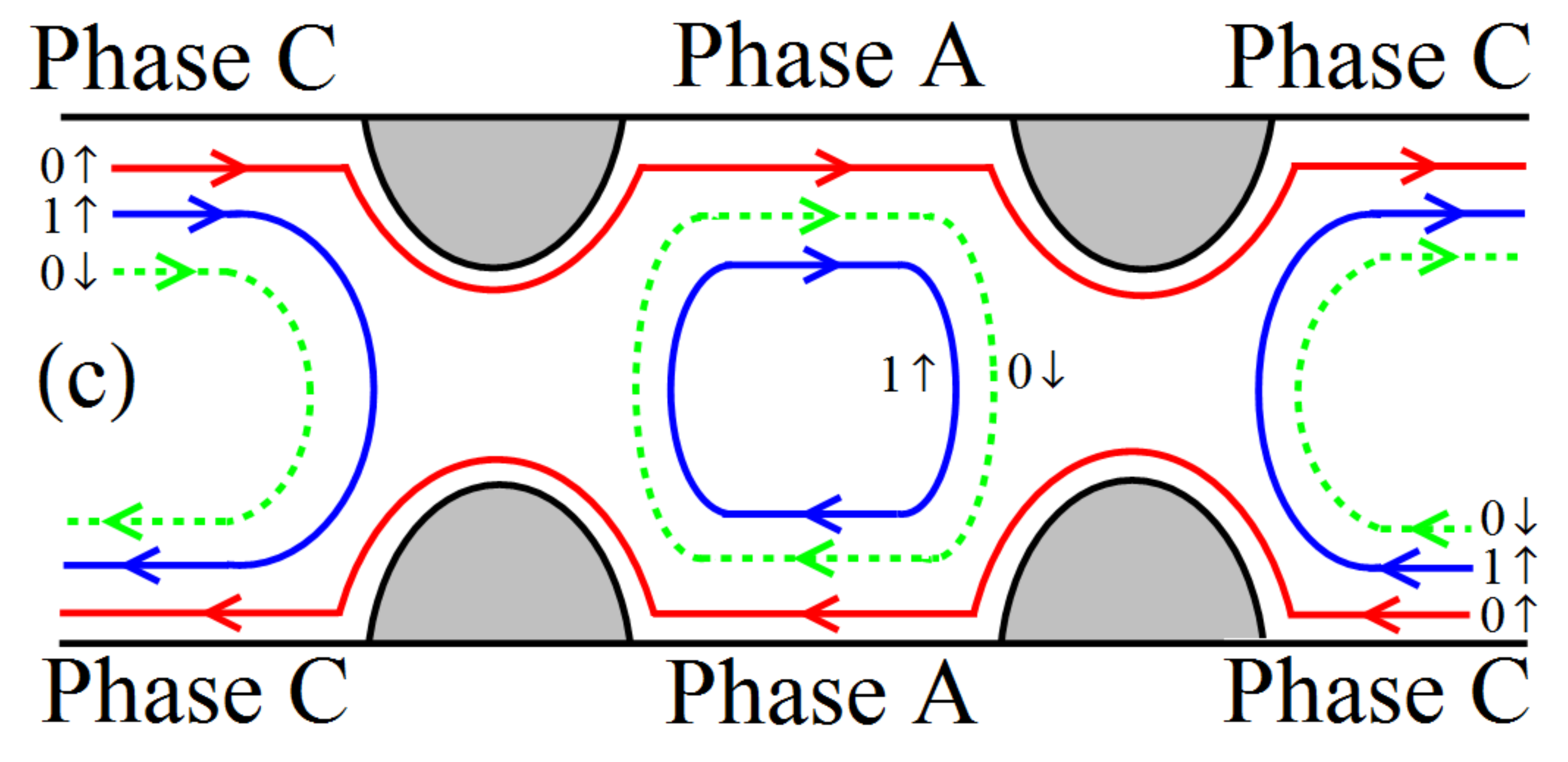}
\includegraphics[width=0.33\textwidth]{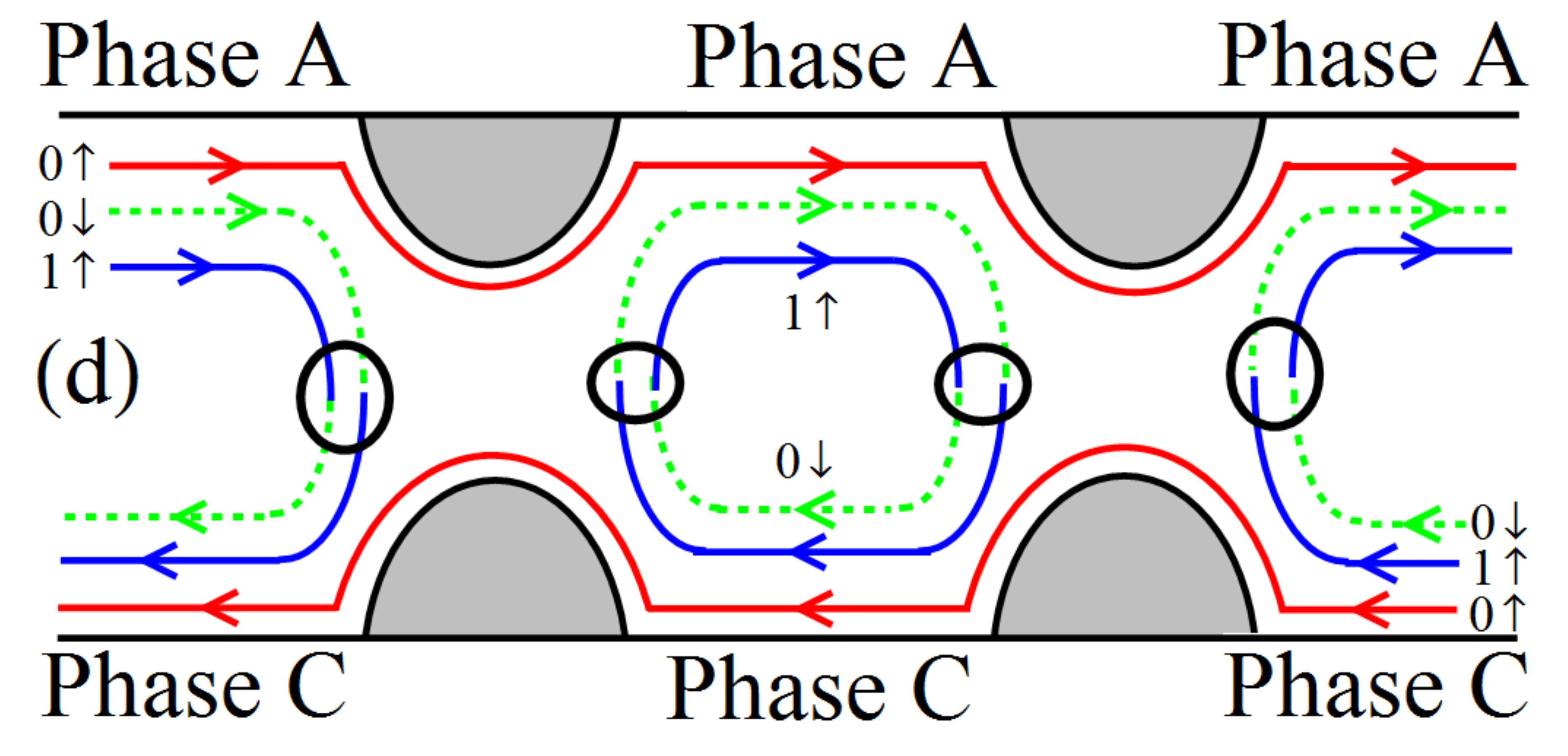}
\caption{(Color online) A two-QPC setup tuned at the $g_2=1$ plateau
  connecting regions with different confining potentials. Here
  $\tEc>2.13$, which implies that the smooth edges are in Phase C. 
  All 3 incoming modes on the left or right section are biased while in
  the middle section only the outer ($0\ua$) mode is
  biased. Following the reasoning presented in the text, we see that
  in (a,b) disorder induced backscattering is possible, while in (c,d)
  edge-to-edge backscattering cannot take place. }
\label{fig:FigS11}
\end{figure*}
%------------------

%------ Fig 12 ------
\begin{figure*}[ht]
\includegraphics[width=0.33\textwidth]{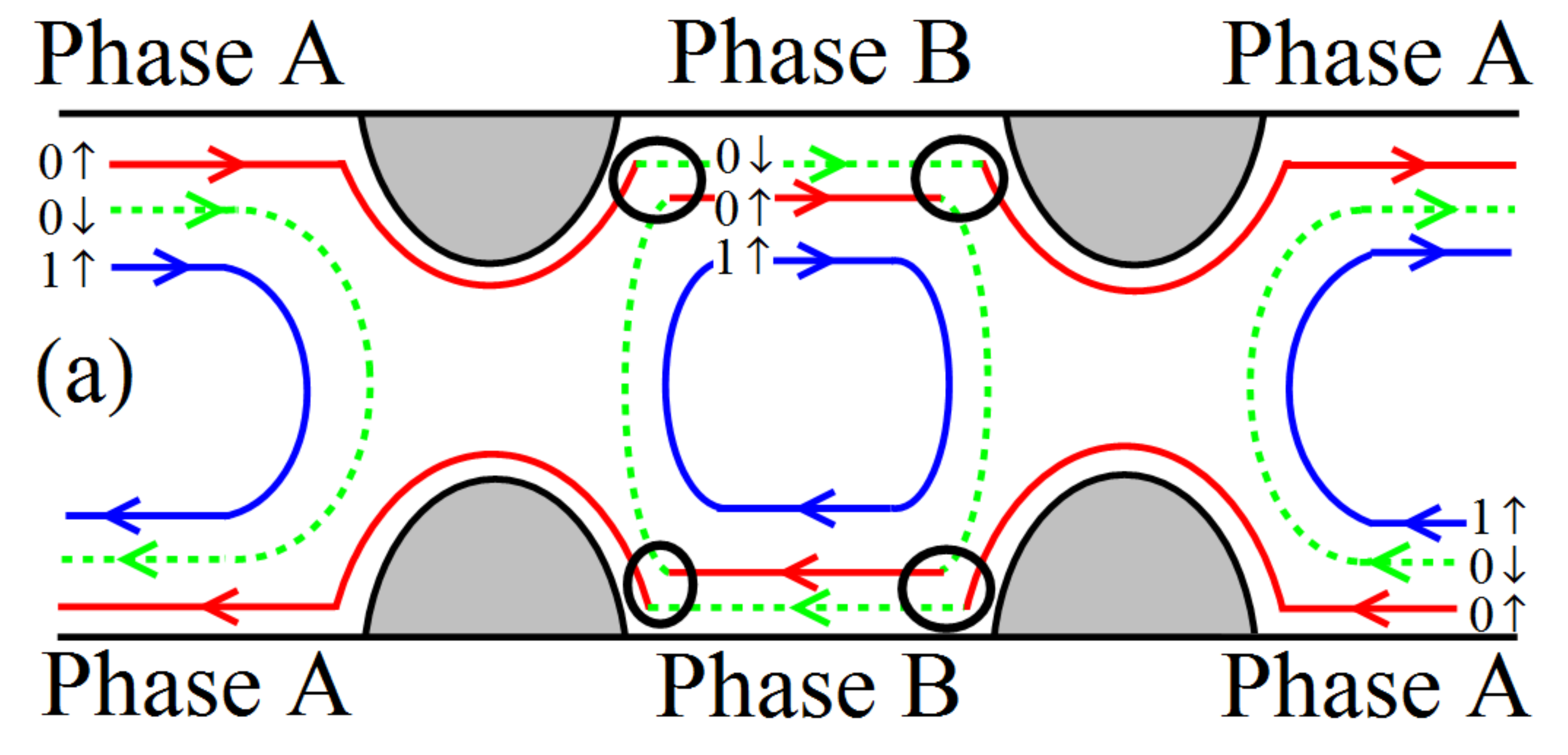}
\includegraphics[width=0.33\textwidth]{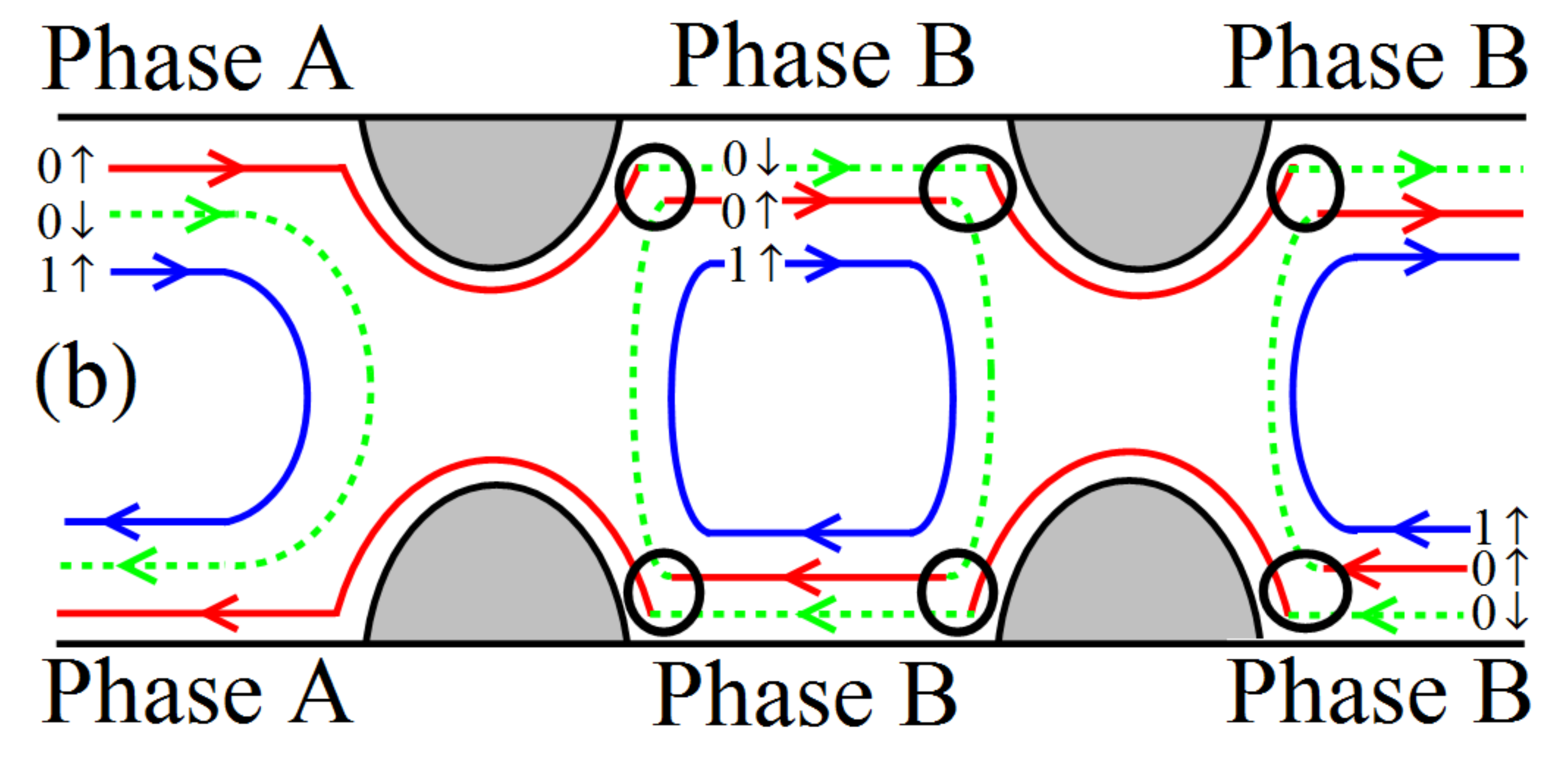}
\includegraphics[width=0.33\textwidth]{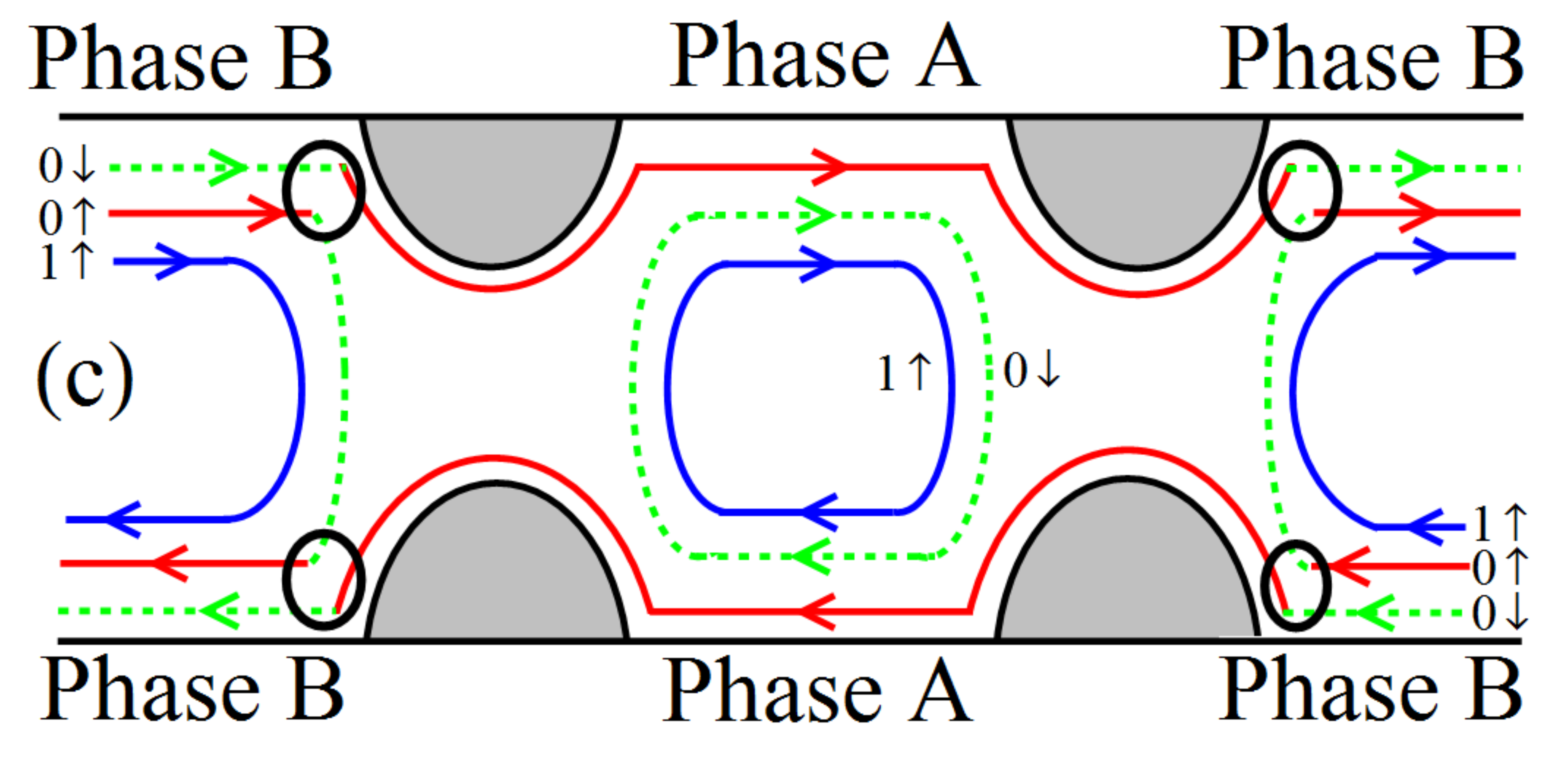}
\includegraphics[width=0.33\textwidth]{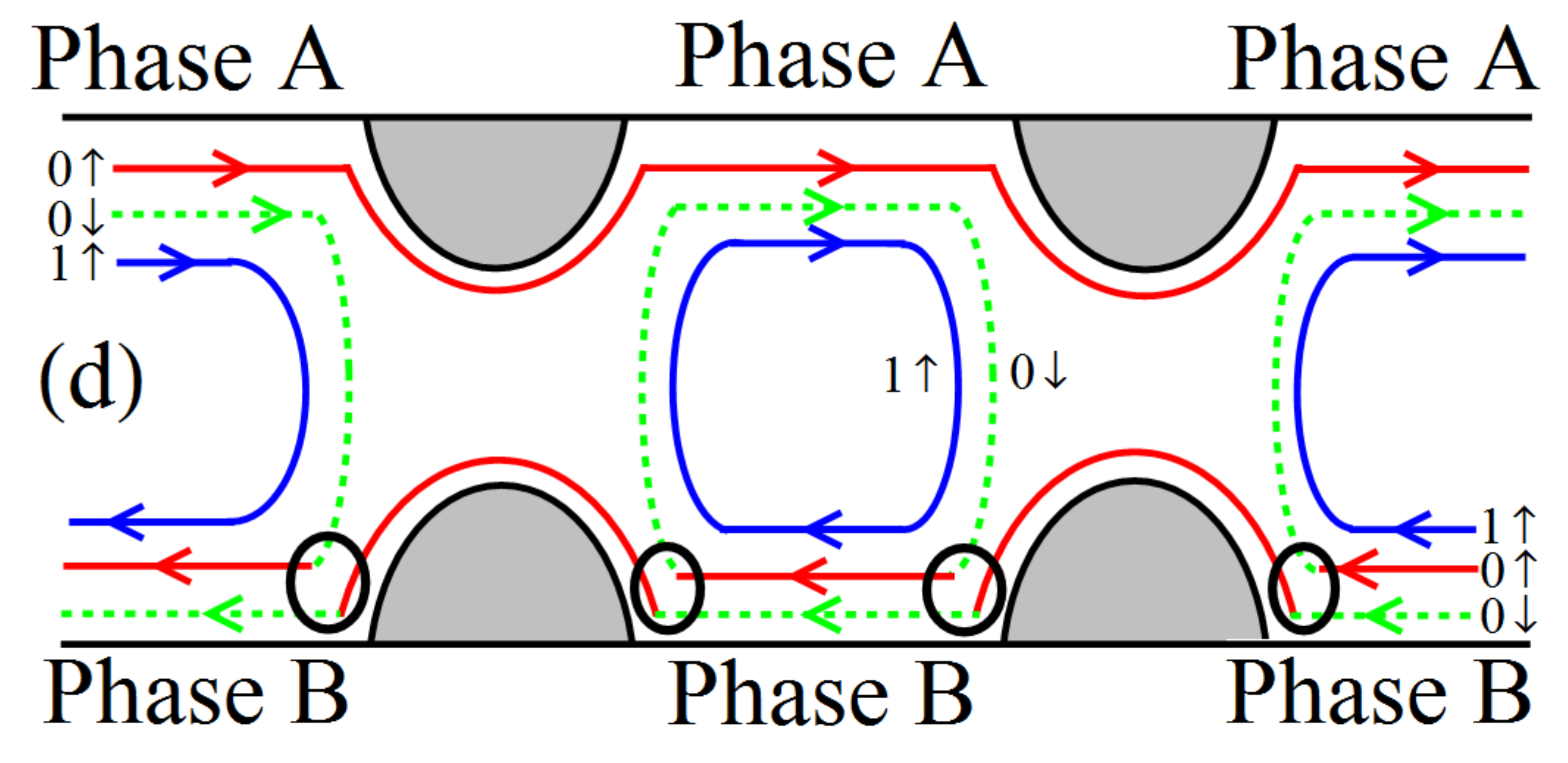}
\caption{(Color online) A two-QPC setup tuned at the $g_2=1$ plateau
  connecting regions with different confining potentials. Here
  $\tEc<2.13$, implying that the smooth edges are in Phase B. In the
  middle section only the ($0\ua$) mode is biased. (a,b)
  Disorder induced backscattering is possible. (c,d) Edge-to-edge
  backscattering cannot take place. }
\label{fig:FigS12}
\end{figure*}
%------------------

%------ Fig 13 ------
\begin{figure*}[ht]
\includegraphics[width=0.33\textwidth]{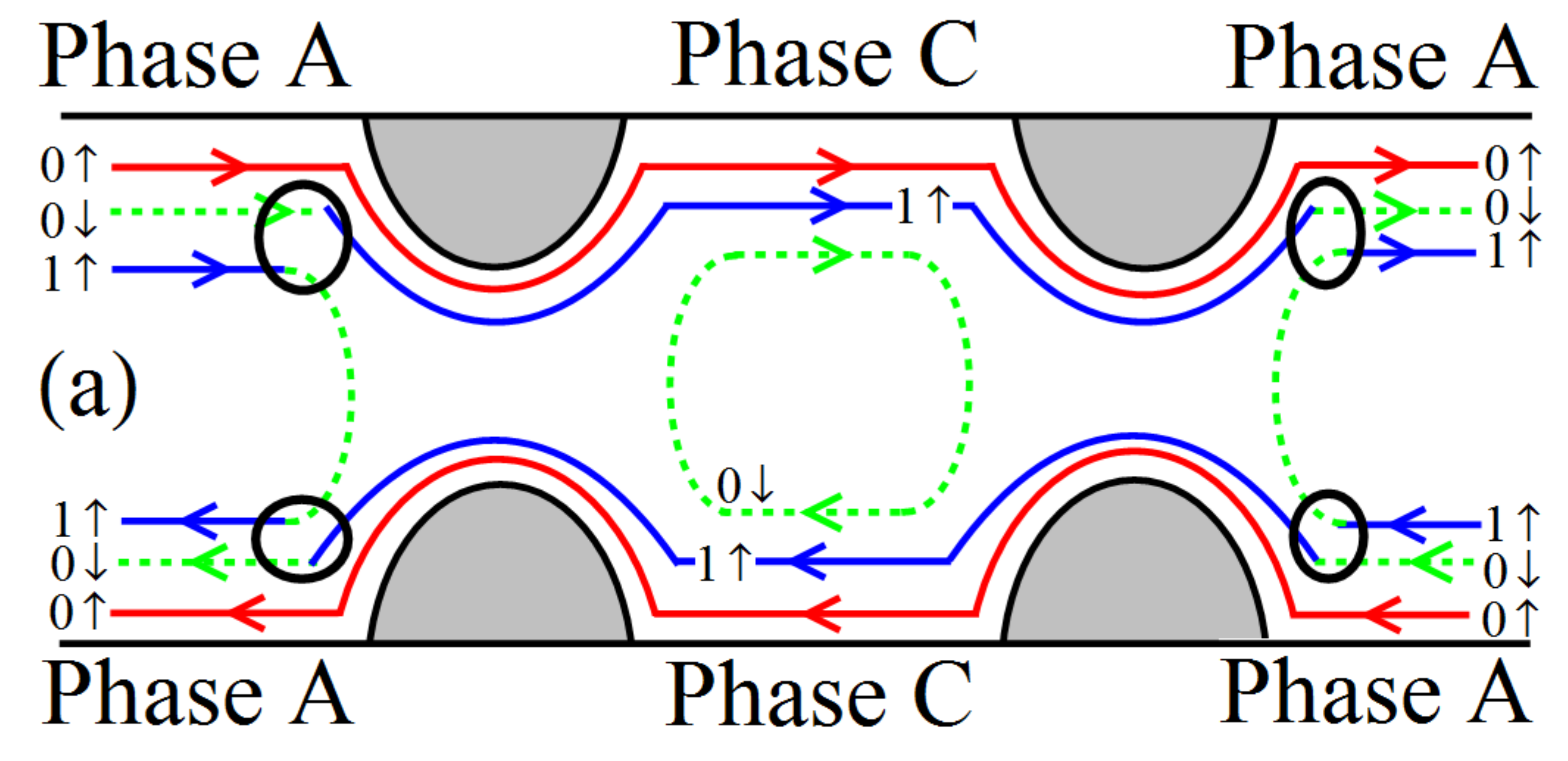}
\includegraphics[width=0.33\textwidth]{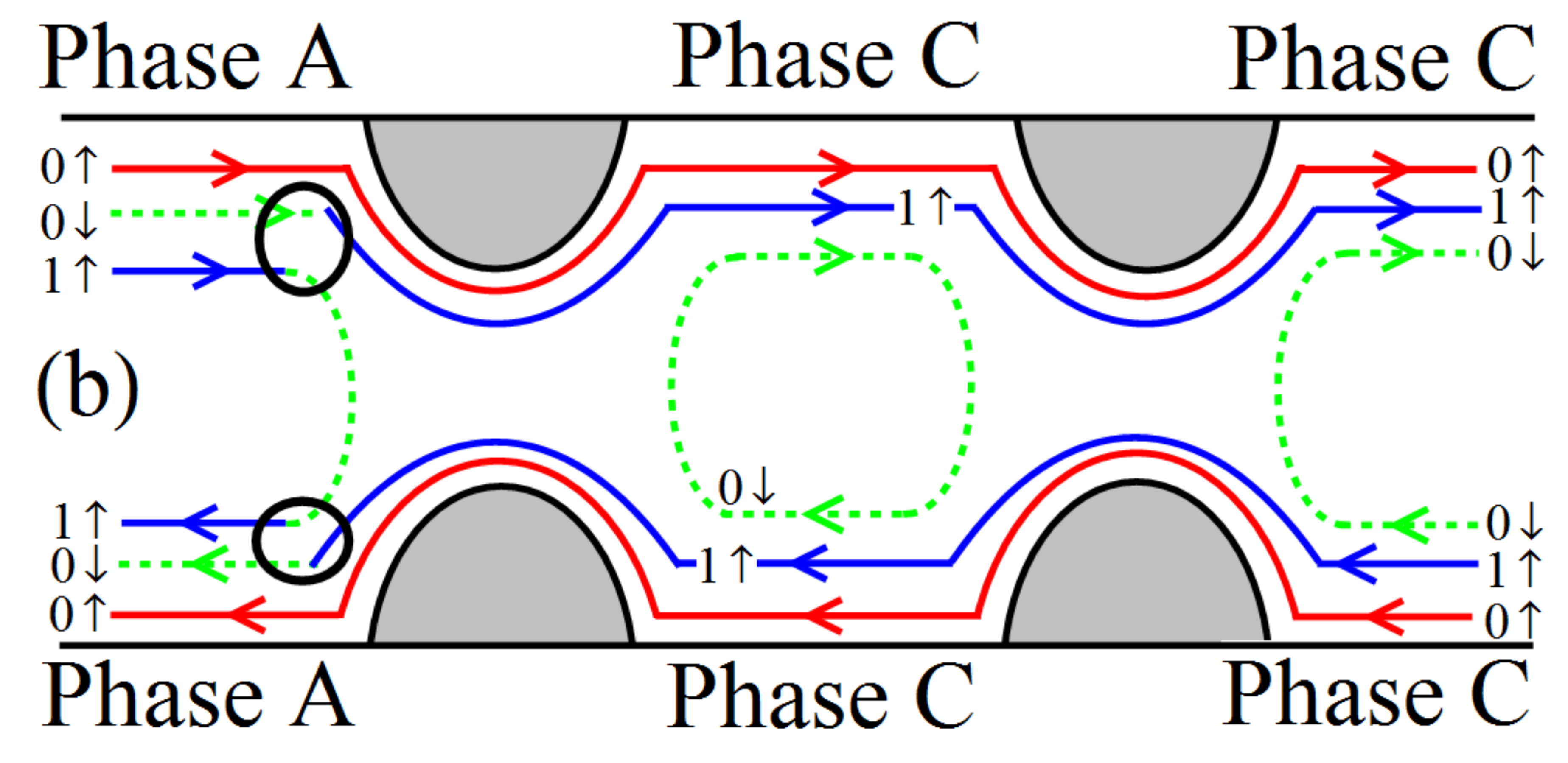}
\includegraphics[width=0.33\textwidth]{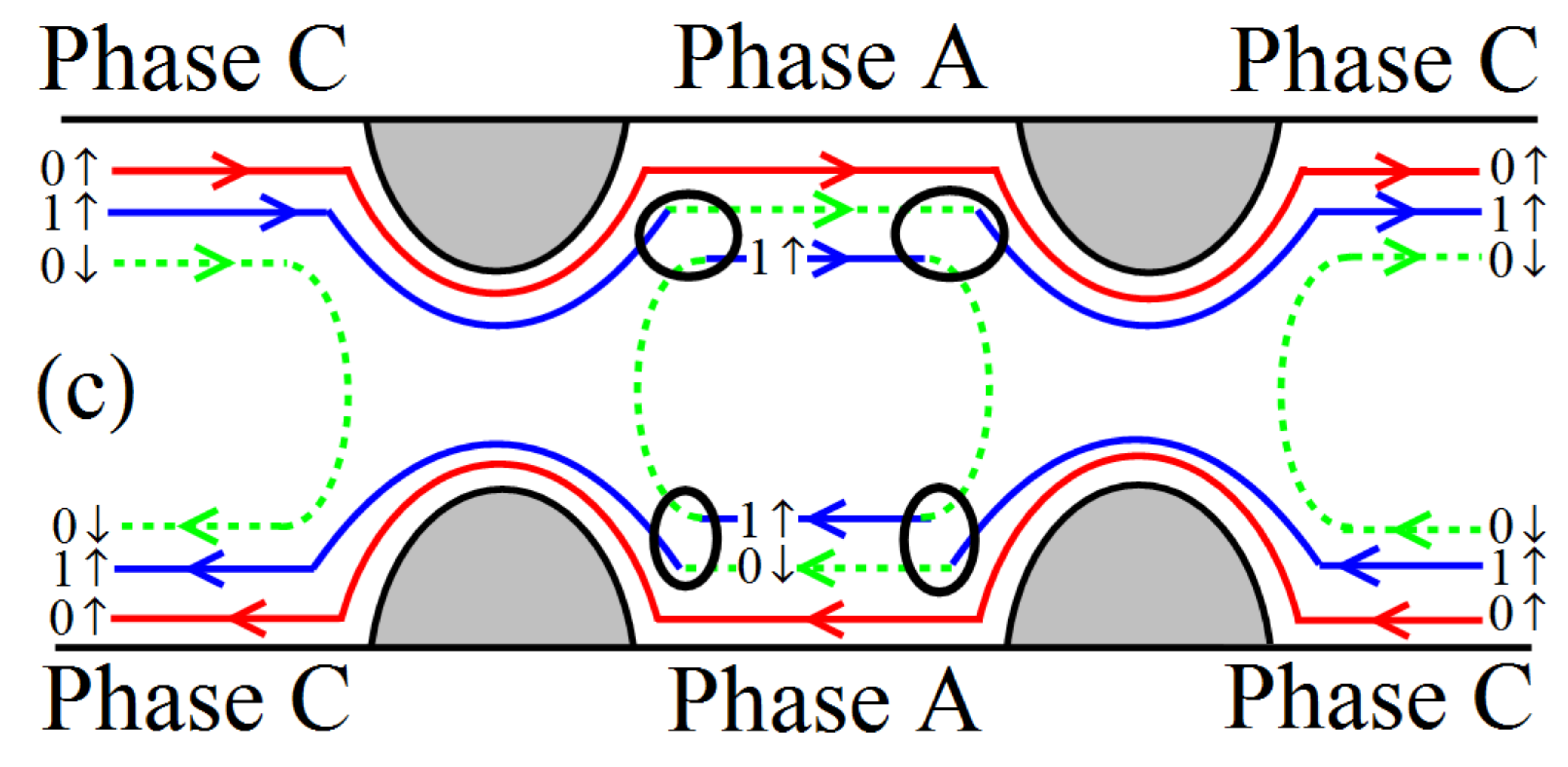}
\includegraphics[width=0.33\textwidth]{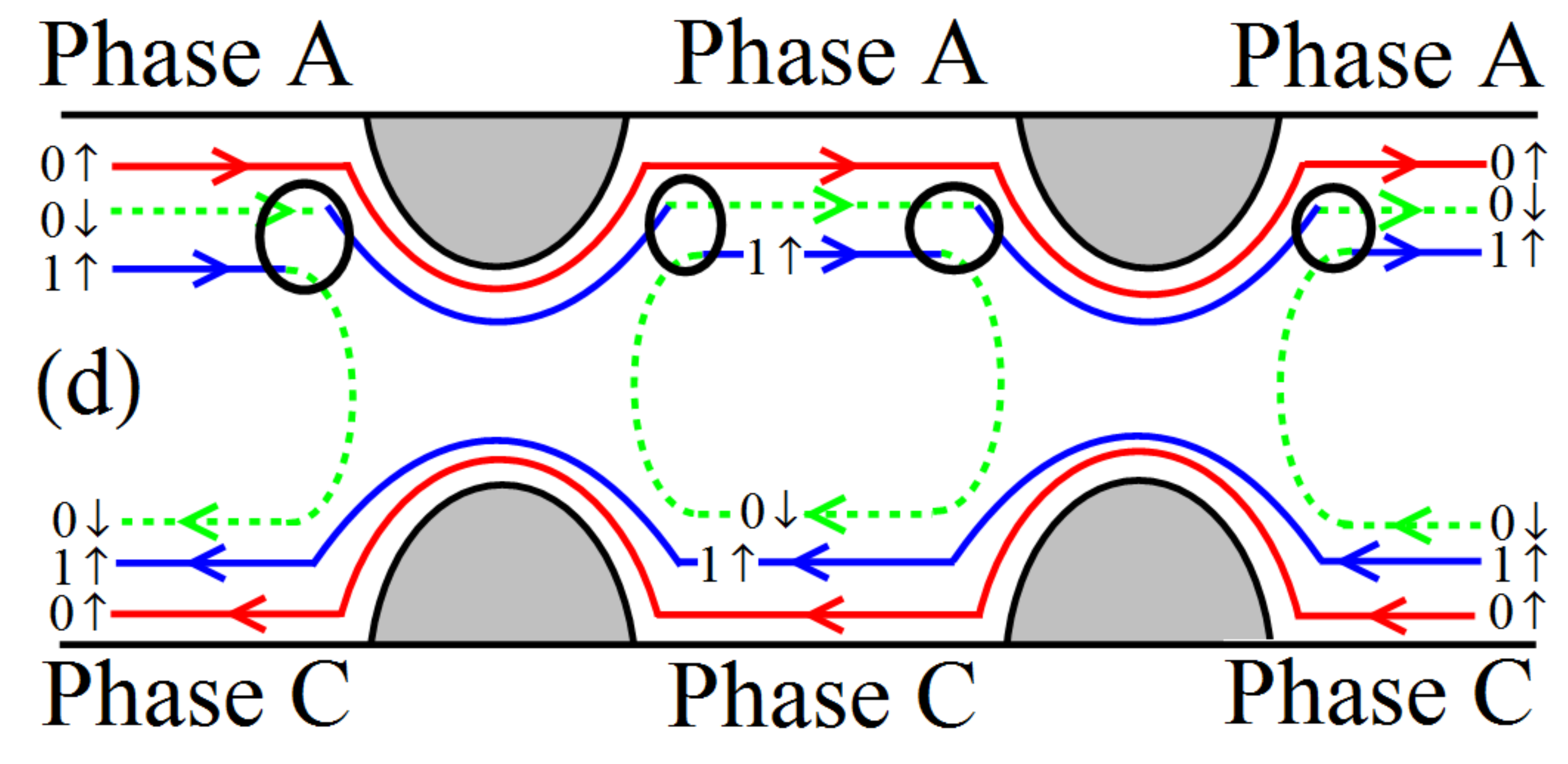}
\caption{(Color online) The same potential configuration as in
  Fig.~\ref{fig:FigS11}(a), at the $g_2=2$ plateau. Here $\tEc>2.13$,
  implying that the smooth edges are in Phase C, and the QPC regions
  are fully polarized. The reasoning explained in the text indicates
  that disorder-induced backscattering cannot take place.  }
\label{fig:FigS13}
\end{figure*}
%------------------

%------ Fig 14 ------
\begin{figure*}[ht]
\includegraphics[width=0.33\textwidth]{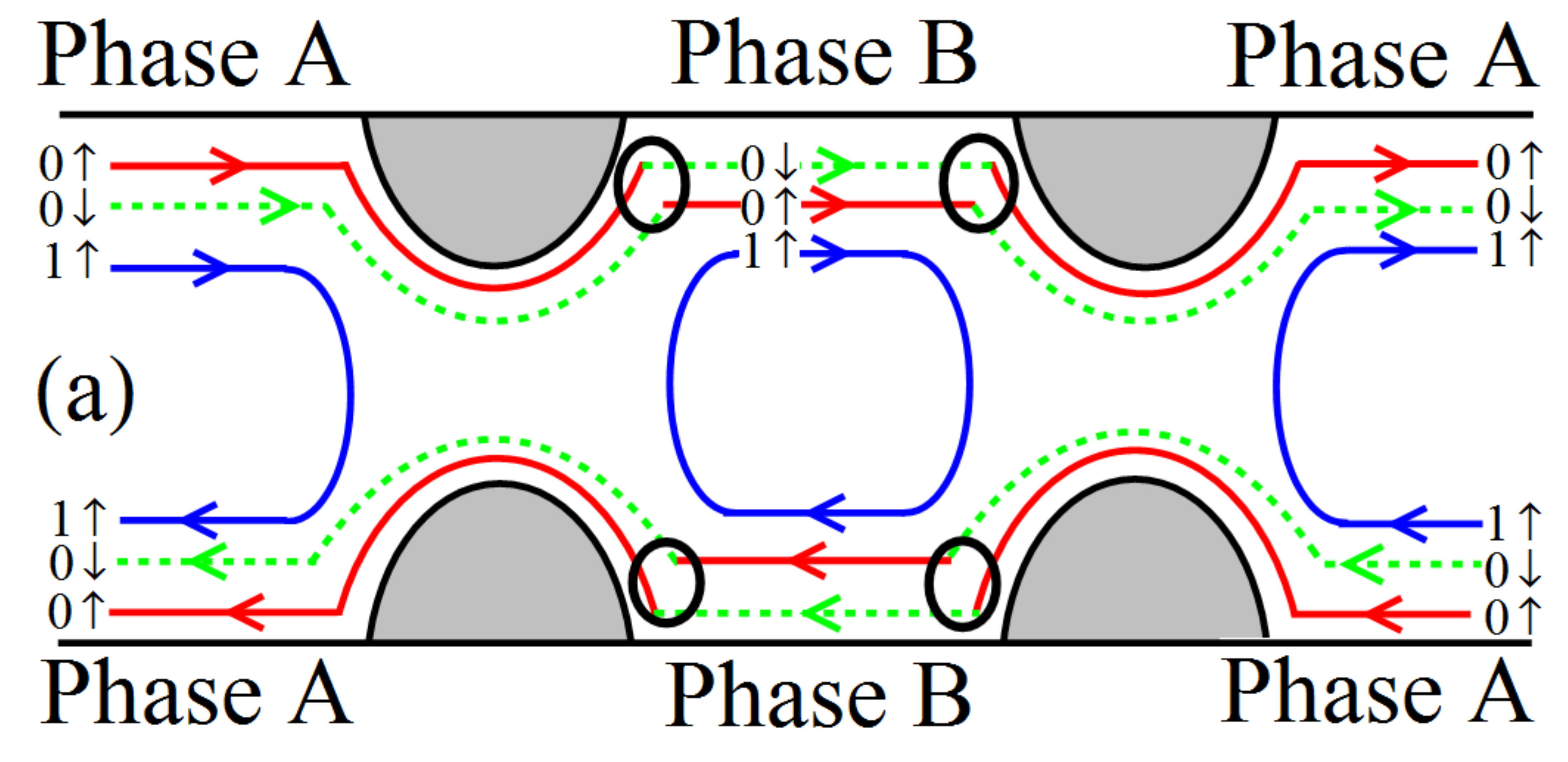}
\includegraphics[width=0.33\textwidth]{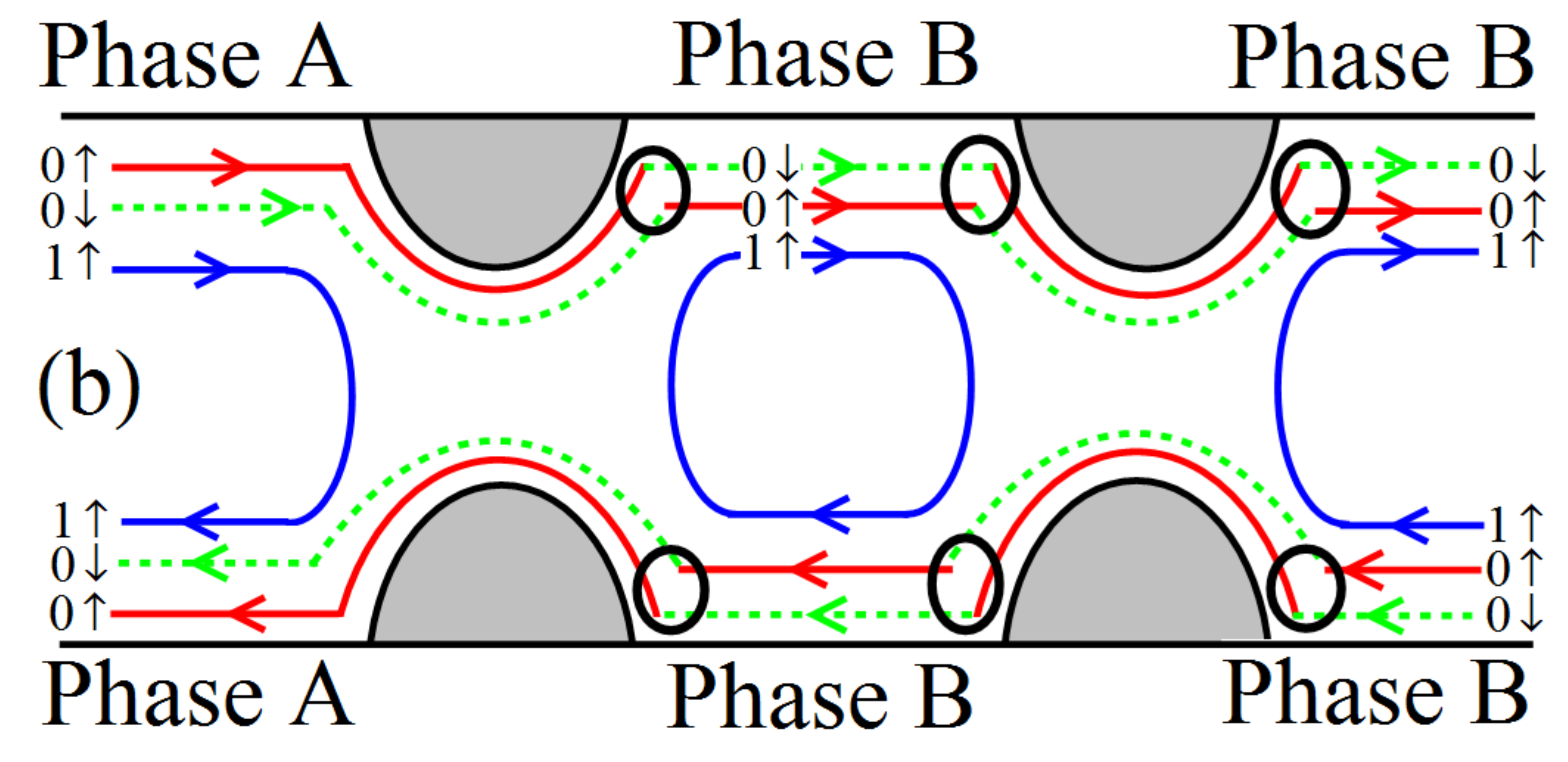}
\includegraphics[width=0.33\textwidth]{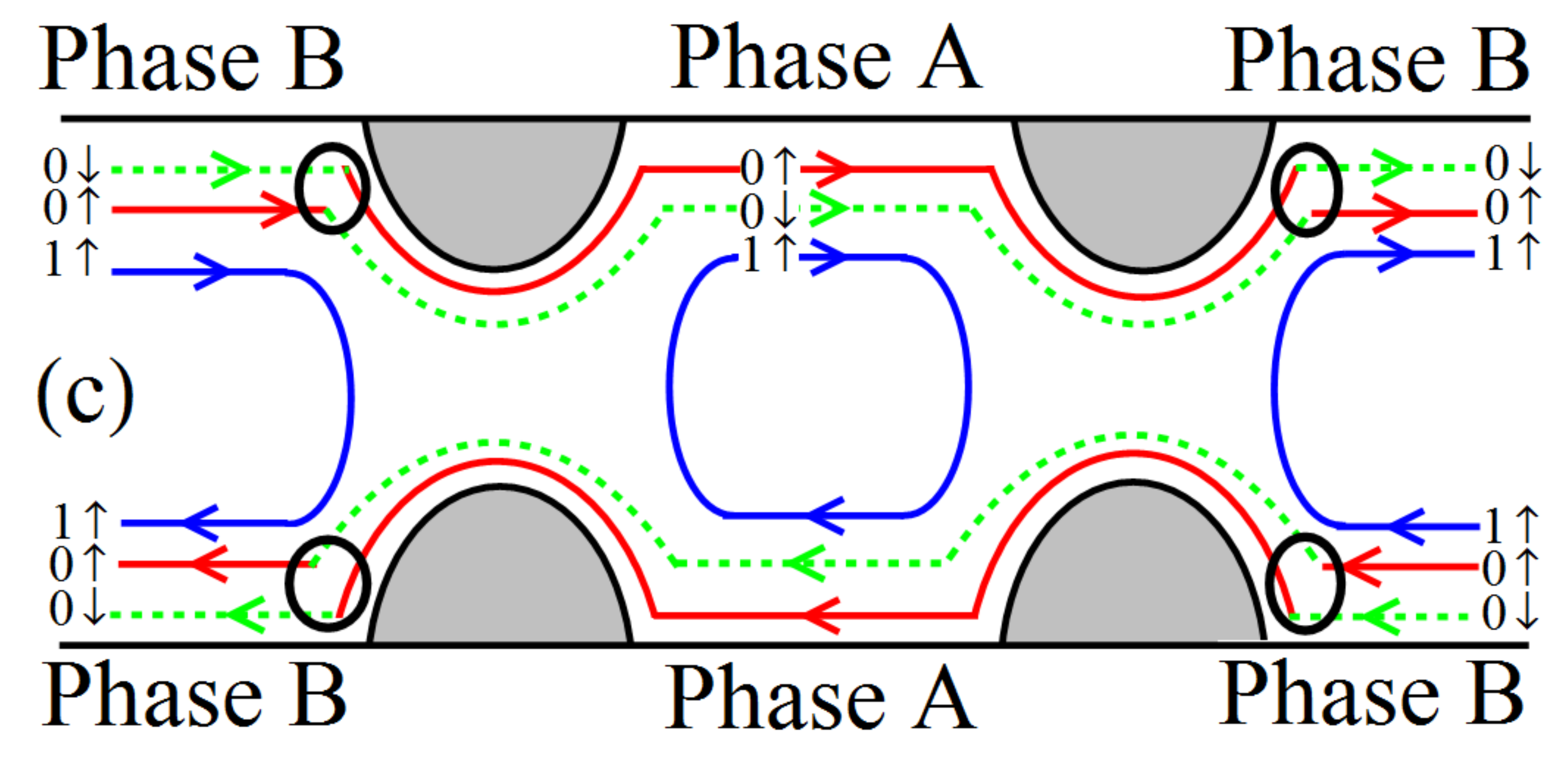}
\includegraphics[width=0.33\textwidth]{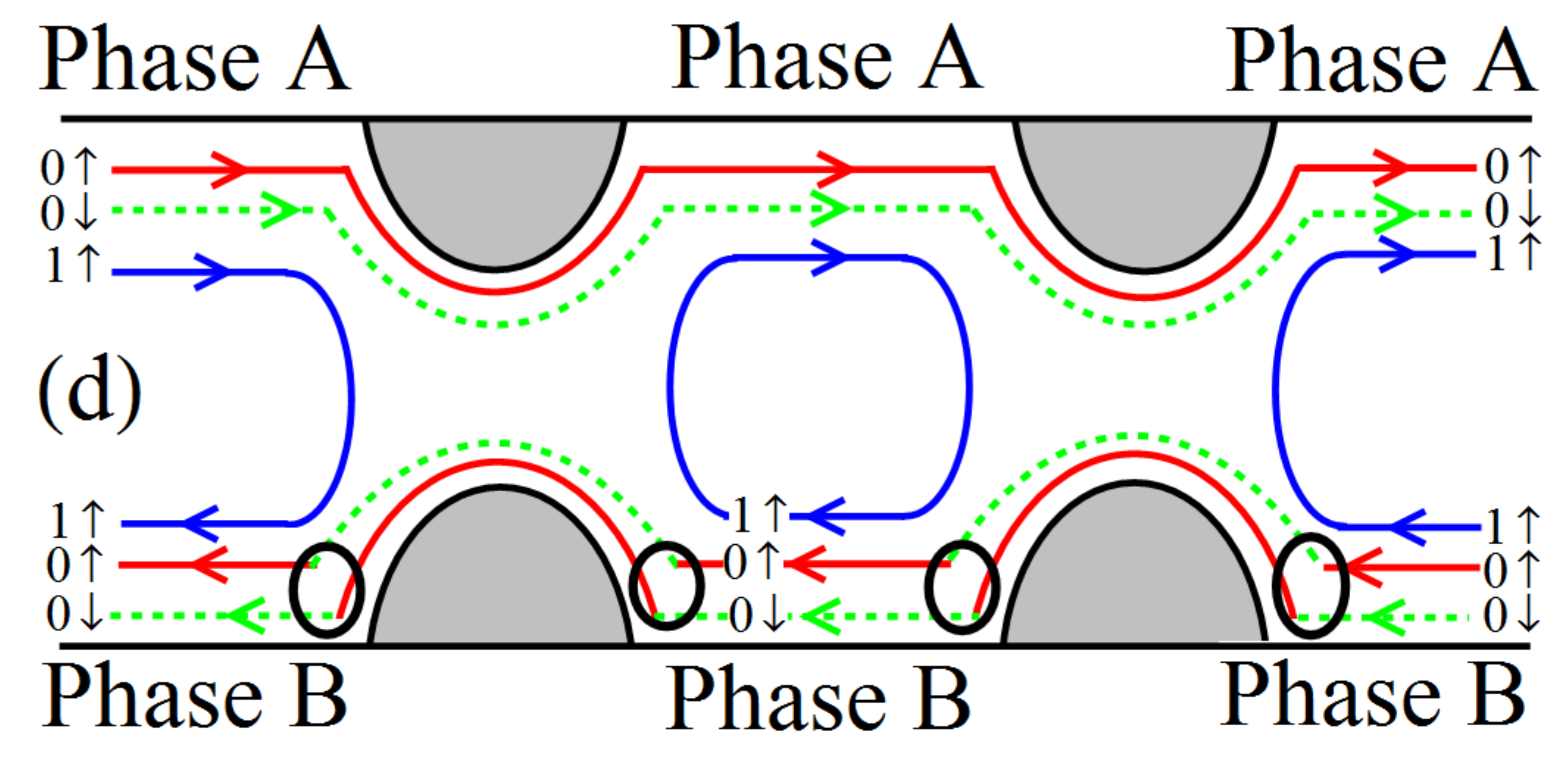}
\caption{(Color online) The same potential configurations as in
  Fig.~\ref{fig:FigS12}, at the $g_2=2$ plateau. Here $\tEc<2.13$,
  implying that the smooth edges are in Phase B, and that the QPC
  regions are unpolarized. The reasoning explained in the text
  indicates that disorder-induced backscattering can take place in the setups 
  shown in (a,b).  }
\label{fig:FigS14}
\end{figure*}
%------------------

\section{IV.\,\,\,\,\,\,    Experimental Signatures}

Edge modes play a crucial role in transport in quantum Hall states. It
is therefore expected that signatures of spin-mode-switching will be
manifested in transport experiments.  While the total Hall conductance
of the $\nu = 3$ state is fixed by the bulk topology, signatures of
mode-switching can be expected to show up in constrained geometries
with quantum point contacts (QPCs) when only some of the edge modes
are allowed to go through. Here again there are two classes of
signatures, those that require measuring the spin-polarization of the
current, and those that do not. 

The spin-dependent signatures have been discussed sufficiently in the
main text, so we will focus here on signatures of spin-mode-switching
that depend on the effects of disorder in one- and two-QPC setups. As
mentioned in the main text, nonmagnetic disorder allows tunneling
between neighboring chiral channels of the same spin. We make the
simplifying assumption that if two $\ua$ channels separated in space
have a $\da$ chiral channel between them, no tunneling occurs between
the $\ua$ channels.

We will show that disorder can induce a left-right asymmetry in the
plateau of the tunneling conductance in a single-QPC geometry when
spin-mode-switching has occurred on (say) the right side of the QPC,
but not the left. The asymmetry pertains to the current source being
either in edge Phase A (not mode-switched), or in the edge spin-mode
switched phase.

Consider a Hall bar with a single QPC connecting regions of the sample
below (Phase A) and above (Phase B/C) the switching transition as shown in
Fig.~\ref{fig:FigS7} - \ref{fig:FigS10}. When the QPC is open (no constriction), the
2-terminal conductance is $g_2 = 3$ (in units of $ \frac{e^2}{h}$).
Let us now pinch off the QPC potential, and consider first $g_2 = 1$
configurations, shown in Fig.~\ref{fig:FigS7} and \ref{fig:FigS8},
which depict the cases $\tEc>2.13$ and $\tEc<2.13$ respectively. For some of these
configurations spin hybridization at the QPC is required (it is marked
by a circle): there is no spin hybridization in
Fig.~\ref{fig:FigS7}(a), a single one in Fig.~\ref{fig:FigS7}(d),~\ref{fig:FigS8}(d), and
two in Fig.~\ref{fig:FigS7}(b),~\ref{fig:FigS7}(c),~\ref{fig:FigS8}(a)-(c) each. Let us now
introduce weak (non-magnetic) static disorder which allows for
tunneling between neighbouring same-spin edge modes. We then voltage
bias, for each configuration, the outermost incoming mode, 
either on the top left or on the bottom right. For the
setup of Fig.~\ref{fig:FigS7}(a), when the bias put on the left-moving
chiral $(0\ua)$ (lower right), the current can partially tunnel
to the neighbouring $(1\ua)$ mode and eventually be
backscattered, resulting in degradation of the quantized $g_2 = 1$
value. This will not be the case if the bias is put on the
right-moving $(0\ua)$ mode (upper left). The result is a
disorder-induced breaking of left-right symmetry of electric
transport. That will apply to the setup depicted in
Fig.~\ref{fig:FigS7}(b) as well, but not Fig.~\ref{fig:FigS7}(c) and
Fig.~\ref{fig:FigS7}(d). On the other hand, for all the 
setups in Fig.~\ref{fig:FigS8}, the outermost and middle incoming modes
have opposite spins. Therefore, disorder induced tunneling between like spin modes is not possible 
in this case and the quantized $g_2 = 1$ value will not suffer from presence 
of disorder.

By somewhat opening the QPC, we may tune the system to be at the $g_2
= 2$ plateau. First we consider the case when $\tEc>2.13$ (depicted in Fig.~\ref{fig:FigS9}), 
which implies that the smooth edges are in Phase C, 
and that the QPC region is fully polarized. 
Such configurations require that spin hybridization takes place at two or 
more different points near the QPC.  Following the same logic as in
the previous paragraph, we see that disorder-generated tunneling
between like-spin channels will not give rise to back-scattering,
regardless of where the source and drain are. This means that
conductance quantization is robust and has no left-right asymmetry.

Still staying with $g_2=2$, we now consider the case when $\tEc<2.13$ 
(depicted in Fig.~\ref{fig:FigS10}), which implies that the smooth 
edges are in Phase B, and that the QPC
region is unpolarized. In this case,
when the source is located in the region where the edge is in Phase A,
no disorder-induced backscattering takes place. However, if the source
is in the region where the edge is in Phase B, disorder-induced
backscattering can degrade the conductance. Thus, depending on the
particular configuration of edges, there can be a left-right asymmetry
in the quality of the conductance plateau.

Next, we analyze a 2-QPC setup, depicted in Fig.~\ref{fig:FigS11}-\ref{fig:FigS14}. 
We consider the case where the QPCs are pinched so as to obtain
a $g_2 = 1$ plateau (Fig.~\ref{fig:FigS11} and \ref{fig:FigS12}). 
In the region between the two QPCs only the
($0\ua$) mode is biased. If the potential on both edges in
the middle section is made smooth, disorder induced tunneling from
($0\ua$) to the ($1\ua$) channel on one edge, and from
($1\ua$) to the ($0\ua$) on the other edge, may take place
(Fig.~\ref{fig:FigS11}(a),(b) and \ref{fig:FigS12}(a),(b)).  
The quality of the $g_2 = 1$ plateau is undermined due 
to this edge-to-edge backscattering. There is no backscattering when either of the middle 
section edges is sharp (Phase A) (Fig.~\ref{fig:FigS11}(c),(d) and \ref{fig:FigS12}(c),(d)). 
For the same configurations, with the QPCs slightly 
more open ($g_2 = 2$, cf. Fig.~\ref{fig:FigS13},~\ref{fig:FigS14}), we note that backscattering will 
only take place in the setups shown in Fig.~\ref{fig:FigS14}(a),(b).


\begin{thebibliography}{235}
\bibitem{footnote-charge} Coulomb interactions have both
  direct and exchange components, so there can be confusion about what
  is meant by ``charge'' versus ``exchange'' effects. Our working
  definition is that when a reorganization of the charge density
  occurs as a consequence of a edge reconstruction it is essentially a
  ``charge'' effect, while any reconstruction without such a
  reorganization of charge density is essentially an ``exchange''
  effect.
\bibitem{footnote-spatial-crossing} Our work so far has focused on
  translation invariant states along the edge. In the situations we
  consider for experimental signatures, the slope of $V_{\text{edge}}$
  changes along the edge. We assume that as long as the changes occur
  over a scale long compared to $\ell$ our analysis remains valid. 
  We also assume that in this region, there is a mechanism which allows the spin to relax 
  to its equilibrium value.
\end{thebibliography}
\end{document}